\documentclass[a4paper,11pt]{article}
\pdfoutput=1 % if your are submitting a pdflatex (i.e. if you have
\usepackage{jheppub} % for details on the use of the package, please
\usepackage[T1]{fontenc} % if needed
\usepackage{tikz}  
\usetikzlibrary{arrows.meta,positioning,calc,decorations.markings,shapes.misc}
\usepackage{tkz-graph}
\usepackage{array,amsfonts,mathtools}
\usepackage{enumerate}
\usepackage{adjustbox}

\usepackage{etoolbox}
\makeatletter
\patchcmd{\l@section}{1.0em}{0.8em}{}{}
\makeatother

\title{Goldstone bosons on celestial sphere and conformal soft theorems}
\author[a]{Karol Kampf,}\emailAdd{karol.kampf@mff.cuni.cz}
\author[a]{Ji\v{r}\'{i} Novotn\'{y},}\emailAdd{jiri.novotny@mff.cuni.cz}
\author[a,b]{Jaroslav Trnka,}\emailAdd{trnka@ucdavis.edu}
\author[a]{Petr Va\v{s}ko}\emailAdd{petr.vasko@mff.cuni.cz}

\affiliation[a]{Institute of Particle and Nuclear Physics, Charles University, V Hole\v{s}ovi\v{c}k\'ach 2,\\Prague, Czech Republic}
\affiliation[b]{Center for Quantum Mathematics and Physics (QMAP) and Department of Physics,\\ University of California, Davis, California, USA}

\abstract{In this paper, we study celestial amplitudes of Goldstone bosons and conformal soft theorems. Motivated by the success of soft bootstrap in momentum space and the important role of the soft limit behavior of tree-level amplitudes, our goal is to extend some of the methods to the celestial sphere. The crucial ingredient of the calculation is the Mellin transformation, which transforms four-dimensional scattering amplitudes to correlation functions of primary operators in the celestial CFT. The soft behavior of the amplitude is then translated to the singularities of the correlator. Only for amplitudes in ``UV completed theories'' (with sufficiently good high energy behavior) the Mellin integration can be properly performed. In all other cases, the celestial amplitude is only defined in a distributional sense with delta functions. We provide many examples of celestial amplitudes in UV-completed models, including linear sigma models and $Z$-theory, which is a certain completion of the $SU(N)$ non-linear sigma model. We also comment on the BCFW-like and soft recursion relations for celestial amplitudes and the extension of soft bootstrap ideas.}

\begin{document} 
\maketitle
\flushbottom

%%%%%%%%%%%%%%%%%%%%%%%%%%%%
\section{Introduction}
%%%%%%%%%%%%%%%%%%%%%%%%%%%%%

In the last two decades, there has been remarkable progress in the study of on-shell scattering amplitudes and the development of efficient computational tools. This has led to many new insights and discoveries completely invisible in the standard formulation of particle physics. This ranges from the efficient on-shell computational methods \cite{Britto:2004ap,Britto:2005fq,Arkani-Hamed:2010zjl,Bern:1994zx,Bern:1994cg,Bern:2005iz,Bern:2009kd,Bern:2008ap,Bern:2012uc,Bern:2018jmv,Carrasco:2021otn,Bourjaily:2016evz,Bourjaily:2017wjl,Henn:2014qga} to the to discovery of hidden symmetries and new profound connections to mathematics \cite{Drummond:2006rz,Drummond:2008vq,Drummond:2009fd,Arkani-Hamed:2009ljj,Arkani-Hamed:2009nll,Arkani-Hamed:2009kmp,Arkani-Hamed:2009pfk,Mason:2009qx,Arkani-Hamed:2012zlh,Arkani-Hamed:2014bca,Franco:2015rma,Bourjaily:2016mnp,Heslop:2016plj,Herrmann:2016qea,Farrow:2017eol,Armstrong:2020ljm,Arkani-Hamed:2013jha,Arkani-Hamed:2013kca,Arkani-Hamed:2017vfh,Damgaard:2019ztj,Ferro:2020ygk,Herrmann:2022nkh,Paranjape:2022ymg,Brown:2022wqr}. In most of these developments, the central objects are on-shell scattering amplitudes in four-dimensional Minkowski spacetime written in momentum space which make the translational invariance manifest. 

Recently, it was proposed to express amplitudes in a boost eigenbasis that was derived within the celestial holography program by the requirement of conformal covariance of the scattering amplitudes viewed as correlators in a putative celestial conformal field theory (CCFT) living on the celestial sphere at asymptotic infinity. If successful, this would be a major progress in the search for flat space holography as one of the most important open challenges in theoretical physics. We refer to these correlation functions as {\it celestial amplitudes} and they possess some very special properties. While the key question about the existence of CCFT is an open question, we have learned in recent years a lot about various properties of celestial amplitudes. This includes the study of collinear and soft limits \cite{Fan:2019emx,Pate:2019mfs,Adamo:2019ipt,Puhm:2019zbl,Guevara:2019ypd,Nandan:2019jas,Guevara:2021abz}, color-kinematics duality and double copy relations \cite{Casali:2020vuy,Campiglia:2021srh,Casali:2020uvr,Monteiro:2022lwm}, the role of $w_{1+\infty}$ structures \cite{Strominger:2021lvk,Himwich:2021dau,Adamo:2021lrv,Ball:2021tmb,Mago:2021wje}, celestial operator product expansion \cite{Pate:2019lpp,Banerjee:2020kaa,Adamo:2022wjo}. Furthermore, the celestial avatar of the dual superconformal symmetry was studied in \cite{Hu:2021lrx} and explicit formulas for $n$-pt MHV amplitudes and integrable 2d $S-$matrices were obtained in \cite{Schreiber:2017jsr} and \cite{Kapec:2022xjw}. In this paper, we focus on a complementary direction and study celestial duals for amplitudes in effective field theories (EFT) for scalar particles and soft theorems. 

Already in the late 50s and middle 60s, Low~\cite{Low:1958sn} and then Weinberg~\cite{Weinberg:1965nx} asked what happens to a scattering amplitude when one null momentum of a massless particle (they considered photons and gravitons) approaches zero. The answer turned out to be remarkably universal, the amplitude factorizes into a theory-independent soft factor parametrized just by conserved charges of the particles involved (e.g. momenta, electric charges, spins and angular momenta), and a lower point amplitude with the soft particle erased. In a particular case of Goldstone bosons, the soft limit of the amplitude in certain cases vanishes, also referred to as \emph{Adler zero}. This property has been used in the context of modern amplitudes methods in the soft recursion relations \cite{Cheung:2015ota,Kampf:2019mcd,Luo:2015tat}. The tree-level amplitudes in a number of quantum field theories can be reconstructed from the residues on poles via BCFW recursion relations \cite{Britto:2004ap,Britto:2005fq} (and generalizations \cite{Risager:2005vk,Cheung:2008dn,Elvang:2008na,Cohen:2010mi,Cheung:2015cba}). This is based on simple properties of locality and unitarity of momentum space amplitudes, which translate into factorization properties. In effective field theories, this procedure can not be used as the amplitudes have additional poles at infinite momenta, which are not known from first principles. However, as shown in \cite{Kampf:2013vha,Cheung:2014dqa,Cheung:2016drk,Cheung:2018oki,Kampf:2019mcd,Kampf:2021bet,Bartsch:2022pyi,Kampf:2020tne,Elvang:2018dco} this lack of information can be traded for the knowledge of the behavior in the soft limit. This then allowed us to fully fix and then reconstruct amplitudes in a variety of EFTs, including the SU(N) non-linear sigma model (NLSM), Dirac-Born-Infeld theory (DBI), vector Born-Infeld theory (BI) and a newly discovered special Galileon. The same theories also appear in the context of scattering equations and CHY formula \cite{Cachazo:2013hca,Cachazo:2013iea,Cachazo:2014xea}, ambitwistor strings \cite{Mason:2013sva,Geyer:2014fka,Casali:2015vta} and the color-kinematics duality\cite{Bern:2008qj,Bern:2010ue,Bern:2019prr}. 

The universality property of soft theorems has long been an inspiration for further studies and searching for symmetry arguments behind the scenes. However, the plane wave basis in which the soft limits were derived was not a well-adapted framework for this task. It took until 2013 when Strominger~\cite{Strominger:2013lka,Strominger:2013jfa} observed a crucial connection between soft theorems and asymptotic symmetries. At that point the celestial holography program was born. This fresh viewpoint allowed to derive soft theorems as consequences of asymptotic symmetries and express them as Ward identities among correlation functions. An immediate prediction of the newly developed formalism was the existence of a sub-sub-leading soft graviton theorem, which was indeed confirmed in~\cite{Cachazo:2014fwa}. It didn't take long to discover that the correlation functions related by ``soft'' Ward identities have many properties appropriate for 2D CFT correlators. Inspired by these findings, a basis transform from translation eigenstates (plane waves) to boost eigenstates that make the conformal properties and ``soft'' Ward identities manifest was proposed soon after in~\cite{Pasterski:2016qvg}.
As a matter of fact, the above-described relation between soft theorems and asymptotic symmetries works best for gauge theories (gravity included as a gauge theory of diffeomorphisms). In that case, soft theorems take a transparent form of Ward identities for Kac--Moody or Virasoro (formed from the stress tensor) currents. These conserved currents and associated charges require the symmetry transformations to depend on the position on the celestial sphere. Such transformations get naturally interpreted as large gauge transformations, which exist only in gauge theories. For this reason, it is fair to say that for scalar theories not embedded into gauge theories, the asymptotic symmetry interpretation of soft theorems is less understood, despite some progress in this direction~\cite{Campiglia:2017dpg,Campiglia:2017xkp,Hamada:2017atr,Campiglia:2018see}. Nevertheless, effective theories (EFTs) of scalars provide a highly intriguing web of soft theorems.

In this paper, we investigate the form of soft theorems for theories of Goldstone bosons in the celestial basis. We don't attempt to make a connection with asymptotic symmetries here, this is left to experts or to future analysis. The aim of this work is to collect a representative selection of soft behaviors of scalar particles, transform it to the celestial basis and thus provide a representative database of conformally soft theorems for scalar Goldstone bosons. 
In particular (see section~\ref{sec:6}), we derive soft theorems whose right hand is a highly non-trivial linear combination of lower point amplitudes. 

The soft behavior of Nambu--Goldstone EFTs is captured by non-linear sigma models. Their effective action (up to two derivatives and all orders in fields) is completely determined in terms of a (pseudo)-Riemannian metric on the target space of the NLSM. However, celestial amplitudes of NLSMs are not very well-behaved. For massless particles, the transform to a celestial basis is a Mellin transform (see section~\ref{sec:3}) that exchanges light-cone energies in Minkowski spacetime with scaling dimensions in CCFT. An essential object associated with any Mellin transform is its fundamental strip, where it converges. For NLSMs the integrals are marginally convergent (i.e. their fundamental strip is empty), which results in generalized delta functions (with complex arguments) of scaling dimensions. The left border of the fundamental strip is controlled by the IR expansion of the amplitude, while the right border by the UV expansion. When we want to analyze a particular soft behavior, we consider the NLSM as fixed and thus cannot improve the IR expansion (the left border of the fundamental strip is fixed). However, to gain better control over the celestial amplitude, we can improve the high-energy behavior. This is done by moving one step up (against the RG-flow) in the hierarchy of EFTs. For this reason, we always construct a pair of NLSM and its UV completion.
This is either the linearized\footnote{The standard lore says that they are only partial UV completions. None of these models are actually asymptotically free or conformal in the far UV.} version of the corresponding NLSM or some stringy UV completion. These completions always have non-empty fundamental strips, and therefore, their celestial amplitudes don't have a distributional character in scaling dimensions. 

The pairs of NLSMs and their linearized versions allow us to illustrate the subtleties of the interrelations between the corresponding celestial amplitudes.
In momentum space, the amplitudes of the NLSM are formally obtained as the leading order of the low-energy expansion of the amplitude within the linearized theory. Equivalently, it can be achieved by means of an appropriate limit sending the masses of the massive particles to infinity. 
This can be obtained in various ways in the parametric space of the linearized theory.
This approach is, however, rather subtle: a naive limiting procedure may lead to apparent paradoxes when applied on the level of celestial amplitudes. 
We will give several explicit examples and discuss the related phenomena in detail.

Another issue connected with the soft theorems is their interpretation within the CCFT.
Some examples of the Goldstone bosons soft theorems allow us to interpret them as Ward identities of some symmetries acting on the conformal primaries, which are explicitly or anomalously broken. 
We also provide an example (the pion-dilaton model), which obeys rather exotic conformal soft theorems, where such interpretation seems not to be possible.

The paper is organized as follows: In section~\ref{sec:2}, we review soft theorems for scalar EFTs and some recent developments on their momentum space on-shell scattering amplitudes. In sections~\ref{sec:3} and~\ref{sec:4}, we review the basic properties of celestial amplitudes for scalars, their conformal properties and discuss in detail four-point and five-point cases. In section~\ref{sec:5}, we discuss soft limits and provide celestial analogues of Adler zero and enhanced soft behavior. In section~\ref{sec:6}, we provide many explicit examples of Goldstone boson models, linear and non-linear sigma models and their UV completions. In section~\ref{sec:7}, we briefly comment on the celestial recursion relations and the possibility of the soft bootstrap program. We give some concluding remarks in section~\ref{sec:8}. The more technical parts and calculations are moved to appendices.  
%%%%%%%%%%%%%%%%%%%%%%%%%%%%%%%%%%%
\section{Review of soft limits in momentum space}\label{sec:2}
%%%%%%%%%%%%%%%%%%%%%%%%%%%%%%%%%%%

In quantum field theory, scattering amplitudes describe the probability amplitudes for particle scatterings. They are important because by squaring them, one gets the probabilities of the corresponding particle scatterings, which can be compared with the experimental data. The given scattering amplitude can be obtained by summing over all Feynman diagrams that connect the initial and final states for the given process. The Feynman diagrams calculation is a straightforward way how to get to the result. However, there are different techniques for calculating the scattering amplitudes. Namely, the BCFW \cite{Britto:2005fq} recursion enables reducing a full amplitude into a sum of products of two simpler amplitudes by changing (shifting) two momenta. This process continues until the entire scattering amplitude has been expressed in terms of the seed amplitudes, typically the three-point vertices. The BCFW recursion is particularly useful for theories with massless particles, such as gauge theories like QCD. These theories can have amplitudes with many Feynman diagrams, but the BCFW allows one to simplify the process and calculate them more efficiently.
Application of the recursion procedure on the effective theories of Goldstone bosons is spoiled by bad high-energy behavior of the amplitudes. This can be overcome by the information on the amplitude at some different kinematical point(s). Typically, effective field theories focus on the low energy limit and it is thus natural to consider the behavior at the zero momentum or momenta. We will summarize below the main results of the so-called soft bootstrap program first introduced in \cite{Cheung:2014dqa, Cheung:2015ota,Cheung:2016drk} and developed further e.g. in \cite{Cheung:2018oki,Elvang:2018dco,Kampf:2019mcd,Low:2019ynd,Kampf:2020tne,Kampf:2021bet,Kampf:2021tbk,Bartsch:2022pyi}. 
We will start with the well-known example, the single-shift limit, the Adler zero, connected with the $SU(N)$ NLSM \cite{Kampf:2012fn,Kampf:2013vha}.
%==================================
\subsection*{Shift symmetry and Adler zero}
%==================================

Shift symmetry and, consequently, Adler zero are important concepts in particle physics. In fact, the so-called Alder zero, i.e. the vanishing of the amplitude with one external soft particle, was discovered in connection with the strong interaction, more precisely with the $SU(N)$ NLSM.

 Let us consider the invariance of a theory at the quantum level with corresponding non-anomalous Noether current $N^\mu(x)$. Now we will assume that this symmetry is spontaneously broken. The necessary and sufficient condition for spontaneous symmetry breaking is the coupling of the Noether current to the set of Goldstone bosons $\phi^a$ as
\begin{equation}
    \langle0| N^{a\mu}(x) | \phi^b(p)\rangle = i \delta^{ab}p^\mu F^b {\rm e}^{-ip\cdot x}\,.
\end{equation}
Sandwiching the current between the in- and out- states ($\alpha$ and $\beta$, respectively), we find out that this matrix element develops a pole for $p^2\to0$ with the residue of the amplitude of the Goldstone boson emission:
\begin{align}
    \langle \beta|N^{a\mu}(0)|\alpha\rangle &= \frac{i}{p^2}\sum_b \langle 0 | N^{a\mu}(0)| \phi^b(p)\rangle \langle \beta + \phi^b(p)| \alpha\rangle  + R^{a\mu}(p)\notag\\
    &= -\frac{p^\mu}{p^2} F^a \langle \beta+\phi^a(p)|\alpha\rangle + R^{a\mu}(p)\,,
\end{align}
where $p=p_\alpha-p_\beta$. Assuming the current conservation and regular remnant $R^{a\mu}$ in the soft Goldstone boson limit, we get
\begin{equation}
    \lim_{p\to0}  F^a \langle \beta + \phi^a(p)|\alpha\rangle = \lim_{p\to0} p_\mu R^{a\mu}(p) = 0.
\end{equation}
We can verify this generic statement which is based solely on symmetry arguments and is valid non-perturbatively for explicit tree-level amplitudes. The leading order Lagrangian of the massless theory, which encodes the $SU(N)\times SU(N)/SU(N)$ chiral symmetry breaking, can be written as:
\begin{equation}
{\cal L} = \frac{F^2}{4}\langle u_\mu u^\mu \rangle, \quad \text{where }\; u_\mu= i (u^\dagger\partial_\mu u - u \partial_\mu u^\dagger),\quad
u = {\rm e}^{i \Phi/\sqrt2F},\quad \Phi=t^a\phi^a,
\end{equation}
where $t^a$ are generators of $SU(N)$. We have introduced a multiplet of Goldstone bosons (pions) $\phi^a$ and can easily calculate their scattering amplitudes. We can only focus on the ordered amplitudes and thus strip the flavor indices. For the 4pt we get (in our convention when we assume all momenta incoming):
\begin{equation}
A_4^{NLSM}  = -\frac{1}{2F^2} s_{13}
\end{equation}
with $s_{ij\ldots} = (p_i + p_j + \ldots)^2 = 2 p_i\cdot p_j + \ldots$. This amplitude corresponds to the seed 4pt vertex. The 6pt scattering is given by
\begin{equation}
A_6^{NLSM} = -\frac{1}{4 F^4}\Bigl( \frac{s_{13}s_{46}}{s_{123}}+\frac{s_{26}s_{35}}{s_{345}}+\frac{s_{15}s_{24}}{s_{234}} - \frac12 (s_{13}+s_{15}+s_{24}+s_{26}+s_{35}+s_{46})\Bigr)\,.
\end{equation}
The structure is clear; we have the factorization terms corresponding to the 4pt-vertex insertions and the last term, linear in $s_{ij}$, corresponding to the 6pt vertex. One can easily verify that setting, for example, $p_6\to 0$ the factorization terms collapse to  $s_{24}$ and together with the leftover from the last term, we get indeed 0. The 6pt amplitude was calculated from the above Lagrangian, and all terms were fixed. But  we can get to the result without the Lagrangian, just using the information of the Adler zero. Starting with the generic 4pt vertex, we would get the factorization terms. The requirement of the vanishing amplitude in the soft limit will force us to include the 6pt contact term of exactly the same form as above. This is the basis of the soft-bootstrap program or bottom-up reconstruction (see \cite{Bijnens:2019eze} and \cite{Kampf:2021jvf} for more details and higher orders).
%==================================
\subsection*{Enhanced soft limits and exceptional EFTs}
%==================================

In the above example, we saw that the structure of the theory could be reconstructed solely based on the given powercounting and the information of the Adler zero. We will briefly describe here how it can be extended for more generic situations (for more details, please refer to \cite{Cheung:2014dqa,Cheung:2015ota,Cheung:2016drk}). 

The powercounting can be connected with the following parameter
\begin{equation}
\rho = \frac{m-2}{n-2}\,,
\end{equation}
defined for each $n$-pt vertex with $m$-derivatives. In the above NLSM example, $\rho=0$ for all vertices. It is easier to focus first on the single-$\rho$ theories. 

The spontaneous symmetry breaking above is connected with the shift symmetry, for the single field given by
\begin{equation}
    \phi \to \phi + a\,.
\end{equation}
We can assume the following generalization of the shift symmetry
\begin{equation}
\phi \to \phi + \theta_{\mu_1\ldots \mu_r} x^{\mu_1}\ldots x^{\mu_r} + O[\phi](x)\,,
\end{equation}
where $\theta$ is traceless tensor and the local operator $O[\phi](x)$ is at least quadratic in the field $\phi$ and its derivatives. If we assume that the Lagrangian is invariant with respect to such a polynomial symmetry and there are no trilinear vertices, we get that the amplitudes will vanish as
\begin{equation}
    A(p_1,\ldots,p_n) = {\cal O}(p_i^\sigma),\qquad \text{with}\quad p_i \to 0\quad \text{and}\quad \sigma\equiv r+1\,.
\end{equation}
I.e. we have obtained the Adler zero of the $\sigma\equiv(r+1)$-order. This order $\sigma$ is an important parameter and together with $\rho$ enables classification of effective theories. In the $(\rho,\sigma)$ parametric space we get the so-called exceptional single-$\rho$ theories -- $(0,1)$: NLSM, $(1,2)$: DBI, $(2,2)$: Galileon and finally $(2,3)$: special Galileon (sGal). 

Possible generalizations can address the above assumptions or limitations, namely the description of solely scalar particles, the absence of the trilinear interactions, or the focus only on the single $\rho$ theories. Few attempts beyond these cases can be found for example in the following works:  \cite{Cheung:2018oki,Kampf:2019mcd,Kampf:2020tne,Kampf:2021bet,Kampf:2021tbk}.

%%%%%%%%%%%%%%%%%%%%%%%%%%%%%%%%%%%%%%%%%%%%
\section{Fundamentals of celestial amplitudes for scalars}\label{sec:3}
%%%%%%%%%%%%%%%%%%%%%%%%%%%%%%%%%%%%%%%%%%%%

A fundamental piece in any holographic correspondence is the matching between bulk and boundary symmetries. In the celestial holography context, it amounts to the isomorphism between the bulk Lorentz group $\mathrm{SO}(1,3)$ and the global conformal group $\mathrm{PSL}(2,\mathbb{C})$ of the boundary celestial sphere. Bulk momentum space S-matrix makes manifest translation invariance, as it is expressed in its eigenstates (plane waves). Lorentz boost (in $z$-direction) gets mapped to a dilatation under the above isomorphism. It is the basis of boost/dilatation eigenstates that makes manifest conformal covariance of the boundary CCFT correlators. Thus the kinematic part of the holographic dictionary is implemented by a change of basis between scattering states (see~\cite{deBoer:2003vf,Strominger:2017zoo,Cheung:2016iub} for early results and observations)
\begin{align*}
\begin{array}{|c|}
\hline
\text{plane wave} \\
\text{(translation eigenstate)} \\ \hline
\end{array}
\Longleftrightarrow
\begin{array}{|c|}
\hline
\text{conformal primary wavefunction} \\
\text{(boost/dilatation eigenstate)} \\ \hline
\end{array}
\end{align*}
The precise form of the transformation was worked out in~\cite{Pasterski:2016qvg,Pasterski:2017kqt} (see also~\cite{Banerjee:2018gce} for an alternative presentation). For massive particles it is realized as an integral transform on a hyperbolic slice $H_3$ of the dual momentum space to Minkowski spacetime defined by the constraint $\hat{p}^2=1$, with an integration kernel given by the $H_3$ bulk-to-boundary propagator $G_{\Delta}\left(\hat{p}\vert (w,\overline{w})\right)$
\begin{align}\label{eq:massive_transf}
&\Phi_{\Delta}^\varepsilon\left(X\vert(w,\overline{w})\right)=\int_{H_3} \widetilde{\mathrm{d}}\widehat{p}\,G_{\Delta}\left(\hat{p}\vert (w,\overline{w})\right) e^{\varepsilon i m \hat{p}\cdot X}; && G_{\Delta}\left(\hat{p}\vert q(w,\overline{w})\right)=\left(\frac{1}{2\hat{p}\cdot q(w,\overline{w})}\right)^\Delta,
\end{align}
where $\Phi$ is the massive conformal primary wavefunction (of a scalar particle for simplicity). It depends on a spacetime point $X$  and an auxiliary null momentum 
$$q(w,\overline{w})=\tfrac{1}{2}(1+w\overline{w},w+\overline{w},i(w-\overline{w}),w\overline{w}-1)),$$ which has the interpretation of a point at the asymptotic boundary  of a  mass-shell hyperboloid $H_3$. The bulk-to-boundary propagator acts in momentum space and captures propagation between this auxiliary boundary point $q(w,\overline{w})$  and a point $\hat{p}$ on the upper sheet of a unit momentum mass-shell hyperboloid (that is being integrated over in~\eqref{eq:massive_transf}). Equivalently, a particle with momentum $q(w,\overline{w})$ would pierce the future celestial sphere (i.e. the boundary of the  Minkowski spacetime) in a point $(w,\overline{w})$ defined in stereographic coordinates. Outgoing/incoming particles are distinguished by $\varepsilon=\pm$ and the unitary~\footnote{Unitarity is required with respect to the Klein--Gordon scalar product inherited from bulk, not the usual scalar product used for Euclidean 2D CFTs that is equivalent to radial quantization, see e.g.~\cite{Crawley:2021ivb} for a discussion.} irreducible representation of $\mathrm{PSL}(2,\mathbb{C})$ is characterized by a scaling dimension $\Delta$, lying on the principal continuous series $\Delta=1+i\mathbb{R}_+$, as required by a normalization condition~\cite{Pasterski:2016qvg,Pasterski:2017kqt}. 

For massless particles this expression reduces to a Mellin transform with respect to the particle's light front energy $\omega$, see Appendix~\ref{app:A} for summary of definitions. It can be achieved either by direct computation or a limiting procedure (for a review see~\cite{Raclariu:2021zjz})
\begin{align}\label{eq:massless_transf}
\phi_{\Delta}^\varepsilon\left(X\vert(w,\overline{w})\right)=\int_{0}^{\infty} \mathrm{d}\omega \omega^{\Delta-1} e^{\varepsilon i \omega q\cdot X}, 
\end{align}
where $\phi$ is the massless conformal primary wavefunction, $\omega q(w,\overline{w})$ the null momentum of a particle that pierces the celestial sphere at a point $(w,\overline{w})$, and $\Delta=1+i\lambda,\;\lambda\in\mathbb{R}$.

%==================================
\subsection{Celestial amplitudes}
%==================================

The formulas~\eqref{eq:massive_transf}, \eqref{eq:massless_transf} constitute an essential piece of the holographic dictionary and allow to express any momentum space S-matrix element as a celestial amplitude (CCFT correlator). The map is very simple, one just needs to do an appropriate integral transform for every external leg based on its mass~\footnote{Here we are interested in scalar particles. For massless particles the 4D helicities get identified with 2D spins and the map remains to be a Mellin transform. In the massive case, the 4D spin associated with irrep of the $\mathrm{SO}(3)$ little group must be decomposed into 2D $\mathrm{SO}(2)$ spins, while the scalar bulk-to-boundary propagator gets generalized to its spinning version (see~\cite{Law:2020tsg} for details).}
\begin{align}\label{correlator_definition}
&\left\langle O_{\Delta _{1}}^{\varepsilon _{1}}\left( z_{1},\overline{z%
}_{1}\right) \ldots O_{\Delta _{n}}^{\varepsilon _{n}}\left( z_{n},%
\overline{z}_{n}\right) \right\rangle = \\
&\Big(\prod_{i:\mathrm{massless}}\int_{0}^{\infty} \mathrm{d}\omega
_{i}\omega _{i}^{\Delta _{i}-1}\Big)\Big(\prod_{j:\mathrm{massive}}\int_{H_3} \widetilde{\mathrm{d}}\widehat{p}_j G_{\Delta}\left(\hat{p}_j\vert (z_j,\overline{z}_j)\right)\Big)\mathcal{A}_{n}\left( \varepsilon _{i}\omega
_{i}q\left( z_{i},\overline{z}_{i}\right); \varepsilon _{j}m_j\hat{p}
_{j} \right) ,\notag 
\end{align}
where $i\cup j=\{1,\ldots,n\}$ is a division of external particles into massless/massive and
\begin{equation}
\mathcal{A}_{n}=\delta ^{(4)}\Bigl(
\sum_i\varepsilon _{i}\omega _{i}q(z_{i},\overline{z}
_{i})+\sum_j\varepsilon_jm_j\hat{p}_j \Bigr) A_{n}\bigl( \varepsilon _{i}\omega
_{i}q( z_{i},\overline{z}_{i}); \varepsilon _{j}m_j\hat{p}
_{j} \bigr)
\end{equation}
is the momentum space scattering amplitude including momentum conservation.
Here we parameterize the on shell momenta of massless and massive particles as
\begin{equation}
    p_i=\omega_i q(z_i,\bar{z}_i),\,\,\,\,\,p_i=m_i \hat{p}_i
\end{equation}
respectively, where $\hat{p}_i^2=1$ and
\begin{equation}
    q(z,\bar{z})=\frac{1}{2}(1+|z|^2,z+\bar{z},{\rm{i}}(z-\bar{z}),|z|^2 -1).
\end{equation}
For external particles
$\varepsilon_i=\pm 1$ for outgoing and incoming particles, respectively.
For our conventions and further details see Appendix \ref{app:A}.

From now on, we will concentrate on tree-level amplitudes with massless external states~\footnote{We preferred to state the general celestial dictionary since later a recursion formula based on~\eqref{residue} will be presented and lower point celestial amplitudes with one massive external leg will be needed.}. In that case, the on-shell amplitude $A_{n}$ is a rational function of $\omega _{i}$. In accord with~\cite{Pasterski:2017ylz}, it is useful to perform a change of variables 
\begin{equation}
\sum\limits_{i=1}^{n}\omega _{i}=\sqrt{u},~~~~~\omega _{i}=\sigma _{i}\sqrt{u%
} \label{masslesskin}
\end{equation}
and obtain
\begin{eqnarray}
&&\hspace{-1.5cm}\left\langle O_{\Delta _{1}}^{\varepsilon _{1}}\left( z_{1},%
\overline{z}_{1}\right) \ldots O_{\Delta _{n}}^{\varepsilon
_{n}}\left( z_{n},\overline{z}_{n}\right) \right\rangle  \notag \\
&&\hspace{-0.5cm} =\int_{\langle 0,1\rangle ^{n}}\prod\limits_{i=1}^{n}d\sigma _{i}\sigma
_{i}^{\Delta _{i}-1}\delta \left( 1-\sum\limits_{i=1}^{n}\sigma _{i}\right)
\delta ^{(4) }\left( \sum\limits_{i=1}^{n}\varepsilon _{i}\sigma
_{i}q\left( z_{i},\overline{z}_{i}\right) \right)  \notag \\
&&\hspace{1.7cm}\times \frac{1}{2}\int_{0}^{\infty }\mathrm{d}u~u^{\Delta -1}A_{n}\left( 
\sqrt{u}\varepsilon _{1}\sigma _{1}q\left( z_{1},\overline{z}_{1}\right)
,\ldots ,\sqrt{u}\varepsilon _{n}\sigma _{n}q\left( z_{n},\overline{z}%
_{n}\right) \right),  \label{mellin_transform_complete}
\end{eqnarray}
where in the last integral $\Delta$ is given by
\begin{equation}\label{eq:def_delta}
\Delta =\frac{1}{2}\sum\limits_{i=1}^{n}\Delta _{i}-2\,.
\end{equation}
Note that for massive particles the kinematics does not factorize like in (\ref{masslesskin}). Let us look at the expression (\ref{mellin_transform_complete}) in more details. For general $n$-point correlators there are $n$ integrals over parameters $\sigma_i$ which satisfy 5 conditions from delta functions. This means that we can fix $\sigma_1,{\dots},\sigma_5$ and integrate over remaining $\sigma_6,\dots,\sigma_n$, together with the $u$ integral. For the future reference to simplify the notation, we define the measure of the $\sigma$-integrations,
\begin{equation}
[{\rm d}\sigma,\Delta] \equiv \prod\limits_{i=1}^{n}d\sigma _{i}\sigma
_{i}^{\Delta _{i}-1}\delta \left( 1-\sum\limits_{i=1}^{n}\sigma _{i}\right),
\end{equation}
and we define for fixed external states (and signs $\epsilon_i$)
\begin{equation}
\widetilde{\mathcal{A}}_{n}\left(
\left\{ \Delta_{i}\right\} ,z_{1},\ldots ,z_{n}\right)\equiv \left\langle O_{\Delta_{1}}^{\varepsilon_{1}}\left( z_{1},\overline{z}_{1}\right)\ldots O_{\Delta_{n}}^{\varepsilon_{n}}\left(z_{n},\overline{z}_{n}\right) \right\rangle
\end{equation}
and will refer to $\widetilde{\mathcal{A}}_{n}$ as a \emph{celestial amplitude}. 

%==================================
\subsection{Homogeneous amplitude}
%==================================

If the amplitude $A_{n}$ is a homogeneous function of momenta with degree of homogeneity $2d$,
\begin{equation}
A_{n}\left( 
\sqrt{u}\varepsilon _{1}\sigma _{1}q\left( z_{1},\overline{z}_{1}\right)
,\ldots ,\sqrt{u}\varepsilon _{n}\sigma _{n}q\left( z_{n},\overline{z}%
_{n}\right) \right) = u^d A_{n}\left( 
\varepsilon _{1}\sigma _{1}q\left( z_{1},\overline{z}_{1}\right)
,\ldots ,\varepsilon _{n}\sigma _{n}q\left( z_{n},\overline{z}%
_{n}\right) \right)
\end{equation}
and after plugging into (\ref{mellin_transform_complete}) we can perform the $u$-integral explicitly. The integral does not converge in the usual sense but it could be understood in the sense of distributions as a generalized Mellin transform of the distribution $f(u) =u^{d}$ leading to the identification (e.g. (2.20) in~\cite{Pasterski:2017ylz}) 
\begin{equation*}
\int_{0}^{\infty }\mathrm{d}u~u^{\Delta +d-1}=2\pi\delta \left( \Delta +d\right) \,,
\end{equation*}
where $\Delta $ is allowed to be complex. As a result we get,
\begin{multline}
\widetilde{\mathcal{A}}_{n}(\{ \Delta_{i}\} ,z_{1},\ldots ,z_{n}) =\pi\delta( \Delta +d)\\
\times \int_{\langle 0,1\rangle
^{n}} [{\rm d}\sigma,\Delta]\,\delta^{(4)}\Bigl( \sum\limits_{i=1}^{n}\varepsilon _{i}\sigma _{i}q( z_{i},\overline{z}_{i}) \Bigr) A_{n}\left( \varepsilon _{1}\sigma _{1}q(z_{1},\overline{z}_{1}) ,\ldots ,\varepsilon _{n}\sigma _{n}q( z_{n},\overline{z}_{n}) \right), \label{celest1}
\end{multline}
where only the integration over $\sigma_i$ is left. Note that the 4D bulk dilatation operator acts on $k$-th external leg as $-i(\Delta_k-1)=\lambda_k$~\cite{Stieberger:2018onx}. Thus the amplitude is classically scale invariant if the delta function in scaling dimensions reduces to $\delta(\sum_{k=1}^n\lambda_k)$. This happens precisely when the degree of homogeneity satisfies $2d=4-n$.

The amplitudes in exceptional EFTs are homogeneous and the simplification does apply to them. In particular (cf. section~\ref{sec:2}):
\begin{center}
NLSM: $d=1$,\qquad DBI: $d=\frac{n}{2}$,\qquad sGal: $d=n{-}1$\,.
\end{center}
But it is worth emphasizing that the celestial correlators (\ref{celest1}) for these exceptional EFTs are only defined in the distribution sense, as indicated by the presence of the delta function $\delta(\Delta+d)$. This is a general feature of any homogeneous amplitude, in order to perform the Mellin transformation in the usual sense we need an amplitude in a UV completed theory, as discussed further.

%==================================
\subsection{Non-homogeneous amplitudes and UV completion\label{Non-homogeneous amplitudes and UV completion}}
%==================================

In the general case of non-homogeneous amplitude such a simplification of the $u-$integration is not possible and the integral is entangled with the $\sigma_i$-integration. In that case we are left with the general formula (\ref{mellin_transform_complete}).  

\begin{figure}[ht]
\centering
\begin{tikzpicture}
\tikzset{cross/.style={cross out, draw=black, minimum size=2*(#1-\pgflinewidth), inner sep=0pt, outer sep=0pt},
%default radius will be 1pt. 
cross/.default={2.7pt}}
% configurable parameters
\def\bigradius{3}

\draw [fill=gray!20,draw=none] (-2,-.5*\bigradius) rectangle (2,.5*\bigradius);
% Axes
\draw [help lines,->] (-1.6*\bigradius, 0) -- (1.6*\bigradius,0);
\draw [help lines,->] (0, -.5*\bigradius) -- (0, .5*\bigradius);

% The labels
	\node at (5,-0.35){$\mathrm{Re}(\Delta)$};
	\node at (0.69,1.5) {$\mathrm{Im}(\Delta)$};
	\node at (2,-0.25) {\small$\frac{n}{2}-2$};
   	\node at (-2,-0.25) {\small$-\alpha$};
   
\draw (2,0) node[cross,blue,thick] {};
\draw (3,0) node[cross,blue,thick] {};
\draw (4,0) node[cross,blue,thick] {};
\draw (-2,0) node[cross,red,thick] {};
\draw (-3,0) node[cross,red,thick] {};
\draw (-4,0) node[cross,red,thick] {};
\end{tikzpicture}
	\caption{Positions and residues of red poles encode the low energy (IR) EFT expansion of the amplitude (i.e. Wilson coefficients). Blue ones correspond to the high energy (UV) asymptotic expansion. Gray strip (outer edges excluded) represents the fundamental strip for the Mellin transform: the celestial amplitude is guaranteed to be holomorphic there.} 
\label{fig:position}
\end{figure}
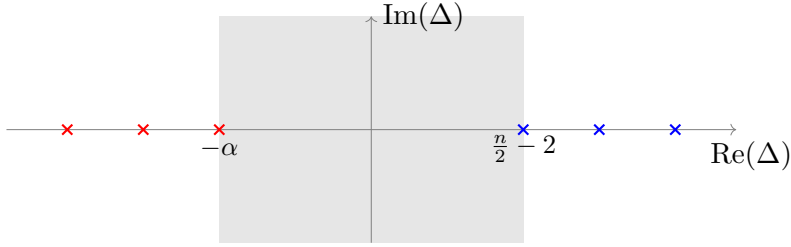

For renormalizable theories, we have further constraints on the function $A_{n}\left( u\right) \equiv A_{n}\left( \ldots,\sqrt{u}\varepsilon _{i}\sigma _{i}q\left( z_{i},\overline{z}_{i}\right),\ldots \right) $ from Weinberg's theorem, namely for $u\rightarrow \infty$ the amplitude generally behaves as
\begin{equation}
A_{n}\left( \ldots ,\sqrt{u}\varepsilon _{i}\sigma _{i}q\left( z_{i},%
\overline{z}_{i}\right) ,\ldots \right) =O\left( u^{2-n/2}\right), 
\label{asymptotics}
\end{equation}
which determines the upper bound on ${\rm{Re}}\,\Delta$ for which the
integral converges in the UV
\begin{equation}
{\rm{Re}}\,\Delta <\frac{n}{2}-2.
\end{equation}
On the other hand, the limit $u\rightarrow 0$ yields the leading term of the low energy expansion of the amplitude $A_{n}$. It might be calculated
directly using the effective low energy theory obtained by integrating out
all but the massless degrees of freedom. Provided $A_{n}=O\left( u^{\alpha
}\right) $ in this limit, the $u$ integral converges in the IR for 
\begin{equation}
{\rm{Re}}\Delta >-\alpha .
\end{equation}
Therefore, for the real part of $\Delta$ restricted to the fundamental strip $-\alpha <{\rm{Re}}\Delta <n/2-2$, the Mellin transform is a holomorphic function of $\Delta$ (see Fig.~\ref{fig:position}).

\begin{figure}[ht]
\centering
\scalebox{.8}{
\begin{tikzpicture}[auto,node distance=2cm,thick,main node/.style={circle,fill=blue!20,draw,font=\sffamily\Large\bfseries}, scale=2, pin distance=0.8cm]
            \tikzset{every pin edge/.append style={black, thick}}
            \node[main node,pin=90:$i_{l}$,pin=110:$i_{l-1}$,pin=-110:$i_{2}$,pin=-90:$i_{1}$] (F) at (0,0) {$\mathcal{F}_{\phantom{c}}$};
            \node[main node,pin=90:$i_{l+1}$,pin=70:$i_{l+2}$,pin=-70:$i_{n-1}$,pin=-90:$i_{n}$] (Fc) at (2,0) {$\mathcal{F}_c$};
            \draw[dotted] (2.5,-0.4) arc (-30:30:0.8) ;
            \draw[dotted] (-0.5,0.4) arc (150:210:0.8) ;
            \path [-] (F) edge node {$u_{\mathcal{F}}Q_{\mathcal{F}}^2-M_{\mathcal{F}}^2=0$} (Fc);
\end{tikzpicture}}
        \caption{Schematic figure of a factorization channel $\mathcal{F}$ for a tree-level $n$-point amplitude with a massive scalar exchange.} \label{fig:fact_channel}
\end{figure}
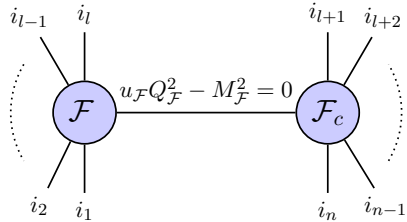

 For $\Delta$ in the fundamental strip we can calculate the $u$ Mellin integral using the complex Hankel contour, more details can be found in Appendix \ref{App:Hankel}. As a result, we get
\begin{equation}
\int_{0}^{\infty }\mathrm{d}u~u^{\Delta -1}A_{n}\left( u\right) 
= -\frac{\pi }{\sin(\pi\Delta)} \sum\limits_{\mathcal{F}}\mathrm{e}^{\pi \mathrm{i}\Delta \cdot \mathrm{
sign}\left( u_{\mathcal{F}}\right) }u_{\mathcal{F}}^{\Delta -1}\mathrm{res}
\left( A_{n}\left( u\right) ,u_{\mathcal{F}}\right),
\label{residue_formula_for_A_n}
\end{equation}
where the contour integral was calculated as a sum over residues in $u$. These residues correspond to massive factorization channels $\cal{F}$ of the tree-level amplitude $A_n$, and they localize $u$ to $u_{\cal F}$ given by
\begin{equation}\label{eq:u_poles}
u_{\mathcal{F}}=\frac{M_{\mathcal{F}}^{2}}{Q_{\mathcal{F}}^{2}}=\frac{M_{%
\mathcal{F}}^{2}}{\sum\limits_{\left( i<j\right) \in \mathcal{F}}\varepsilon
_{i}\varepsilon _{j}\sigma _{i}\sigma _{j}\left\vert z_{ij}\right\vert ^{2}}\,.
\end{equation}
We schematically draw the factorization diagram in Fig.~\ref{fig:fact_channel}. 

Plugging into (\ref{mellin_transform_complete}) we obtain a partial expression for the celestial amplitude after $u$-integration,
\begin{eqnarray}
\widetilde{\mathcal{A}}_{n}\left(
\left\{ \Delta_{i}\right\} ,z_{1},\ldots ,z_{n}\right)  
&=&-\frac{\pi }{\sin \pi \Delta }\int \left[ \mathrm{d}\sigma ,\Delta \right]
\delta^{(4)}\left( \sum\limits_{i=1}^{n}\varepsilon _{i}\sigma
_{i}q\left( z_{i},\overline{z}_{i}\right) \right)   \notag \\
&&\hspace{1.5cm}\times \sum\limits_{\mathcal{F}}\mathrm{e}^{\pi \mathrm{i}\Delta \cdot\mathrm{sign}\left( u_{\mathcal{F}}\right) }u_{\mathcal{F}}^{\Delta -1}\mathrm{res}%
\left( A_{n}\left( u\right) ,u_{\mathcal{F}}\right).
\label{residue_formula}
\end{eqnarray}
Note that this formula involves the calculation of residues of $A_n(u)$ on the pole $u_{\cal F}$. Even if all external particles are massless, the pole corresponds to the exchange of the massive particle. Hence the residue depends on amplitudes of particles with both massive and massless external legs.

%=============================================
\subsection{General four-point and five-point amplitudes}
%=============================================

The case $n=4$ and $n=5$ are special since the integration over the
measure $\left[ \mathrm{d}\sigma ,\Delta \right] $ is saturated by the $\delta$-functions. The number of delta functions is five so for $n=5$ we exactly saturate all $\sigma$-integrations. The values of $\sigma_i$ are reduced to the solutions of the set of linear equations of the form
\begin{equation}
\mathbf{A}_{n}\cdot \boldsymbol{\sigma }=\mathbf{b}  \qquad \mbox{where}\qquad 
\mathbf{A}_{n}=\left( 
\begin{array}{ccc}
1 & \ldots & 1 \\ 
\varepsilon _{1}q_{1} & \ldots & \varepsilon _{n}q_{n}%
\end{array}%
\right) ,~~~\mathbf{b}=\left( 
\begin{array}{c}
1 \\ 
0%
\end{array}%
\right) ,
\label{sigma_equations}
\end{equation}
where $\mathbf{A}_{n}$ is $(5\times n)$ matrix and $\mathbf{\sigma}=(\sigma_1\dots \sigma_n)$, while $\mathbf{b}$ is a 5-vector. 

%=================================
\subsubsection*{Five-point amplitudes}
%=================================

For $n=5$, $\mathbf{A}_{5}$ is a square matrix with determinant\footnote{Here the hat means omission of the corresponding momentum from the list.}
\begin{align}
\det \mathbf{A}_{5} &=\sum\limits_{i=1}^{5}(-1)^{i+1}\det( q_{1},{\ldots},\widehat{q_{i}},\ldots ,q_{5})
\prod\limits_{j\neq i}\varepsilon _{j}\notag\\ &= \sum\limits_{i=1}^{5}( -1)^{i}\varepsilon _{\mu _{1}\ldots 
\widehat{\mu _{i}}\ldots\mu _{5}}q_{1}^{\mu_{1}}\ldots\widehat{q_{i}^{\mu_{i}}}\ldots q_{5}^{\mu _{5}}\prod_{j\neq i}\varepsilon _{j}\,,
\end{align}
which is nonzero for generic $q_{k}$. Explicitly
\begin{equation}
\det \left( q_{i},q_{j},q_{k},q_{l}\right) =-\frac{\mathrm{i}}{4}\left(
z_{ij}z_{kl}\overline{z}_{ik}\overline{z}_{jl}-z_{ik}z_{jl}\overline{z}_{ij}%
\overline{z}_{kl}\right) =-\frac{\mathrm{i}}{4}\left\vert z_{ik}\right\vert
^{2}\left\vert z_{jl}\right\vert ^{2}\left( z_{ijkl}-\overline{z}%
_{ijkl}\right) \,, \label{4pt_determinant}
\end{equation}
where $z_{ij}\equiv z_i-z_j$ is a difference of two positions on the celestial sphere. Here we introduced the conformally invariant cross-ratio
\begin{equation}
z_{ijkl}=\frac{z_{ij}z_{kl}}{z_{ik}z_{jl}}. \label{cross}
\end{equation}
The solution $\sigma _{\ast i}$ of (\ref{sigma_equations}) is thus unique
and equal to
\begin{equation}
\sigma _{\ast i}=\frac{\left( -1\right) ^{i+1}\det \left( \varepsilon
_{1}q_{1},\ldots ,\widehat{\varepsilon _{i}q_{i}},\ldots ,\varepsilon
_{5}q_{5}\right) }{\det \mathbf{A}_{5}}.  \label{sigma_i*_solution}
\end{equation}
The product of delta functions can be then rewritten as 
\begin{equation}
    \delta^{(5)}(\mathbf{A}_{5}\cdot\mathbf{\sigma}-\mathbf{b})=\left\vert \det 
\mathbf{A}_{5}\right\vert^{-1}\prod_{i{=}1}^5\delta(\mathbf{\sigma}-\mathbf{\sigma}_{*i}).
\end{equation}
The integration over $\sigma$ then localizes $\sigma_i=\sigma_{*i}$ and we get from (\ref{mellin_transform_complete})
\begin{equation}
\widetilde{\mathcal{A}}_{5}\left(
\left\{ \Delta_{i}\right\} ,z_{1},\ldots ,z_{5}\right) =\frac{\prod\limits_{i=1}^{5}\sigma _{\ast i}^{\Delta _{i}-1}\chi
_{\langle 0,1\rangle }\left( \sigma _{\ast i}\right) }{2\left\vert \det 
\mathbf{A}_{5}\right\vert }\int_{0}^{\infty }\mathrm{d}%
u~u^{\Delta -1}A_{5}\left( \left\{ \sqrt{u}\varepsilon _{j}\sigma _{\ast
j}q\left( z_{j},\overline{z}_{j}\right) \right\} _{j=1}^{5}\right) ,
\label{5pt_general_form}
\end{equation}
where $\chi _{\langle 0,1\rangle }$ is a characteristic function of the
interval $\langle 0,1\rangle $. Hence everything reduces to the $u$-integration. As discussed earlier, for homogeneous amplitudes this integration can be carried out explicitly (\ref{celest1}) and for  non-homogeneous amplitudes within  renormalizable theories we can use partial results (\ref{residue_formula}). For example, for homogeneous amplitudes with degree of homogeneity $2d$ after performing the $u$-integral we get 
\begin{equation}
\widetilde{\mathcal{A}}_{5}\left(
\left\{ \Delta_{i}\right\} ,z_{1},\ldots ,z_{5}\right)  =\frac{\prod\limits_{i=1}^{5}\sigma _{\ast i}^{\Delta _{i}-1}\chi
_{\langle 0,1\rangle }\left( \sigma _{\ast i}\right) }{\left\vert \det 
\mathbf{A}_{5}\right\vert }\pi\delta(\Delta {+}d)\,A_{5}\left(\{ \varepsilon _{j}\sigma _{\ast
j}q\left( z_{j},\overline{z}_{j}\right)\} _{j=1}^{5}\right),
\end{equation}
where $\sigma_{\ast j}$ are given by (\ref{sigma_i*_solution}). 

%=================================
\subsubsection*{Four-point amplitudes}
%=================================

In the four-point case since the rank of the matrix $\mathbf{A}_{4}$ is at most four. The necessary condition for the existence of the solution of (\ref{sigma_equations}) is that $\mathrm{rank}\left( \mathbf{A}_{4},\mathbf{b}\right) \leq 4$, i.e.
\begin{equation}
\det \left( \mathbf{A}_{4},\mathbf{b}\right) =\det \left( q_{1},\ldots
,q_{4}\right) \prod_{j=1}^{4}\varepsilon _{j}=0,
\end{equation}
or explicitly, using (\ref{4pt_determinant})
\begin{equation}
-\frac{\mathrm{i}}{4}\left\vert z_{13}\right\vert ^{2}\left\vert
z_{24}\right\vert ^{2}\left( z_{1234}-\overline{z}_{1234}\right) =0.
\label{4pt_constraint}
\end{equation}
Hence we have to be a bit careful when solving for $\sigma_{\ast i}$ and performing $\sigma$-integrals. The constraints from momentum conservation can be written in the form
\begin{equation}
\sum_{i=1}^{4}\varepsilon _{i}\sigma _{i}|i\rangle \lbrack i|=0,\qquad \mbox{where}\qquad
|i\rangle =\left( 
\begin{array}{c}
z_{i} \\  1
\end{array}\right) ,~~[i|=\left( \overline{z}_{i},1\right) ,
\end{equation}
and can be solved for three $\sigma_i$ in terms of one free $\sigma _{j}$ with $j$ fixed as 
\begin{equation}
\sigma _{\ast i}\left( \sigma _{j}\right) =-\varepsilon _{i}\varepsilon
_{j}\sigma _{j}\frac{\langle k,j\rangle \lbrack j,l]}{\langle k,i\rangle
\lbrack i,l]}=-\varepsilon _{i}\varepsilon _{j}\sigma _{j}\frac{z_{jk}%
\overline{z}_{jl}}{z_{ik}\overline{z}_{il}} =-\varepsilon _{i}\varepsilon
_{j}\sigma _{j}\frac{z_{jl}\overline{z}_{jk}}{z_{il}\overline{z}_{ik}},\label{sigma_i_solution}
\end{equation}
where $k,l\neq i,j$ label the complementary two particles. Note that we get again
\begin{equation}
\frac{z_{jk}\overline{z}_{jl}}{z_{ik}\overline{z}_{il}}=\frac{z_{jl}%
\overline{z}_{jk}}{z_{il}\overline{z}_{ik}},  \quad\mbox{or}\quad 
\frac{z_{jk}z_{il}}{z_{jl}z_{ik}}=\frac{\overline{z}_{jk}\overline{z}_{il}}{%
\overline{z}_{jl}\overline{z}_{ik}},
\label{4pt_constraint_1}
\end{equation}
in accord with (\ref{4pt_constraint}). The corresponding four-dimensional $\delta -$functions can be then rewritten
as
\begin{equation}
\delta ^{(4) }\left( \sum\limits_{i=1}^{4}\varepsilon _{i}\sigma
_{i}q\left( z_{i},\overline{z}_{i}\right) \right) =\frac{4}{\sigma _{j}}%
\frac{1}{\left\vert z_{13}\right\vert ^{2}\left\vert z_{24}\right\vert ^{2}}%
\delta \left( -\mathrm{i}\left( z_{1234}-\overline{z}_{1234}\right) \right)
\prod_{i\neq j}\delta \left( \sigma _{i}-\sigma _{\ast i}\left(
\sigma _{j}\right) \right) .  \label{momentum_delta}
\end{equation}
From the fifth constraint (\ref{momentum_delta}) we obtain the solution for the fixed $\sigma_{\ast j}$,
\begin{equation}
\sum\limits_{i=1}^{4}\sigma _{i}=1\quad \rightarrow\quad 
\sigma _{\ast j}=\frac{1}{a_{j}}  
\quad \mbox{where}\quad
a_{j}=1+\sum_{\substack{\mathrm{cycl}(i,k,l)\\\{i,k,l\}\neq j}}\varepsilon _{i}\varepsilon
_{j}\frac{z_{jk}\overline{z}_{jl}}{z_{ik}\overline{z}_{il}}.\label{sigma_j*}
\end{equation}
Note also that on the support of momentum conservation $\delta -$function (\ref{momentum_delta}) we have
\begin{equation}
\delta \left( 1-\sum\limits_{i=1}^{n}\sigma _{i}\right) =\delta \left(
1-a_{j}\sigma _{j}\right) =\frac{1}{\left\vert a_j\right\vert }\delta \left(
\sigma _{j}-\sigma _{\ast j}\right) =\sigma _{\ast j}\delta \left( \sigma
_{j}-\sigma _{\ast j}\right) \mathrm{sign}\sigma _{\ast j}.
\end{equation}
Inserting (\ref{sigma_j*}) in (\ref{sigma_i_solution}) and using (\ref%
{4pt_constraint_1}), the final solutions for remaining $\sigma _{i\ast }$
have the same form as (\ref{sigma_j*}) (up to the appropriate permutation of indices) and for any $j=1,\ldots ,4$ can be written again as
\begin{equation}
\sigma _{\ast j}=\frac{1}{a_{j}}\,,
\end{equation}
which is now true for all $j=1,\dots,5$. As a result we get
\begin{eqnarray}\label{eq:cel_massless_4}
\widetilde{\mathcal{A}}_{4}\left(
\left\{ \Delta_{i}\right\} ,z_{1},\ldots ,z_{4}\right) &=&\frac{4}{\left\vert
z_{13}\right\vert ^{2}\left\vert z_{24}\right\vert ^{2}}\delta \left( -%
\mathrm{i}\left( z_{1234}-\overline{z}_{1234}\right) \right)
\prod\limits_{i=1}^{4}\sigma _{\ast i}^{\Delta _{i}-1}\chi _{\langle
0,1\rangle }\left( \sigma _{\ast i}\right)  \notag \\
&&\hspace{1.3cm}\times \frac{1}{2}\int_{0}^{\infty }\mathrm{d}u~u^{\Delta -1}A_{4}\left(
\left\{ \sqrt{u}\varepsilon _{i}\sigma _{\ast i}q\left( z_{i},\overline{z}%
_{i}\right) \right\} _{i=1}^{4}\right) ,
\end{eqnarray}
which principally differs from the five-point case by the presence of the delta function $\delta\left(-\mathrm{i}\left( z_{1234}-\overline{z}_{1234}\right) \right)$. Formulae for a generic massless celestial 4pt amplitude were derived e.g. in~\cite{Arkani-Hamed:2020gyp,Mizera:2022sln} (see also reviews~\cite{Raclariu:2021zjz,Pasterski:2021rjz,McLoughlin:2022ljp}). In fact, we see that the delta function inherited from bulk translation invariance forces the cross-ratio to be real (i.e. the celestial operators lie on a big circle on the celestial sphere as derived in~\cite{Arkani-Hamed:2020gyp}), while the characteristic function $\chi$ requires an in-out-in-out configuration of celestial insertions as proven in~\cite{Mizera:2022sln} (where the reader can find a detailed analysis of implications of bulk kinematics on the geometry of celestial $n$-pt amplitudes). Thus the 4pt CCFT correlator is a distribution, which is not a usual property encountered in 2D CFT and complicates application of standard CFT techniques, e.g. the conformal block expansion. For this reason many authors tried to relax it, either by performing a shadow/light transform on some of the CCFT operators or by coupling the bulk theory to external sources breaking translational invariance. See~\cite{Fan:2021isc,Fan:2021pbp,Fan:2022vbz,Fan:2022kpp,Stieberger:2022zyk,Gonzo:2022tjm,Hu:2022syq,Jorge-Diaz:2022dmy} for a sample of works in this direction.

To summarize, for 4pt and 5pt amplitudes, the Mellin transform reduces to
the calculation of a single $u$-integral and generic massless celestial 4pt and 5pt amplitudes can be computed according to~\eqref{eq:cel_massless_4} and~\eqref{5pt_general_form}. 

%%%%%%%%%%%%%%%%%%%%%%%%%%%%%%%%%%%%%%%
\section{Conformal structure of celestial amplitudes}\label{sec:4}
%%%%%%%%%%%%%%%%%%%%%%%%%%%%%%%%%%%%%%%

The structure of the celestial amplitudes can be better understood by taking their transformation properties with respect to the global conformal group into account. For simplicity, we will assume that all particles have zero helicity.

%==============================
\subsection{Five-point amplitudes} 
%==============================

By appropriate M\"{o}bius transformation, we can fix the gauge for the 5pt amplitude 
\begin{equation}
\left( z_{1},z_{2},z_{3},z_{4},z_{5}\right) \rightarrow \left(
1,0,\varepsilon ^{-1},t_{4}^{-1},t_{5}^{-1}\right)\,,
\end{equation}
where we want to take the limit $\epsilon\rightarrow0$ in the end. For all scalar particles we get the following representation
\begin{equation}
\widetilde{\mathcal{A}}_{5}\left(
\left\{ \Delta _{i}\right\} ,z_{1},\ldots ,z_{5}\right) 
=\prod_{j=1}^{5}\left\vert cz_{j}+d\right\vert ^{-2\Delta _{j}}%
\widetilde{\mathcal{A}}_{5}\left( \left\{ \Delta _{i}\right\}
,1,0,\varepsilon ^{-1},t_{4}^{-1},t_{5}^{-1}\right),\label{transf}\\
\end{equation}
where the parameters of the transformation
satisfy 
\begin{equation}
\frac{az_{1}+b}{cz_{1}+d}=1,~~\frac{az_{2}+b}{cz_{2}+d}=0,~~\frac{az_{3}+b}{%
cz_{3}+d}=\varepsilon ^{-1}.\ 
\end{equation}
The points $t_{4,5}$ are then given by
\begin{equation}\label{eq:t_def}
t_{4} =\frac{cz_{4}+d}{az_{4}+b}=(1-\varepsilon )z_{1234}+\varepsilon , \qquad 
t_{5} =\frac{cz_{5}+d}{az_{5}+b}=(1-\varepsilon )z_{1235}+\varepsilon ,
\end{equation}
where the conformal cross ratios $z_{ijkl}$ is defined as usual (cf. (\ref{cross})). Solving for the parameters $a,b,c,d$, we
get
\begin{eqnarray}
C_{5}\left( \left\{ \Delta _{i},z_{i}\right\}
\right) &\equiv \varepsilon ^{2\Delta _{3}} &\prod\limits_{j=1}^{5}\left\vert cz_{j}+d_{j}\right\vert
^{-2\Delta _{j}}=\left\vert z_{12}\right\vert ^{-\Delta _{1}-\Delta
_{2}+\Delta _{3}-\Delta _{4}-\Delta _{5}}\left\vert z_{13}\right\vert
^{-\Delta _{1}+\Delta _{2}-\Delta _{3}+\Delta _{4}+\Delta _{5}}  \notag \\ &&\hspace{3.3cm}\times \left\vert z_{23}\right\vert ^{\Delta _{1}-\Delta _{2}-\Delta
_{3}+\Delta _{4}+\Delta _{5}}\left\vert z_{34}\right\vert ^{-2\Delta
_{4}}\left\vert z_{35}\right\vert ^{-2\Delta _{5}}  \label{C_5}
\end{eqnarray}
and taking the limit $\varepsilon \rightarrow 0$ we obtain  for the celestial amplitude
\begin{equation}
\widetilde{\mathcal{A}}_{5}\left( \left\{ \Delta _{i}\right\} ,z_{1},\ldots
,z_{5}\right) =C_{5}\left( \left\{ \Delta _{i},z_{i}\right\} \right)
\lim_{\varepsilon \rightarrow 0}\varepsilon ^{-2\Delta _{3}}\widetilde{%
\mathcal{A}}_{5}\left( \left\{ \Delta _{i}\right\} ,1,0,\varepsilon
^{-1},t_{4}^{-1},t_{5}^{-1}\right) .
\end{equation}
Note that the gauge fixed amplitude $\lim_{\varepsilon\to 0}\varepsilon ^{-2\Delta _{3}}\widetilde{%
\mathcal{A}}_{5}\left( \left\{ \Delta _{i}\right\} ,1,0,\varepsilon
^{-1},t_{4}^{-1},t_{5}^{-1}\right)$  depends only on the invariants $t_{4,5}$
and is therefore conformally invariant by itself. 
Therefore, only the factor $C_{5}\left( \left\{ \Delta _{i},z_{i}\right\} \right) $ is responsible for the proper conformal covariance of the full amplitude. From the representation (\ref{5pt_general_form}) and using the explicit solution for $\sigma _{\ast i}$ (see Appendix~\ref{5pt_sigmas}), we observe that for the configuration $(1,0,\varepsilon ^{-1},t_{4}^{-1},t_{5}^{-1})$, we have 
\begin{equation}
\left\vert \det \mathbf{A}_{5}\right\vert ^{-1}=O\left( \varepsilon
^{2}\right) ,~~~\sigma _{\ast 3}=O\left( \varepsilon ^{2}\right) ,~~~\sigma
_{\ast i}=O(1) ,~~i\neq 3
\end{equation}
and thus the limit
\begin{equation}
f_{5}\left( \left\{ \Delta _{i}\right\} ,t_{4},t_{5}\right) \equiv
\lim_{\varepsilon \rightarrow 0}\frac{\varepsilon ^{-2\Delta
_{3}}\prod\limits_{i=1}^{5}\sigma _{\ast i}^{\Delta _{i}-1}\chi _{\langle
0,1\rangle }\left( \sigma _{\ast i}\right) }{2\left\vert \det \mathbf{A}%
_{5}\right\vert } \label{f_5}
\end{equation}
exists and is finite and conformally invariant. Note that the limit in (\ref{f_5}) can be performed explicitly using the solutions for $\sigma_{\ast i}$. Therefore, the celestial amplitude can be written in the form
\begin{equation}
\widetilde{\mathcal{A}}_{5}\left( \left\{ \Delta _{i}\right\} ,z_{1},\ldots
,z_{5}\right) =C_{5}\left( \left\{ \Delta _{i},z_{i}\right\} \right)
f_{5}\left( \left\{ \Delta _{i}\right\} ,t_{4},t_{5}\right) \mathcal{F}%
_{5}\left( \Delta ,t_{4},t_{5}\right) ,
\label{A_5_celestial}
\end{equation}
where the first two factors $C_5$ and $f_5$ are universal and given by (\ref{C_5}) and (\ref{f_5}) while the last term
\begin{equation}
\mathcal{F}_{5}\left( \Delta ,t_{4},t_{5}\right) \equiv \lim_{\varepsilon
\rightarrow 0}\int_{0}^{\infty }\mathrm{d}u~u^{\Delta -1}\left. A_{5}\left(
\left\{ \sqrt{u}\varepsilon _{j}\sigma _{\ast j}q\left( z_{j},\overline{z}%
_{j}\right) \right\} _{j=1}^{5}\right) \right\vert _{\left( 1,0,\varepsilon
^{-1},t_{4}^{-1},t_{5}^{-1}\right) }
\end{equation}
reflects a particular theory.

%=================================
\subsection{Four-point amplitude} 
%=================================

Similar consideration can be applied to the 4pt amplitude. In this case the reference configuration is $\left( 1,0,\varepsilon ^{-1},t_{4}^{-1}\right) $
where again $t_{4}=(1-\varepsilon )z_{1234}+\varepsilon$, and
\begin{equation}
\widetilde{\mathcal{A}}_{4}\left( \left\{ \Delta _{i}\right\} ,z_{1},\ldots
,z_{4}\right) =C_{4}\left( \left\{ \Delta _{i},z_{i}\right\} \right)
\lim_{\varepsilon \rightarrow 0}\varepsilon ^{-2\Delta _{3}}\widetilde{%
\mathcal{A}}_{4}\left( \left\{ \Delta _{i}\right\} ,1,0,\varepsilon
^{-1},t_{4}^{-1}\right), 
\end{equation}
where now
\begin{equation}
C_{4}\left( \left\{ \Delta _{i},z_{i}\right\} \right) =\left\vert
z_{12}\right\vert ^{-\Delta _{1}-\Delta _{2}+\Delta _{3}-\Delta
_{4}}\left\vert z_{13}\right\vert ^{-\Delta _{1}+\Delta _{2}-\Delta
_{3}+\Delta _{4}}\left\vert z_{23}\right\vert ^{\Delta _{1}-\Delta
_{2}-\Delta _{3}+\Delta _{4}}\left\vert z_{34}\right\vert ^{-2\Delta _{4}}\,.
\label{C_4}
\end{equation}
For the reference configuration we get
\begin{equation}
z_{13}=1-\varepsilon ^{-1},~~~z_{24}=-t_{4}^{-1},~~~z_{1234}=\frac{%
t_{4}-\varepsilon }{1-\varepsilon }
\end{equation}
and thus
\begin{align}
\widetilde{\mathcal{A}}_{4}\left( \left\{ \Delta _{i}\right\}
,1,0,\varepsilon ^{-1},t_{4}^{-1}\right)  &=\frac{4\varepsilon ^{2}t_{4}^{2}%
}{\left( 1-\varepsilon \right)^2 }\delta \left( -\mathrm{i}\left( t_{4}-%
\overline{t_{4}}\right) \right) \prod\limits_{i=1}^{4}\sigma _{\ast
i}^{\Delta _{i}-1}\chi _{\langle 0,1\rangle }\left( \sigma _{\ast i}\right) \\
&\times\frac{1}{2}\int_{0}^{\infty }\mathrm{d}u~u^{\Delta -1}\left. A_{4}\left(
\left\{ \sqrt{u}\varepsilon _{i}\sigma _{\ast i}q\left( z_{i},\overline{z}%
_{i}\right) \right\} _{i=1}^{4}\right) \right\vert _{\left( 1,0,\varepsilon
^{-1},t_{4}^{-1}\right) }.\notag
\end{align}
Again, in this formula (see Appendix~\ref{4pt_sigmas}), we have 
$\sigma _{\ast 3}=O\left( \varepsilon ^{2}\right)$ and $\sigma _{\ast
i}=O(1)$ for $i\neq 3$. Therefore we can write
\begin{equation}
\widetilde{\mathcal{A}}_{4}\left( \left\{ \Delta _{i}\right\} ,z_{1},\ldots
,z_{4}\right) =C_{4}\left( \left\{ \Delta _{i},z_{i}\right\} \right)
f_{4}\left( \left\{ \Delta _{i}\right\} ,t_{4}\right) \mathcal{F}_{4}\left(
\Delta ,t_{4}\right) ,
\label{A_4_celestial}
\end{equation}
where $C_4$ and $f_4$ are universal factors. The latter is given by
\begin{equation}
f_{4}\left( \left\{ \Delta _{i}\right\} ,t_{4}\right) =\delta \left( -%
\mathrm{i}\left( t_{4}-\overline{t_{4}}\right) \right) \lim_{\varepsilon
\rightarrow 0}\frac{2\varepsilon ^{2}t_{4}^{2}}{\left( 1-\varepsilon \right)^2 
}\varepsilon ^{-2\Delta _{3}}\prod\limits_{i=1}^{4}\sigma _{\ast i}^{\Delta
_{i}-1}\chi _{\langle 0,1\rangle }\left( \sigma _{\ast i}\right)
\label{limit1}
\end{equation}
and the theory-specific term is defined as
\begin{equation}
\mathcal{F}_{4}\left( \Delta ,t_{4}\right) \equiv \lim_{\varepsilon
\rightarrow 0}\int_{0}^{\infty }\mathrm{d}u~u^{\Delta -1}\left. A_{4}\left(
\left\{ \sqrt{u}\varepsilon _{j}\sigma _{\ast j}q\left( z_{j},\overline{z}%
_{j}\right) \right\} _{j=1}^{5}\right) \right\vert _{\left( 1,0,\varepsilon
^{-1},t_{4}^{-1}\right) }\,.
\label{limit2}
\end{equation}
Both  limits on the right hand sides of (\ref{limit1})and (\ref{limit2}) are finite and conformal invariant.  

The calculation of the nonuniversal factor $\mathcal{F}_{4}\left( \Delta
,t_{4}\right) $ can be further simplified. The amplitude $A_{4}\left(
\left\{ p_{i}\right\} _{i=1}^{4}\right) $ for scalar particles is a function of  Mandelstam variables
\begin{equation}
s=\left( p_{1}+p_{2}\right) ^{2},~~~t=\left( p_{1}+p_{3}\right)
^{2},~~~u=\left( p_{1}+p_{4}\right) ^{2}.
\label{Mandelstam_variables}
\end{equation}
Only two of them are independent, since $s+t+u=0$, let us write therefore
$A_{4}\equiv A_{4}\left( s,t;u\right)$. Inserting here $p_{i}=\sqrt{U}%
\varepsilon _{j}\sigma _{\ast j}q\left( z_{j},\overline{z}_{j}\right)$ at
the reference point, we get for $\varepsilon \rightarrow 0$
\begin{equation}
s=U\frac{t_{4}-1}{\left[ t_{4}\varepsilon _{1}\left( \varepsilon
_{2}-\varepsilon _{4}\right) +\varepsilon _{4}\left( \varepsilon
_{1}-\varepsilon _{2}\right) \right] ^{2}},~~~t=-\frac{s}{t_{4}}%
,~~~u=-\left( 1-\frac{1}{t_{4}}\right) s
\end{equation}
and for the theory-specific function $\mathcal{F}_{4}$ we get
\begin{equation}
\mathcal{F}_{4}\left( \Delta ,t_{4}\right) =\left\{ \frac{t_{4}-1}{\left[
t_{4}\varepsilon _{1}\left( \varepsilon _{2}-\varepsilon _{4}\right)
+\varepsilon _{4}\left( \varepsilon _{1}-\varepsilon _{2}\right) \right] ^{2}%
}\right\} ^{-\Delta }\int_{0}^{\infty }\mathrm{d}s~s^{\Delta -1}A_{4}\left(
s,-\frac{s}{t_{4}};-\left( 1-\frac{1}{t_{4}}\right) s\right) .
\end{equation}
For particles $1$ and $2$ incoming and $3$ and $4$ outgoing and for $\varepsilon \rightarrow 0$ we obtain
\begin{equation*}
\sigma _{\ast 1}=\frac{1}{2t_{4}},~~~\sigma _{\ast 2}=\frac{t_{4}-1}{2t_{4}}%
,~~~\sigma _{\ast 3}=\varepsilon ^{2}\frac{t_{4}-1}{2t_{4}^{2}},~~~\sigma
_{\ast 4}=\frac{1}{2}
\end{equation*}
and the universal factor $f_4$ takes the form\footnote{In the following formula, $\theta$ is the Heaviside step function.},
\begin{equation}
f_{4}\left( \left\{ \Delta _{i}\right\} ,t_{4}\right) =2^{-2\Delta +1}\delta
\left( -\mathrm{i}\left( t_{4}-\overline{t_{4}}\right) \right)
t_{4}^{6-\Delta _{1}-\Delta _{2}-2\Delta _{3}}\left( t_{4}-1\right) ^{\Delta
_{2}+\Delta _{3}-2}\theta(t_4-1),
\end{equation}
while for the function ${\cal F}_4$ we get
\begin{equation}
\mathcal{F}_{4}\left( \Delta ,t_{4}\right) =\left( \frac{t_{4}-1}{4t_{4}^{2}}%
\right) ^{-\Delta }\mathcal{G}_{4}\left( \Delta ,t_{4}\right) ,  \label{F_4}
\end{equation}
where we denoted
\begin{equation}
\mathcal{G}_{4}\left( \Delta ,t_{4}\right) \equiv \int_{0}^{\infty }\mathrm{d%
}s~s^{\Delta -1}A_{4}\left( s,-\frac{s}{t_{4}};-\left( 1-\frac{1}{t_{4}}%
\right) s\right) .  \label{G_4}
\end{equation}
Finally, the 4pt massless celestial amplitude of scalars takes the standard form (up to the delta function in the conformal cross ratio) dictated by conformal covariance
\begin{align}
\widetilde{\mathcal{A}}_{4}\left( \left\{ \Delta _{i}\right\} ,z_{1},\ldots
,z_{4}\right) &=2\delta \left( -\mathrm{i}\left( z_{1234}-\overline{z}_{1234}\right)
\right) \times \left\vert z_{12}\right\vert ^{-\Delta _{1}-\Delta _{2}+\Delta
_{3}-\Delta _{4}}\left\vert z_{13}\right\vert ^{-\Delta _{1}+\Delta
_{2}-\Delta _{3}+\Delta _{4}}  \notag \\
&\hspace{-2.5cm}\times \left\vert z_{23}\right\vert ^{\Delta
_{1}-\Delta _{2}-\Delta _{3}+\Delta _{4}}\left\vert z_{34}\right\vert
^{-2\Delta _{4}} z_{1234}^{2-(\Delta _{1}-\Delta _{2}+\Delta _{3}-\Delta
_{4})/2}z_{1423}^{(-\Delta _{1}+\Delta _{2}+\Delta _{3}-\Delta _{4})/2}\notag\\%
&\hspace{-2.5cm}\times\theta(z_{1234}-1)\mathcal{G}_{4}\left( \Delta ,z_{1234}\right), 
\label{A_4_simplified}
\end{align}
where we used that in the $\varepsilon\rightarrow0$ limit $t_{4}=z_{1234}$ and $t_4-1=z_{1234}z_{1423}$.

%======================================
\subsection{Explicit examples of exceptional EFT amplitudes}\label{eft_amps}
%======================================

Let us now look at amplitudes in the special class of theories reviewed in section~\ref{sec:2}, which have special soft limit behavior: $SU(N)$ NLSM, DBI and special Galileon. At 4pt these amplitudes take a simple form,
\begin{align}
A_4^{\rm NLSM} &= t\\
A_4^{\rm DBI} &= s^2 + t^2 + u^2\\
A_4^{\rm sGal} &= s^3+t^3+u^3\,.
\end{align}
In fact, these are unique kinematical invariants with $O(p^2),O(p^4)$ and $O(p^6)$ powercounting. For the $A_4=O(p^2)$ the amplitude is cyclic rather than permutational invariant as $s+t+u=0$. We can now calculate the celestial 4pt amplitude for all three cases, at least formally\footnote{Of course, all these theories are effective, i.e. physically meaningful only up to some intrinsic energy scale. Integration over full range of energies in the Mellin transform therefore does not make much sense. We quote the results here for illustrative purposes only.}. In fact, using the results of (\ref{A_4_simplified}) it is enough to calculate only the ${\cal G}_4(\Delta,t_4)$ function using (\ref{G_4}). Note that all three amplitudes are homogeneous and the integration over $s$ trivializes,
\begin{eqnarray}\label{nlsm}
    {\cal G}_4^{\rm NLSM}&=&-\frac{2\pi}{t_4}\delta(\Delta+1)\\
      {\cal G}_4^{\rm DBI}&=&4\pi\left(1-\frac{1}{t_4}+\frac{1}{t_4^2}\right)\delta(\Delta+2)\\
      {\cal G}_4^{\rm sGal}&=& \frac{6\pi}{t_4}\left(1-\frac{1}{t_4}\right)\delta(\Delta+3).\label{sGal}
\end{eqnarray}
We can also consider the 5pt amplitude in a (general) Galileon theory,
\begin{equation}
    A_5^{\rm Gal} =\det \{s_{ij}\}_{i,j=1}^4,
\end{equation}
where $s_{ij}=(p_i+ p_j)^2$.
This is again a homogeneous amplitude and we can calculate 
\begin{equation}
 \mathcal{F}_{5}\left( \Delta ,t_{4},t_5\right)  = 2\pi\delta(\Delta+4)\lim_{\varepsilon\to 0}\prod_{k=1}^{4} \sigma_{*k}\det\{|z_i-z_j|^2\}_{i,j=1}^4\vert _{\left( 1,0,\varepsilon
^{-1},t_{4}^{-1},t_5^{-1}\right) },
\end{equation}
where $\sigma_{*i}$'s are listed in Appendix~\ref{5pt_sigmas}. The explicit result is not much illustrative, let us note only that for $\overline{t_4}\to t_4$ it behaves as
\begin{equation}
  \mathcal{F}_{5}\left( \Delta ,t_{4},t_5\right)=O((\overline{t_4}-t_4)^2).
  \label{F_5_behavior_for_t_4_real}
\end{equation}
which mimics the ${\cal O}(t^2)$ behavior of the (general) Galileon amplitudes in the soft limit.

%===========================
\subsection{Conformal soft limit for the universal factors}\label{sec:cs5pt}
%===========================

Up to this point we have obtained expressions for 4pt and 5pt massless celestial amplitudes of scalars. Since our main aim is to study the conformal soft theorems, we make a short digression and, as a next step, we will consider the limit $\Delta_5\to-k$ of the universal factors $f_5$ and $C_5$ of the 5pt amplitude. The latter limit  corresponds to the case when the fifth particle becomes conformally soft, which in the momentum space reflects the $O(\omega_5^k)$ behavior in the limit $\omega_5\to 0$. For this purpose it is useful to consider the following formal equality
\begin{equation}
\lim_{\Delta _{5}\rightarrow -k}(\Delta _{5}+k)f_{5}\left( \left\{ \Delta
_{i}\right\} ,t_{4},t_{5}\right){\sigma _{\ast
5}^k} =\lim_{\varepsilon \rightarrow 0}\frac{%
\varepsilon ^{-2\Delta _{3}}\prod\limits_{i=1}^{4}\sigma _{\ast i}^{\Delta
_{i}-1}\chi _{\langle 0,1\rangle }\left( \sigma _{\ast i}\right) }{%
2\left\vert \det \mathbf{A}_{5}\right\vert }\delta \left( \sigma _{\ast
5}\right) \,,
\end{equation}
where we used the identity (see (2.8) in~\cite{Pate:2019mfs})\footnote{ In general, $\lim_{\Delta\to -k}\sigma^{\Delta-1}=(-1)^k\delta^{(k)}(\sigma)/k!$. Note also that for $f(\sigma)=O(\sigma^n)$ when $\sigma\to 0$ and for $n\ge k$, we get $\delta^{(k)}(\sigma)f(\sigma)=\delta(\sigma)\lim_{\sigma\to 0}f(\sigma)/\sigma^k$.
\label{footnote} }
\begin{equation}
\lim_{\Delta\rightarrow 0}\Delta \sigma _{\ast 5}^{\Delta
-1}=\delta \left( \sigma _{\ast 5}\right) .
\label{soft_delta_function}
\end{equation}
As a consequence of (\ref{sigma_i*_solution}) and (\ref{4pt_determinant}) we have
\begin{equation}
\delta \left( \sigma _{\ast 5}\right) =\frac{4}{\left\vert z_{13}\right\vert
^{2}\left\vert z_{24}\right\vert ^{2}}\delta \left( -\mathrm{i}\left(
z_{1234}-\overline{z}_{1234}\right) \right) \left\vert \det \mathbf{A}%
_{5}\right\vert =\frac{4\varepsilon ^{2}t_{4}^{2}}{\left( 1-\varepsilon
\right) }\delta \left( -\mathrm{i}\left( t_{4}-\overline{t_{4}}\right)
\right) \left\vert \det \mathbf{A}_{5}\right\vert .
\end{equation}%
At the 5pt reference point it can be also shown (cf. Appendix \ref{appendix_sigmas}) that for $i=1,\dots,4$ we get
\begin{equation}
\left. \sigma _{\ast i}\left( 1,0,\varepsilon
^{-1},t_{4}^{-1},t_{5}^{-1}\right) \right\vert _{\overline{t_{4}}%
=t_{4}}=\sigma _{\ast i}\left( 1,0,\varepsilon ^{-1},t_{4}^{-1}\right) ,
\end{equation}%
where on the right-hand side stays the corresponding solution for 4pt at the 4pt reference point. Therefore
\begin{equation}
\lim_{\Delta _{5}\rightarrow -k}(\Delta _{5}+k)f_{5}\left( \left\{ \Delta
_{i}\right\} ,t_{4},t_{5}\right)\sigma_{*5}^k =f_{4}\left( \left\{ \Delta _{i}\right\}
,t_{4}\right) .
\label{f_5_Delta_to_k_limit}
\end{equation}
Similarly from (\ref{C_5}) and (\ref{C_4}) we see that
\begin{equation}\label{eq:54_conf_soft_prefactor}
\lim_{\Delta _{5}\rightarrow -k}C_{5}\left( \left\{ \Delta _{i},z_{i}\right\}
\right) =C_{4}\left( \left\{ \Delta _{i},z_{i}\right\} \right)\left\vert \frac{z_{12}z_{35}^2}{z_{13}z_{23}}\right\vert^k.
\end{equation}
Thus, in the special case $k=0$, the universal conformal factor of the 4pt amplitude can be obtained formally as the residue of the universal conformal factor of the 5pt amplitude for $\Delta _{5}\rightarrow 0$. This relation is obviously not true in general for the full celestial amplitude including the theory-specific part ${\cal F}_5$. 

%=============================
\subsubsection*{General $n$-pt amplitude}
%=============================

The considerations performed above for a 4pt and 5pt amplitude can be generalized to an arbitrary $n$-pt amplitude of
massless particles. We again fix the gauge according to
\begin{equation}
\left( z_{1},\ldots ,z_{n}\right) =\left( 1,0,\varepsilon
^{-1},t_{4}^{-1},\ldots ,t_{n}^{-1}\right) ,  \label{n_gauge}
\end{equation}
the $t_{k}$ is then the conformal invariant cross-ratio in the limit $%
\varepsilon \rightarrow 0$
\begin{equation}
t_{k}=\left( 1-\varepsilon \right) z_{123k}+\varepsilon \,\,\rightarrow\,\, z_{123k}=\frac{z_{12}z_{3k}}{z_{13}z_{2k}}.
\end{equation}
The analog of the conformal factor (\ref{C_5}) reads now
\begin{equation}
C_{n}\left( \left\{ \Delta _{i},z_{i}\right\} \right) =\left\vert
z_{12}\right\vert ^{-\Delta _{1}-\Delta _{2}+\Delta
_{3}-\sum\limits_{i=4}^{n}\Delta _{i}}\left\vert z_{13}\right\vert
^{-\Delta _{1}+\Delta _{2}-\Delta _{3}+\sum\limits_{i=4}^{n}\Delta
_{i}}\left\vert z_{23}\right\vert ^{\Delta _{1}-\Delta _{2}-\Delta
_{3}+\sum\limits_{i=4}^{n}\Delta _{i}}\prod\limits_{j=4}^{n}\left\vert
z_{3i}\right\vert ^{-2\Delta _{i}}
\label{conformal_factor_n}
\end{equation}
and the celestial amplitude can be represented in the form
\begin{equation}
\widetilde{\mathcal{A}}_{n}\left( \left\{ \Delta _{i},z_{i}\right\} \right)
=C_{n}\left( \left\{ \Delta _{i},z_{i}\right\} \right) \mathcal{H}_{n}\left(
\left\{ \Delta _{i}\right\} ,t_{4},\ldots ,t_{n}\right) \,.
\label{general_npt}
\end{equation}
The conformal invariant Mellin integral $\mathcal{H}_{n}$ reads 
\begin{align}
&\mathcal{H}_{n}\left( \left\{ \Delta _{i}\right\} ,t_{4},\ldots
,t_{n}\right) \equiv \lim_{\varepsilon \rightarrow 0}\varepsilon ^{-2\Delta
_{3}}\widetilde{\mathcal{A}}_{n}\left( \left\{ \Delta _{i}\right\}
,1,0,\varepsilon ^{-1},t_{4}^{-1},\ldots ,t_{n}^{-1}\right)   \\
&=\lim_{\varepsilon \rightarrow 0}\varepsilon ^{-2\Delta _{3}}\int
[d\sigma,\Delta]\,\delta^{(4)}\Bigl( \sum\limits_{k=1}^{n}\varepsilon _{k}\sigma
_{k}q( z_{k},\overline{z_{k}}) \Bigr)\, \frac{1}{2}\int_{0}^{\infty}\mathrm{d}u~u^{\Delta
-1}A_{n}\left( \sqrt{u}\varepsilon _{1}\sigma _{1}q(z_{1},\overline{z}_{1}) ,\ldots \right) \Bigr\vert _{z_i^\ast},\notag 
\end{align}
where we evaluate $z_i^\ast$ on (\ref{n_gauge}). The first five $\sigma _{i}$'s can be solved in terms of the
remaining ones, the corresponding linear equations have the form of (\ref{sigma_equations}), now with
\begin{equation}
\mathbf{b}=\left( 
\begin{array}{c}
1-\sum\limits_{j=6}^{n}\sigma _{j} \\ 
-\sum\limits_{k=6}^{n}\varepsilon _{k}\sigma _{k}q\left( t_{k}^{-1},%
\overline{t_{k}}^{-1}\right) 
\end{array}%
\right) 
\end{equation}
and the same matrix $\mathbf{A}_{5}$. The nontrivial solution exists,
provided not all $\varepsilon _{k}$'s have the same sign. As above we can
write
\begin{align}
&\mathcal{H}_{n}\left( \left\{ \Delta _{i}\right\} ,t_{4},\ldots
,t_{n}\right) =\lim_{\varepsilon \rightarrow 0}\int \prod\limits_{i=6}^{n}%
\mathrm{d}\sigma _{i}\sigma _{i}^{\Delta _{i}-1}\frac{\varepsilon ^{-2\Delta
_{3}}\prod\limits_{i=1}^{5}\sigma _{\ast i}^{\Delta _{i}-1}\chi _{\langle
0,1\rangle }\left( \sigma _{\ast i}\right) }{2\left\vert \det \mathbf{A}%
_{5}\right\vert }  \notag \\
&\hspace{1.3cm}\times \left. \frac{1}{2}\int_{0}^{\infty }\mathrm{d}u~u^{\Delta
-1}A_{n}\left( \sqrt{u}\varepsilon _{1}\sigma _{\ast 1}q\left( 1,1\right)
,\ldots ,\sqrt{u}\varepsilon _{6}\sigma _{6}q\left( t_{6}^{-1},\overline{%
t_{6}}^{-1}\right) ,\ldots \right) \right\vert _{z_i^\ast}.
\end{align}
Note, however, that unlike the 5pt case the right-hand side does not
factorize, since $\sigma _{\ast i}$'s depend on the integration variables $\sigma _{j}$, $j=6,\ldots ,n$. Let us note that in analogy to \eqref{eq:54_conf_soft_prefactor} we have
for $j=6,\ldots ,n$
\begin{equation}
\lim_{\Delta _{j}\rightarrow -k}C_{n}\left( \left\{ \Delta _{i},z_{i}\right\}
_{i=1}^{n}\right) =C_{n-1}\left( \left\{ \Delta _{i},z_{i}\right\}
_{i=1}^{n-1}\right)\left\vert \frac{z_{12}z_{3j}^2}{z_{13}z_{23}}\right\vert^k ,
\label{C_n_to_C_n-1}
\end{equation}
and clearly for $i=1,\ldots ,5$ and $j=6,\ldots ,n$
\begin{equation}
\lim_{\sigma _{j}\rightarrow 0}\sigma _{\ast i}^{(n) }=\sigma
_{\ast i}^{(n-1) },
\label{sigma_n_to_sigma_n-1}
\end{equation}
where the superscript refers to $n-$point and $(n-1)-$point kinematics corresponding to the omission of the $j-$th particle
respectively.

%%%%%%%%%%%%%%%%%%%%%%%%%%%%%%%%%%%%%%%%%%%%%%%%
\section{Soft limits of celestial amplitudes}\label{sec:5}
%%%%%%%%%%%%%%%%%%%%%%%%%%%%%%%%%%%%%%%%%%%%%%%

In this section we will discuss the soft limits of the momentum space amplitudes $A_n$ and their relation to the {\emph{conformal}} soft limits of celestial amplitudes $\widetilde{{\cal A}}_n$ . Momentum space soft theorems are written as a power series in energy of the soft particle $i$. Soft theorem of order $k$ in energy gets translated via the Mellin transform to a conformal soft theorem for celestial amplitudes, encoded as a residue on the pole $\Delta _{i}=-k$. 
We have then the following general picture\footnote{In fact, this statement is more subtle. It is valid only when $\Delta_i=-k$ is situated on the left of the {\emph{fundamental strip}} where the Mellin integral is well defined. We will give explicit examples of these subtleties in what follows. }
\begin{align*}
\begin{array}{|c|}
\hline
\text{${\cal O}(t^k)$ term in the soft expansion of $A_n(p_i)$} \\
\text{for $p_i\rightarrow t p_i$ and $t\rightarrow0$} \\ \hline
\end{array}
\Longleftrightarrow
\begin{array}{|c|}
\hline
\text{residue of $\widetilde{\cal A}_n(\{\Delta_i,z_i\})$} \\
\text{on the pole $\Delta_i=-k$} \\ \hline
\end{array}
\end{align*}
This result was established and checked in many papers, for a selection see~\cite{Donnay:2018neh,Pate:2019mfs,Nandan:2019jas,Guevara:2019ypd,Fan:2019emx,Pasterski:2021dqe,Puhm:2019zbl,Banerjee:2021cly}. 
%

%=========================================
\subsection*{Soft theorem for 5pt amplitude}
%========================================

The case of 5pt amplitude is special, since it allows to formulate the conformal soft theorem  differently, solely in terms of the theory-specific conformal invariant functions $\mathcal{F}_{k}\left( \left\{ \Delta _{i},z_i\right\} \right)$ introduced in section \ref{sec:4}.

As an illustration, let us take the limit $\Delta _{5}\rightarrow0$ of a general 5pt celestial amplitude $\widetilde{{\cal{A}}}_5$ of massless scalars, assuming that it indeed probes the leading soft behavior, which implies for the momentum space soft theorem an expansion in energy (of the soft external leg) starting with a constant term. 
As we will see in what follows, in particular theories such a constant term coincides with a linear combination of the lower-point (i.e. 4pt) amplitudes in the same theory, i.e.
\begin{equation}
    \lim_{p_5\to 0}A_5(p_1,\dots,p_5)=\sum_j a_j A_4^{(j)}(p_1,\dots,p_4)\,.
\label{model_soft_theorem}
\end{equation} 
Recall that the 4pt and 5pt amplitudes can be written in the form 
(\ref{A_4_celestial}), (\ref{A_5_celestial}) as%
\begin{equation}
\widetilde{\mathcal{A}}_{k}\left( \left\{ \Delta _{i},z_i\right\} \right)  = C_{k}
f_{k} \mathcal{F}_{k} ,\,\,\,\,k=4,5,
\end{equation}
where $C_{k}$'s are the universal conformally covariant factors, $f_{k}$'s are universal conformal invariants, while $\mathcal{F}_{k}$'s are functions specific for the given theory.  As we have shown in section~\ref{sec:cs5pt}, the following limits hold
\begin{eqnarray}
\lim_{\Delta _{5}\rightarrow 0}\Delta _{5}f_{5}\left( \left\{ \Delta
_{i}\right\} ,t_{4},t_{5}\right)  &=&f_{4}\left( \left\{ \Delta _{i}\right\}
,t_{4}\right) \sim \delta \left( \sigma _{\ast 5}\right) \sim \delta \left( -%
\mathrm{i}\left( t_{4}-\overline{t_{4}}\right) \right) ,  \notag \\
\lim_{\Delta _{5}\rightarrow 0}C_{5}\left( \left\{ \Delta _{i},z_{i}\right\}
\right)  &=&C_{4}\left( \left\{ \Delta _{i},z_{i}\right\} \right) ,
\end{eqnarray}
and therefore the residue of $\widetilde{\mathcal{A}}_{5}\left(
\left\{ \Delta _{i}\right\} ,z_{1},\ldots ,z_{5}\right)$ on the $\Delta_5=0$ pole can be written as
\begin{equation}
\lim_{\Delta _{5}\rightarrow 0}\Delta _{5}\widetilde{\mathcal{A}}_{5}\left(
\left\{ \Delta _{i}\right\} ,z_{1},\ldots ,z_{5}\right) =C_{4}\left( \left\{
\Delta _{i},z_{i}\right\} \right) f_{4}\left( \left\{ \Delta _{i}\right\}
,t_{4}\right) \lim_{t_{4}\rightarrow \overline{t_{4}}}\mathcal{F}_{5}\left(
\Delta ,t_{4},t_{5}\right).
\label{Delta_to_0_A_5_residue}
\end{equation}
Notice that on the right-hand side the total scaling dimension $\Delta$ is summed over scaling dimensions of four particles as appropriate for a 4pt function 
\begin{equation}
\Delta =\frac{1}{2}\sum\limits_{i=1}^{4}\Delta _{i}-2.
\end{equation}
For the theory-dependent part ${\cal F}_5$ we get
\begin{equation}
\mathcal{F}_{5}\left( \Delta ,t_{4},t_{5}\right) =\lim_{\varepsilon
\rightarrow 0}\int_{0}^{\infty }\mathrm{d}u~u^{\Delta -1}\left. A_{5}\left(
\left\{ \sqrt{u}\varepsilon _{j}\sigma _{\ast j}q\left( z_{j},\overline{z}%
_{j}\right) \right\} _{j=1}^{5}\right) \right\vert _{\left( 1,0,\varepsilon
^{-1},t_{4}^{-1},t_{5}^{-1}\right) },  \label{F_5}
\end{equation}
and since $\sigma _{\ast 5}\rightarrow 0$ for $t_{4}\rightarrow \overline{t_{4}}$, the limit $t_{4}\rightarrow \overline{t_{4}}$ probes the soft behavior of the fifth particle, and the result does not depend on $t_{5}$. Also, the $t_{4}\rightarrow \overline{t_{4}}$ limit of $\left\{ \sigma
_{\ast j}\right\} _{j=1}^{4}$ coincides with the corresponding solution of the 4pt kinematics. Therefore, purely kinematically, $\lim_{t_{4}\rightarrow 
\overline{t_{4}}}\mathcal{F}_{5}\left( \Delta ,t_{4},t_{5}\right) $
resembles the function $\mathcal{F}_{4}\left( \Delta ,t_{4}\right) $ for
some 4pt amplitudes. 
On the other hand, assuming the momentum space soft
theorem in the form (\ref{model_soft_theorem}), we   get the $t_{4}\rightarrow \overline{%
t_{4}}$ limit of the integrand of (\ref{F_5}) just in
terms of linear combinations of 4pt amplitudes and thus the conformal soft theorem reads
\begin{equation}
 \lim_{t_{4}\rightarrow \overline{%
t_{4}}}\mathcal{F}_{5}\left( \Delta ,t_{4},t_{5}\right)=\sum_j a_j   \mathcal{F}_{4}^{j}\left( \Delta ,t_{4}\right).
\label{general_F_5_soft}
\end{equation}
Therefore, in such a case, the conformal soft theorem for a 5pt function can be formulated solely in terms of the functions $\mathcal{F}_{5}\left( \Delta ,t_{4},t_{5}\right) $
and $\mathcal{F}_{4}\left( \Delta ,t_{4}\right)$.

Note that the ${\cal O}(1)$ term did not have to be a leading term in the soft expansion. For example, if the amplitude behaved as $A_5\sim {\cal O}(\frac{1}{t})$ our calculation would correspond to the subleading soft theorem while the leading soft behavior would be captured by the residue of $\widetilde{\cal A}_5$ on $\Delta_i=1$ pole.

As a more explicit example, we can  consider the 5pt amplitude in the Galileon theory. The ${\cal O}(t^2)$ behavior of $A_5(\dots,tp_i,\dots)$ for $t\to 0$ indicates that both the $\Delta_i=0$ and $\Delta_i=-1$ residues vanish. This is of course the case. Using (\ref{Delta_to_0_A_5_residue}) and the behavior of ${\cal{F}}_5$ given by (\ref{F_5_behavior_for_t_4_real}), we have immediately
\begin{equation}
   \lim_{t_{4}\rightarrow \overline{%
t_{4}}}\mathcal{F}_{5}\left( \Delta ,t_{4},t_{5}\right)=0, 
\end{equation}
which implies
\begin{equation}
  \lim_{\Delta _{5}\rightarrow 0} \Delta _{5} \widetilde{%
\mathcal{A}}_{5}(\left\{ \Delta _{i},z_{i}\right\} )=0.  
\end{equation}
Similarly, using (\ref{f_5_Delta_to_k_limit}) and (\ref{eq:54_conf_soft_prefactor}) for $k=1$, we get (cf. footnote \ref{footnote})
\begin{equation}
\lim_{\Delta _{5}\rightarrow -1}\left( \Delta _{1}+1\right) \widetilde{%
\mathcal{A}}_{5}(\left\{ \Delta _{i},z_{i}\right\} )=C_{4}(\left\{ \Delta
_{i},z_{i}\right\} )f_{4}(\left\{ \Delta _{i}\right\} ,t_{4})\left\vert 
\frac{z_{12}z_{35}^{2}}{z_{13}z_{23}}\right\vert \lim_{\overline{t_{4}}%
\rightarrow t_{4}}{{\sigma _{\ast 5}}}^{-1}\mathcal{F}_{5}\left( \Delta
,t_{4},t_{5}\right). 
\end{equation}
As shown in Appendix \ref{5pt_sigmas}, we have
$ \sigma_{*5}=O(t_4-\overline{t_4})$,
which together with (\ref{F_5_behavior_for_t_4_real}) gives finally
\begin{equation}
  \lim_{\overline{t_{4}}%
\rightarrow t_{4}}{{\sigma _{\ast 5}}}^{-1}\mathcal{F}_{5}\left( \Delta
,t_{4},t_{5}\right)=0  
\end{equation}
and therefore
\begin{equation}
 \lim_{\Delta _{5}\rightarrow -1}\left( \Delta _{1}+1\right) \widetilde{%
\mathcal{A}}_{5}(\left\{ \Delta _{i},z_{i}\right\} )=0   \,.
\end{equation}

Note that in all cases, any enhanced soft behavior of the momentum amplitude $A_n$, i.e. vanishing in the soft limit like ${\cal O}(t^k)$, is directly translated to vanishing of the celestial amplitudes on the residues $\Delta_i=-j$ for $j=0,{\dots},k{-}1$. This is a rather trivial consequence of the definition of ${\cal F}_n$ which is directly proportional to $A_n$. Hence, we do not expect any non-trivial information about soft behavior that can be translated from $\widetilde{\mathcal{A}}_n$ to $A_n$.

%%%%%%%%%%%%%%%%%%%%%%%%%%%%%%%%%%%%%%%%%%%%%
\section{Celestial amplitudes of UV completed models}\label{sec:6}
%%%%%%%%%%%%%%%%%%%%%%%%%%%%%%%%%%%%%%%%%%%%%

The celestial amplitude in a generic quantum field theory is not well defined. As discussed earlier, the Mellin transformation requires sufficiently good (and correlated) behavior of the scattering amplitude at low and at high energies such that the $u$-integration in (\ref{mellin_transform_complete}) can be performed. The Mellin transformation is well defined and holomorphic as a function of $\Delta$ for $\Delta$ within a \emph{fundamental strip}. The size of the fundamental strip is given by the behavior of the amplitude at both low and high energies.
For an $n$-pt amplitude, we determine the fundamental strip directly from the $u$-integral in~\eqref{mellin_transform_complete}. Concretely, we rescale each external momentum in $A_n$ by $\sqrt{u}$ as in~\eqref{masslesskin}. To fix the left border of the fundamental strip we calculate the leading term in $A_n$ for $u\to0$. The right border is analogously fixed by the $u\to\infty$ leading term. Therefore, if $A_n(u)$ scales as
\begin{align}
A_n(u) \overset{u\rightarrow 0}{=} u^a,\qquad A_n(u) \overset{u\rightarrow \infty}{=} {u^b}  \,,
\end{align}
the fundamental strip is $\langle -a,-b\rangle\equiv \{z\in \mathbb{C}|-a<{\rm {Re}}z<-b\}$.
In particular, if the theory is renormalizable, $b=2-n/2$.
Under some additional assumptions on the assymptotics for $u\to 0$ and $u\to\infty$, the Mellin transform can be analytically continued outside the fundamental strip to a meromorphic function with simple poles and with residues in one-to-one correspondence with the coefficients of the asymptotic expansion (see Appendix \ref{complex delta} and \cite{Flajolet} for more details). In what follows, we will always assume the Mellin transform to correspond to such a maximal possible continuation.

The fundamental strip can be also empty as indicated by the presence of delta function $\delta(\Delta+d)$ with $d=1,2,3$ in the explicit examples discussed earlier, and the Mellin transformation can be then only understood in the distributional sense. This reflects the fact that the corresponding theories are effective ones that cannot be taken seriously above some intrinsic scale, while the Mellin transform takes into account the whole energy range without any constraint.
Therefore, to obtain a well-defined celestial amplitude for a scattering of Goldstone bosons, it is desirable to start with a (partial) UV completion of the  corresponding effective theory.

In what follows,  we look at further particular models for scattering amplitudes. First, we discuss two linear sigma models where the UV completion is provided by a massive particle. Next, we explore the $U(1)$ fibered $\mathbb{C}P^{N{-}1}$ model studied in \cite{Kampf:2019mcd} as an interesting example of a theory of Goldstone bosons without an Adler zero. This model is not UV complete so we provide a simple UV completion which leads to well-defined celestial amplitudes. We compare it with the result in the effective field theory (original model without a UV completion) which yields celestial amplitudes with delta functions in $\Delta$. Then we study a pion-dilaton model and the corresponding conformal soft theorems. Finally, we explore the celestial amplitudes in the NLSM model and a certain UV completion of this theory called $Z$-theory, studied in the context of color-kinematics duality \cite{Carrasco:2016ldy,Carrasco:2016ygv}. This provides us with the most transparent difference between the celestial amplitudes in renormalizable theory vs effective field theory.

%======================================
\subsection{$U(1)$ linear sigma model}
%======================================

The simplest example of a theory with interacting Goldstone bosons we present is the $U(1) $ linear sigma model with the Lagrangian
\begin{equation}
\mathcal{L}=\partial \phi ^{\ast }\cdot \partial \phi -\frac{\lambda }{4}%
\left( \phi ^{\ast }\phi -\frac{\mu ^{2}}{\lambda }\right) ^{2}.
\end{equation}
Despite its simplicity, it allows to illustrate some apparent paradoxes concerning both the conformal soft theorems and the relationship between the celestial amplitudes of the full theory and the corresponding low-energy theory.

Expanding the Lagrangian around the classical vacuum $\langle \phi \rangle =\mu /\sqrt{%
\lambda }$ and using the parametrization
\begin{equation}
\phi =\frac{1}{\sqrt{2}}\left( \sigma +\mathrm{i}\pi \right) +\frac{\mu }{%
\sqrt{\lambda }},
\end{equation}
we identify the field $\pi $ as the Goldstone boson of the spontaneously
broken $U(1)$ symmetry and $\sigma $ as the massive Higgs particle with mass $\mu .$ Explicitly we get
\begin{eqnarray}
\mathcal{L} &\mathcal{=}&\frac{1}{2}\left( \partial \pi \right) ^{2}+\frac{1%
}{2}\left( \partial \sigma \right) ^{2}-\frac{1}{2}\mu ^{2}\sigma
^{2}-V\left( \pi ,\sigma \right)   \notag \\
V\left( \pi ,\sigma \right)  &=&\frac{\lambda }{16}\sigma ^{4}+\frac{\lambda 
}{16}\pi ^{4}+\frac{1}{2\sqrt{2}}\mu \sqrt{\lambda }\sigma ^{3}+\frac{%
\lambda }{8}\pi ^{2}\sigma ^{2}+\frac{1}{2\sqrt{2}}\mu \sqrt{\lambda }\sigma
\pi ^{2}.
\label{eq: Lagrangian pi-sigma}
\end{eqnarray}
The tree-level four Goldstone boson amplitude $\pi \left( p_{1}\right) +\pi
\left( p_{2}\right) \rightarrow \pi \left( p_{3}\right) +\pi \left(
p_{4}\right) $ reads then
\begin{align}\label{eq:U1LSM_4pi}
A_{4}\left( s,t;u\right) &=-\frac{3}{2}\lambda -\frac{\lambda \mu ^{2}}{2}%
\left( \frac{1}{s-\mu ^{2}}+\frac{1}{t-\mu ^{2}}+\frac{1}{u-\mu ^{2}}\right) \notag \\
&=-\frac{\lambda}{2}\left(\frac{s}{s-\mu^2}+\frac{t}{t-\mu^2}+\frac{u}{u-\mu^2}\right),
\end{align}
where the Mandelstam variables are defined as usual
\begin{equation}
s=\left( p_{1}+p_{2}\right) ^{2},~~~t=\left( p_{1}-p_{3}\right)
^{2}=-as,~~~u=\left( p_{1}-p_{4}\right) ^{2}=-(1-a)s,
\end{equation}
and where the parameter $a$ is related to the CMS scattering angle and conformal cross-ratio (cf.~(\ref{eq:t_def})) as
\begin{equation}
a=\sin ^{2}\frac{\theta }{2}=\frac{1}{z_{1234}}\equiv\frac{1}{t_4},\qquad t_4\in (1,\infty).
\end{equation}
The amplitude (\ref{eq:U1LSM_4pi}) obeys Adler zero, i.e. it vanishes whenever any of the external particles becomes soft.
Since the one-particle soft limit  $p_i\rightarrow 0$ means $s\rightarrow 0$,
we expect, that such a behavior implies vanishing of the residue of the corresponding nonuniversal part of the celestial amplitude $\mathcal{G}_{4}\left( \Delta ,t_{4}\right)$ at $\Delta\rightarrow 0$.
However, more detailed analysis shows that this is not the case.

The asymptotic expansion of the amplitudes for $s\rightarrow 0$ with $a$ fixed reads (here we denote for short  $A_{4}\left( s,-as;-(1-a)s\right)\equiv M_{4}\left( s,a\right) $)
\begin{equation}
M_{4}\left( s,a\right) \overset{s\rightarrow 0}{=}\frac{\lambda }{2}%
\sum_{k=2}^{\infty }\frac{s^{k}}{\mu ^{2k}}\left[ 1+\left( -1\right)
^{k}\left( a^{k}+\left( 1-a\right) ^{k}\right) \right] .  \label{M_LSM_s->0}
\end{equation}
Similarly, for $s\rightarrow \infty$
\begin{equation}
M_{4}\left( s,a\right) \overset{s\rightarrow \infty }{=}-\frac{3}{2}\lambda -%
\frac{\lambda }{2}\sum_{k=1}^{\infty }\frac{\mu ^{2k}}{s^{k}}\left[
1+\left( -1\right) ^{k}\left( a^{-k}+\left( 1-a\right) ^{-k}\right) \right] .
\label{M_LSM_s->infty}
\end{equation}
The nonuniversal part $\mathcal{G}_{4}\left( \Delta ,t_{4}\right) $ of the celestial amplitude can be expressed according to the general formulae  (\ref{F_4}), (\ref{G_4})
\begin{equation}
\mathcal{G}_{4}\left( \Delta ,t_{4}\right) \equiv \int_{0}^{\infty }\mathrm{d%
}s~s^{\Delta -1}M_{4}\left( s,\frac{1}{t_{4}}\right) .
\label{G4_duplicate}
\end{equation}
The definition of the conformal cross-ratio $t_4$ was recalled above and for $\Delta$ see~\eqref{eq:def_delta}. By inspection of (\ref{M_LSM_s->0}) and (\ref{M_LSM_s->infty}), the fundamental
strip for this Mellin transform is $\langle-2,0\rangle$. For $\Delta $
inside this strip, we easily find using the residue theorem (for a general formula see (\ref{integral_mellin})) 
\begin{equation*}
\mathcal{G}_{4}\left( \Delta ,t_{4}\right) =\frac{\pi \lambda }{2}\mu
^{2\Delta }\left[ \mathrm{i}+\cot \left( \Delta \pi \right) +\left( \left( 
\frac{1}{t_{4}}\right) ^{-\Delta }+\left( 1-\frac{1}{t_{4}}\right) ^{-\Delta
}\right) \csc \left( \Delta \pi \right) \right] .
\end{equation*}
This result can be analytically continued outside the fundamental strip.
Note, that for $\Delta_i$ corresponding to the principal continuous series, i.e. $\Delta_i=1+{\rm{i}}\lambda_i$, the value $\Delta$, which is defined by (\ref{eq:def_delta}), is purely imaginary, and  it is placed outside the fundamental strip $\langle-2,0\rangle$.

The output of such a continuation is a meromorphic function with simple poles for $\Delta =k\in 
\mathbb{Z}$, $k\neq -1$ and corresponding residues (see Fig.~\ref{fig:U1_LSM_scal_dim})
\begin{equation*}
\mathrm{res}\left( \mathcal{G}_{4},\Delta=k\right) =\frac{\lambda }{2}\mu ^{2k}%
\left[ 1+(-1)^{k}\left( \left( \frac{1}{t_{4}}\right) ^{-k}+\left( 1-\frac{1%
}{t_{4}}\right) ^{-k}\right) \right] .
\end{equation*}
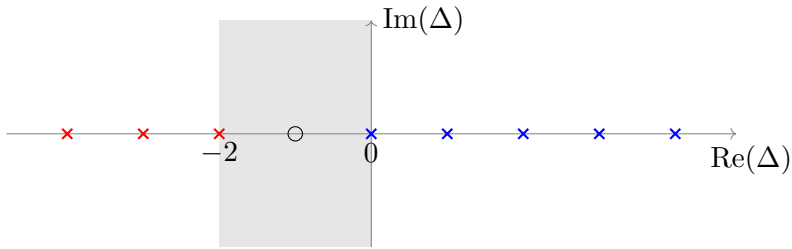
\begin{figure}[ht]
\centering
\begin{tikzpicture}
\tikzset{cross/.style={cross out, draw=black, minimum size=2*(#1-\pgflinewidth), inner sep=0pt, outer sep=0pt},
%default radius will be 1pt. 
cross/.default={2.7pt}}
% configurable parameters
\def\bigradius{3}

\draw [fill=gray!20,draw=none] (-2,-.5*\bigradius) rectangle (0,.5*\bigradius);
% Axes
\draw [help lines,->] (-1.6*\bigradius, 0) -- (1.6*\bigradius,0);
\draw [help lines,->] (0, -.5*\bigradius) -- (0, .5*\bigradius);

% The labels
	\node at (5,-0.35){$\mathrm{Re}(\Delta)$};
	\node at (0.69,1.5) {$\mathrm{Im}(\Delta)$};
	\node at (0,-0.25) {$0$};
	\node at (-2,-0.25) {$-2$};

\draw (0,0) node[cross,blue,thick] {};
\draw (1,0) node[cross,blue,thick] {};
\draw (2,0) node[cross,blue,thick] {};
\draw (3,0) node[cross,blue,thick] {};
\draw (4,0) node[cross,blue,thick] {};
\draw (-1,0) circle (2.7pt);
\draw (-2,0) node[cross,red,thick] {};
\draw (-3,0) node[cross,red,thick] {};
\draw (-4,0) node[cross,red,thick] {};
\end{tikzpicture}
	\caption{Positions and residues of red poles encode the low energy (IR) EFT expansion of the amplitude (i.e. Wilson coefficients). Blue ones correspond to the high energy (UV) asymptotic expansion. Gray strip (outer edges excluded) represents the fundamental strip for the Mellin transform: the celestial amplitude is guaranteed to be holomorphic there. Hollow circle means the absence of a pole for the given value of $\Delta$ (the circle inside the strip implies the vanishing of the $s^1$ IR Wilson coefficient).} \label{fig:U1_LSM_scal_dim}
\end{figure}
These correspond for $k\leq -2$ to the coefficients of the asymptotic
expansions (\ref{M_LSM_s->0}) and for $k\geq 0$ to the coefficients of the asymptotic expansion (\ref{M_LSM_s->infty}). 
Especially,
\begin{equation}
\mathrm{res}\left( \mathcal{G}_{4},\Delta=0\right) =\frac{3 }{2}\lambda\ne 0,
\end{equation}
despite the presence of Adler zero.
Note, however, that the pole at $\Delta =0$ sits \emph{on the right} of the fundamental strip, and therefore it reflects the leading order term for the asymptotics at $s\rightarrow\infty$. This is in apparent contradiction with the usual expectation that a pole at $\Delta =0$ encodes the soft behavior of the amplitude.

Let us now comment on the limit $\mu \rightarrow \infty$, i.e. the limit of infinitely heavy $\sigma $ resonance. 
In principle, there are two ways how to send the mass of $\sigma$ to infinity.
The first one corresponds to performing the limit $\lambda\rightarrow \infty$ while keeping the vacuum expectation value $v\equiv\langle \phi \rangle =\mu /%
\sqrt{\lambda }$ fixed. 
The latter plays a role of the scale of the spontaneous symmetry breaking and sets the strength of the leading order effective derivative interaction of the Goldstone boson $\pi$.
In this case we get
\begin{equation}
    \lim_{\lambda\to\infty, v\,{\rm{fixed}}}A_4(s,t,u)=\frac{1}{v^2}(s+t+u)=0.
    \label{nondecoupling_scenario}
\end{equation}
Due to the spontaneously broken symmetry, this limit is non-decoupling\footnote{Here we use the term ``non-decoupling'' in the sense of Appelquist-Carazzone theorem: in our case, integrating out the heavy $\sigma$, the effective Lagrangian describing the low-energy $\pi$ scattering   {\em{does not}} correspond to the (renormalized) $\sigma$ independent part of the original Lagrangian (\ref{eq: Lagrangian pi-sigma}) plus  non-renormalizable higher dimensional $\pi$ dependent operators with couplings suppressed by inverse powers of $\mu$. The latter scenario would mean {\em{decoupling}} of the dynamical effects of the heavy $\sigma$ in the limit $\mu\to\infty$. In the spontaneusly broken $U(1)$ linear sigma model, the dynamical effect of the heavy $\sigma$ particle does not decouple in this way, instead, it generates nontrivial contributions to the Goldstone boson scattering amplitudes. In this case, the effective Lagrangian is free at the leading order - the trivial $S$-matrix is  a result of the conspiracy between the direct $\pi$ dynamics and the non-decoupling effects of the heavy $\sigma$ exchange. Such a scenario is usually called non-decoupling scenario in this context.  }: the Feynman diagrams with $\sigma$ exchange
do not decouple and their remnant cancels the contribution of the leading $O(\lambda)$ contact
term. The right hand side of (\ref{nondecoupling_scenario}) represents then the subleading $ O(1)$ term, which however vanishes due to the kinematics\footnote{This is in fact the consequence of the Abelian nature of the broken global symmetry. Such a term generally survives in the non-Abelian case, cf. the next subsection.}.

This result can also be understood on the Lagrangian level when we integrate out
the $\sigma $ resonance at the leading order in the derivative expansion.
Indeed, changing the parametrization of the original Lagrangian as 
\begin{equation}
\phi =\frac{\Sigma }{\sqrt{2}}\exp \left( \mathrm{i} \frac{\sqrt{\lambda }%
}{\sqrt{2}\mu }\Pi\right) ,
\end{equation}
we get
\begin{equation}
\mathcal{L}=\frac{1}{2}\partial \Sigma ^{2}+\frac{1}{4}\frac{\lambda }{\mu
^{2}}\Sigma ^{2}\partial \Pi ^{2}-\frac{\lambda }{16}\left( \Sigma ^{2}-%
\frac{2\mu ^{2}}{\lambda }\right) ^{2}.
\end{equation}
In the limit $\lambda\rightarrow\infty$, the field $\Sigma $
freezes at its vacuum expectation value $\langle \Sigma \rangle =\mu %
\sqrt{2/\lambda }=\sqrt{2}v$ and the leading order effective Lagrangian for the
Goldstone boson $\Pi$ is just the free theory\footnote{To be more precise, since the field $\Pi$ is in fact compact scalar with the identification $\Pi\sim \Pi+2\sqrt{2}\pi v$, the resulting theory is rather a nonlinear sigma model,  with the target space corresponding to a circle with radius $\sqrt{2}v$. However, perturbatively, there is no difference between this sigma model and a free theory with target space $\mathbb R$. }.
Of course it does not mean, that the Goldstone bosons do not interact at all when the mass of $\sigma$ is large but finite. The higher order terms in the $1/\lambda $ expansion of the amplitude are nonzero and scale as $(s/v^2)\times (s/\mu^2)^k$, $k>0$.

The second way of sending the mass of $\sigma$ to infinity corresponds to keeping the coupling constant $\lambda$ fixed and taking then the limit $\mu \rightarrow \infty$. This means to send the vacuum expectation value $\langle \phi \rangle =\mu /%
\sqrt{\lambda }$ to infinity at the same time. 
As a result, we immediately obtain
\begin{equation}
\lim_{\mu \rightarrow \infty,\ \lambda\,{\rm{fixed}} }A_{4}\left( s,t;u\right) =0.
\end{equation}
Also in this case, the amplitude vanishes  as a result of the cancellation between the direct contact interaction of the Goldstone bosons and the remnant of the non-vanishing contribution of the $\sigma$ exchange graphs of the original theory. 
However, the physical interpretation is quite different.
The resulting effective theory describing the interaction of Goldstone bosons  is again free\footnote{In contrast to the previous case, this is a free theory with target space $\mathbb R$, since the radius of the circle on which the field $\Pi$ lives is send to infinity.}, but now as a result of the infinite scale of their effective interaction, i.e. the Goldstone bosons effectively decouple. 
The higher order terms in the $1/\mu^2$ 
expansion of the amplitude now scale as $\lambda\times (s/\mu^2)^k$, $k>1$.

Despite the difference of the above two scenarios, in the momentum space, the amplitude within the leading order of effective theory can be recovered by the limit of infinitely heavy Higgs.
For the celestial amplitude, however, one has to be more careful. 
In fact, the effective description cannot be reproduced within the second scenario, since $\mathcal{G}_{4}\left( \Delta ,t_{4}\right) $ is sensitive to the full energy region. No matter how big the mass $\mu$ of the Higgs  is, the energy region $s\approx \mu^2$ contributes non-trivially to the integral on the right hand side of (\ref{G4_duplicate}). 
Indeed, using the formula (cf. Appendix \ref{complex delta} and \cite{Donnay:2020guq}.)
\begin{equation}
\lim_{\mu \rightarrow \infty }\mu ^{2\Delta }=\lim_{\mu \rightarrow \infty }%
\frac{1}{\Gamma \left( \Delta \right) }\int\limits_{0}^{\infty }\mathrm{d}
\alpha\,\alpha ^{\Delta -1}e^{-\alpha /\mu ^{2}}=2\pi \Delta \delta \left(
\Delta \right) ,
\label{delta_representation_limit}
\end{equation}
which is valid in this form for imaginary $\Delta$, i.e. for $\Delta_i$ on the principal continuous series\footnote{However,  a formula of this type can have more general meaning even for general complex $\Delta$, cf. Appendix \ref{complex delta} and \cite{Donnay:2020guq}.}, we obtain (cf. also \cite{Atanasov:2021cje}, where similar considerations were used in the context of different model)
\begin{eqnarray}
\lim_{\mu \rightarrow \infty,\,\lambda\,{\rm{fixed}} }\mathcal{G}_{4}\left( \Delta ,t_{4}\right)
&=&\pi ^{2}\lambda \delta \left( \Delta \right) \lim_{\Delta \rightarrow
0}\Delta \left[ \mathrm{i}+\cot \left( \Delta \pi \right) +\left( \left( 
\frac{1}{t_{4}}\right) ^{-\Delta }+\left( 1-\frac{1}{t_{4}}\right) ^{-\Delta
}\right) \csc \left( \Delta \pi \right) \right] \notag\\
&=&3\pi \lambda \delta \left( \Delta \right) .
\end{eqnarray}
In contrary to the usual intuition, the $\mu \rightarrow \infty $, $\lambda$ fixed limit of the celestial amplitude does not reflect the IR behavior of the amplitude in the momentum space. It is connected with the residue of the celestial amplitude at $\Delta =0$, which, as we have discussed above, corresponds to  the leading order UV behavior of the amplitude $A_{4}\left( s,t;u\right)$. This is a demonstration of a general phenomenon: we can not straightforwardly extract a celestial amplitude in effective theory using the intuition form the momentum space, i.e.  taking blindly the limit $\mu\rightarrow\infty$, from the celestial amplitude in the complete theory. The Wilsonian paradigm is replaced here by the anti-Wilsonian one as discussed in detail in \cite{Arkani-Hamed:2020gyp}.

On the other hand within the first scenario, and using formally (see Appendix \ref{complex delta} for more rigorous treatment)
\begin{equation}
 \lim_{\lambda\rightarrow \infty}\lambda^{\Delta+1} =2\pi(\Delta+1)\delta (\Delta+1),
\end{equation}
we get
\begin{eqnarray}
\lim_{\lambda \rightarrow \infty,\,v\,{\rm{fixed}} }\mathcal{G}_{4}\left( \Delta ,t_{4}\right)
&=&\pi ^{2}{v^{-2}} \delta \left( \Delta+1 \right) \lim_{\Delta \rightarrow
-1}(\Delta+1)\notag\\&&\times \left[ \mathrm{i}+\cot \left( \Delta \pi \right) +\left( \left( 
\frac{1}{t_{4}}\right) ^{-\Delta }+\left( 1-\frac{1}{t_{4}}\right) ^{-\Delta
}\right) \csc \left( \Delta \pi \right) \right]\notag \\
&=&0 .
\end{eqnarray}
and the celestial amplitude of the leading order effective theory is correctly reproduced.

%=======================================
\subsection{$O(N)$ sigma models}
%=========================================

A straightforward generalization of the previous example is represented by
the $O(N)$ linear sigma model with real scalar field $N-$plet $\boldsymbol{\Phi}$ and Lagrangian
\begin{equation}
\mathcal{L}=\partial \boldsymbol{\Phi }\cdot \partial \boldsymbol{\Phi }-%
\frac{\lambda }{8}\left( \boldsymbol{\Phi \cdot \Phi }-\frac{\mu ^{2}}{%
\lambda }\right) ^{2}.
\end{equation}
Expanding around the classical vacuum
\begin{equation}
\langle \boldsymbol{\Phi }\rangle =\left( 
\begin{array}{c}
\boldsymbol{0}_{N-1} \\ 
\frac{\mu }{\sqrt{\lambda }}%
\end{array}%
\right) ,~~~\boldsymbol{\Phi }=\left( 
\begin{array}{c}
\boldsymbol{\pi } \\ 
\frac{\mu }{\sqrt{\lambda }}+\sigma%
\end{array}%
\right)\,,
\end{equation}
we get the Lagrangian for $\boldsymbol{\pi }$ and $\sigma$ fields,
\begin{equation}
\mathcal{L}=\frac{1}{2}\partial \boldsymbol{\pi }\cdot \partial \boldsymbol{%
\pi }+\frac{1}{2}\partial \sigma \cdot \partial \sigma -\frac{1}{2}\mu
^{2}\sigma ^{2}-\frac{\lambda }{8}\left( \boldsymbol{\pi \cdot \pi }+\sigma
^{2}\right) ^{2}-\frac{1}{2}\mu \sqrt{\lambda }\sigma \left( \boldsymbol{\pi
\cdot \pi }+\sigma ^{2}\right) .
\end{equation}
The theory contains $N-1$ massless Goldstone bosons $\pi ^{a}$, $a=1,\ldots
,N-1$, corresponding to the spontaneous symmetry breaking of $O\left(
N\right) $ symmetry down to $O\left( N-1\right) $, and one massive resonance 
$\sigma $. The four Goldstone boson amplitude $\pi ^{a}\pi ^{b}\rightarrow
\pi ^{c}\pi ^{d}$ reads now
\begin{equation}
A_{4}^{abcd}\left( s,t;u\right) = -\lambda \left( \delta ^{ab}\delta
^{cd}+\delta ^{ac}\delta ^{bd}+\delta ^{ad}\delta ^{bc}\right)  -\mu ^{2}\lambda \left( \frac{\delta ^{ab}\delta ^{cd}}{s-\mu ^{2}}+\frac{%
\delta ^{ac}\delta ^{bd}}{t-\mu ^{2}}+\frac{\delta ^{ad}\delta ^{bc}}{u-\mu
^{2}}\right)\,.
\end{equation}
The amplitude obeys the Adler zero but, similarly to the $U(1)$ case, it does not manifest itself by means of vanishing residue of the celestial amplitude at $\Delta\rightarrow 0$. The fundamental strip for the Mellin transform is now $\langle -1,0\rangle $ as evident from the high and low energy limits of the amplitude. The nonuniversal function for $\mathcal{G}_{4}^{abcd}\left( \Delta
,t_{4}\right) $ can be easily obtained for $\Delta$ within the fundamental strip
\begin{eqnarray}
\mathcal{G}_{4}^{abcd}\left( \Delta ,t_{4}\right) &=&\lambda \pi \mu
^{2\Delta }\left\{ \delta ^{ab}\delta ^{cd}\left[ \mathrm{i}+\cot \left( \pi
\Delta \right) \right] \phantom{\frac{1}{2}}\right. \notag \\
&&\hspace{1cm}\left. +\csc \left( \Delta \pi \right) \left[ \delta ^{ac}\delta
^{bd}\left( 1-\frac{1}{t_{4}}\right) ^{-\Delta }+\delta ^{ad}\delta
^{bc}\left( \frac{1}{t_{4}}\right) ^{-\Delta }\right] \right\}.
\end{eqnarray}
The result can be analytically continued to a meromorphic function with simple poles for $\Delta =k\in \mathbb{Z}$, the corresponding residues are
\begin{equation}
\mathrm{res}\left( \mathcal{G}_{4}^{abcd},\Delta=k\right) =\lambda \mu ^{2k}\left[
\delta ^{ab}\delta ^{cd}+(-1)^{k}\left( \delta ^{ac}\delta ^{bd}\left( \frac{%
1}{t_{4}}\right) ^{-k}+\delta ^{ad}\delta ^{bc}\left( 1-\frac{1}{t_{4}}%
\right) ^{-k}\right) \right] .
\end{equation}
Again, the pole $\Delta=0$ is on the right of the fundamental strip and the non-vanishing residue
\begin{equation}
 \mathrm{res}\left( \mathcal{G}_{4}^{abcd},\Delta=0\right) =  
 \lambda \left[
\delta ^{ab}\delta ^{cd}+ \delta ^{ac}\delta ^{bd} +\delta ^{ad}\delta ^{bc} \right]
\end{equation}
reflects
therefore the UV limit of the original amplitude. It is the pole $k=-1$
(which was absent in the previous example) which reflects the leading IR
asymptotics, i.e. the $O\left( s\right) $ behavior of the amplitude for 
$s\rightarrow 0$
\begin{equation}
A_{4}^{abcd}\left(s,-as,-(1-a)s\right)  \label{O(N)_asymptotics} \\
=\lambda \mu ^{-2}\left[ \delta ^{ab}\delta ^{cd}-\left( \delta
^{ac}\delta ^{bd}a+\delta ^{ad}\delta ^{bc}\left( 1-a\right) \right) \right]
s+O\left( s^{2}\right) .
\end{equation}

Also for this model we can  illustrate the difference between the two scenarios of taking the limit of infinitely heavy $\sigma$. The physics is the same as in the case of $U(1)$ linear sigma model, but due to the nonabelian symmetry group,  both scenarios differ already in the limits of the amplitude in the momentum space.
The first scenario, namely $\lambda\rightarrow \infty$ with $v=\mu/\sqrt{\lambda}$ fixed, leads  to
\begin{equation}
\lim_{\lambda\rightarrow\infty,\,v\,{\rm{fixed}}}A_{4}^{abcd}\left( s,t;u\right) =\frac{1}{v^2}\left[ \delta
^{ab}\delta ^{cd}s+\delta ^{ac}\delta ^{bd}t+\delta ^{ad}\delta ^{bc}u\right].
\end{equation}
This coincides with the amplitude calculated within the leading order low energy effective theory
describing the dynamics of the Goldstone bosons only, namely the $O(N)$
nonlinear sigma model. The latter can be obtained by means of integrating out the heavy $\sigma$ at the leading order of the derivative expansion. 
The corresponding $4pt$ amplitude can be got just taking the leading $O\left( p^{2}\right) $ term of the asymptotic
expansion (\ref{O(N)_asymptotics}).
On the other hand, the second scenario, $\mu\rightarrow\infty$ with $\lambda$ fixed, leads to 
 vanishing result,
\begin{equation}
\lim_{\mu \rightarrow \infty,\,\lambda\,{\rm{fixed}} }A_{4}^{abcd}\left( s,t;u\right) =0.
\label{vanishing_limit}
\end{equation}
The resulting theory for the second scenario is just the free theory\footnote{The resulting theory can be also obtained by two-step limiting procedure: first we integrate out the heavy $\sigma$ which corresponds to the first scenario, obtaining the $O(N)$ nonlinear sigma model, and then taking the limit $v\rightarrow \infty$.}, exactly as in the abelian $U(1)$ case.
The  limit of the non-universal part of the celestial amplitude yield for the first scenario\footnote{For formal aspects  of the calculation see the previous section and Appendix \ref{complex delta}. } 
\begin{equation}
\lim_{\lambda\rightarrow\infty,\,v\,{\rm{fixed}}}\mathcal{G}_{4}^{abcd}\left( \Delta
,t_{4}\right) = 2 \pi v ^{-2}\delta \left( \Delta +1\right) \left[ \delta
^{ab}\delta ^{cd}-\left( \delta ^{ac}\delta ^{bd}a+\delta ^{ad}\delta
^{bc}\left( 1-a\right) \right) \right], 
\end{equation}
which correctly reproduces the celestial amplitude of the $O(N)$ nonlinear sigma model: 
\begin{eqnarray}
\mathcal{G}_{4}^{abcd}\left( \Delta ,t_{4}\right) ^{NLSM} &=&\lambda \mu
^{-2}\int_{0}^{\infty }\mathrm{d}s~s^{\Delta -1}\left[ \delta ^{ab}\delta
^{cd}-\left( \delta ^{ac}\delta ^{bd}a+\delta ^{ad}\delta ^{bc}\left(
1-a\right) \right) \right] s  \notag \\
&=&2\lambda \pi \mu ^{-2}\delta \left( \Delta +1\right) \left[ \delta
^{ab}\delta ^{cd}-\left( \delta ^{ac}\delta ^{bd}a+\delta ^{ad}\delta
^{bc}\left( 1-a\right) \right) \right] . 
\end{eqnarray}
On the other hand,
  within the second scenario we get a nonzero limit of the celestial  amplitude, namely
\begin{equation}
\lim_{\mu \rightarrow \infty,\,\lambda\,{\rm{fixed}} }\mathcal{G}_{4}^{abcd}\left( \Delta
,t_{4}\right) =2\lambda \pi \delta \left( \Delta \right) \left[ \delta
^{ab}\delta ^{cd}+\delta ^{ac}\delta ^{bd}+\delta ^{ad}\delta ^{bc}\right] 
\end{equation}
in apparent contradiction with (\ref{vanishing_limit}).
We again conclude that the limit of the celestial amplitude for $\mu \rightarrow \infty$ with $\lambda$ fixed   probes the UV region instead of IR one, since it is connected with the residue at $\Delta=0$ situated on the right of the fundamental strip.
 
%=====================================================================
\subsection{UV completion of the $U(1)$ fibered $\mathbb{C}P^{N-1}$ sigma model}\label{sec:fibred_model}
%=====================================================================

The $U(1) $ fibered $\mathbb{C}P^{N-1}$ model represents a non-trivial example of a theory of Goldstone
bosons with Adler zero violation and with an interesting form of soft
theorems. It describes the dynamics of $(N-1)$ charged and one neutral
Goldstone bosons corresponding to spontaneous symmetry breaking
according to the pattern $U(N)\rightarrow U(N-1)$. In the appropriate
parametrization (see~\cite{Kampf:2019mcd} for more details), the Lagrangian reads
\begin{eqnarray}\label{fibrated_CPN}
\mathcal{L} &=&\partial \boldsymbol{\Phi }^{+}\cdot \partial \boldsymbol{%
\Phi }+\frac{1}{2}\partial \phi ^{2} +\frac{1}{4F^{2}}\frac{\left( \partial \boldsymbol{\Phi }^{+}\cdot
\partial \boldsymbol{\Phi }\right) ^{2}}{1-\frac{\boldsymbol{\Phi }^{+}\cdot 
\boldsymbol{\Phi }}{F^{2}}}-\frac{F_{0}}{4F^{4}}\left( \boldsymbol{\Phi }%
^{+}\cdot \boldsymbol{\Phi }\right) \partial \phi ^{2} \notag\\
&&+\frac{1}{4F^{4}}\left( F^{2}-\frac{1}{2}F_{0}^{2}\right) \left( \partial 
\boldsymbol{\Phi }^{+}\cdot \boldsymbol{\Phi -\Phi }^{+}\cdot \partial 
\boldsymbol{\Phi }\right) ^{2}-\frac{F_{0}}{4F^{8}}\left( F^{2}-\frac{1}{2}F_{0}^{2}\right) \left( 
\boldsymbol{\Phi }^{+}\cdot \boldsymbol{\Phi }\right) ^{2}\partial \phi ^{2}
\notag \\
&&+\mathrm{i}\frac{F_{0}}{2F^{6}}\left( F^{2}-\frac{1}{2}F_{0}^{2}\right)
\left( \boldsymbol{\Phi }^{+}\cdot \boldsymbol{\Phi }\right) \left( \partial 
\boldsymbol{\Phi }^{+}\cdot \boldsymbol{\Phi -\Phi }^{+}\cdot \partial 
\boldsymbol{\Phi }\right) \cdot \partial \phi \,,
\end{eqnarray}
where the field ${\bf \Phi}$ 
\begin{equation}
{\bf \Phi}=\left( 
\begin{array}{c}
\Phi _{1} \\ 
\vdots  \\ 
\Phi _{N-1}%
\end{array}%
\right) 
\end{equation}
is an $(N-1)$-plet of complex scalars and $\phi $ is a real scalar field. The
Lagrangian is invariant with respect to the non-linearly realized $U(N)$
symmetry, the $U(N-1)$ subgroup of which is realized linearly as
\begin{equation}
\boldsymbol{\Phi }^{\prime }=U_{N-1}\cdot \boldsymbol{\Phi },~~~~~\phi
^{\prime }=\phi ,
\end{equation}
where $U_{N-1}\in U(N-1)$ is embedded according to
\begin{equation}
U=\left( 
\begin{array}{cc}
U_{N-1} & 0 \\ 
0 & 1%
\end{array}%
\right) \,,
\end{equation}
while the remaining broken generators give rise to the following infinitesimal
generalized shift symmetry transformations
\begin{eqnarray}
\delta \boldsymbol{\Phi } &=&-\mathrm{i}\boldsymbol{a}\left( 1-{\rm{i}}\frac{F_{0}}{%
2F^{2}}\phi \right) +O\left( \left( \phi ,\boldsymbol{\Phi }\right)
^{2}\right)   \notag \\
\delta \boldsymbol{\Phi }^{+} &=&\mathrm{i}\boldsymbol{a}^{+}\left( 1+{\rm{i}}\frac{%
F_{0}}{2F^{2}}\phi \right) +O\left( \left( \phi ,\boldsymbol{\Phi }\right)
^{2}\right)   \notag \\
\delta \phi  &=&a+\frac{F_{0}}{2F^{2}}\left( \boldsymbol{a}^{+}\cdot 
\boldsymbol{\Phi }+\boldsymbol{\Phi }^{+}\cdot \boldsymbol{a}\right)
+O\left( \left( \phi ,\boldsymbol{\Phi }\right) ^{2}\right), 
\end{eqnarray}
with real parameter $a$ and complex $(N-1)$-plet of parameters $\boldsymbol{a}$.
These symmetries are responsible for the soft theorems of the amplitudes. Let us present them here for the case $N=2$ for simplicity\footnote{The case of general $N$ differs from this special case by appropriate decoration with flavor indices corresponding to the charged Goldstones.}.
The soft neutral Goldstone  bosons $\phi(k)$ obey the usual Adler zero, namely
\begin{equation}
\lim_{k_{1}\rightarrow 0}A\left( p_{1}^{\pm },\ldots ,p_{n}^{\pm
},q_{1}^{\mp },\ldots ,q_{n}^{\mp },k_{1},\ldots ,k_{m}\right) =0\,,
\label{CPN_Adler_zero}
\end{equation}
while the soft charged Goldstones $\phi^{\pm}(p)$ (built from two components of the field ${\bf \Phi}$) violate the Adler zero. In this case, the charged Goldstone boson soft theorem acquires nontrivial right-hand side, which is a linear combination of lower point amplitudes with appropriately changed flavor in accord with charge conservation (in the following formula,  the hat means the omission of the respective particle from the list)
\begin{eqnarray}
&&\lim_{p_{1}\rightarrow 0}A\left( p_{1}^{\pm },\ldots ,p_{n}^{\pm
},q_{1}^{\mp },\ldots ,q_{n}^{\mp },k_{1},\ldots ,k_{m}\right)   \notag \\
&=&\pm \mathrm{i}\frac{F_{0}}{2F^{2}}\sum_{i=1}^{m}A\left( k_{i}^{\pm
},p_{2}^{\pm },\ldots ,p_{n}^{\pm },q_{1}^{\mp },\ldots ,q_{n}^{\mp
},k_{1},\ldots ,\widehat{k_{i}},\ldots ,k_{m}\right)  \notag\\
&&\mp \mathrm{i}\frac{F_{0}}{2F^{2}}\sum_{k=1}^{n}A\left( p_{2}^{\pm
},\ldots ,p_{n}^{\pm },q_{1}^{\mp },\ldots ,\widehat{q_{k}^{\mp }},\ldots
,q_{n}^{\mp },q_{k},k_{1},\ldots ,k_{m}\right) .
\label{CPN_soft_theorem}
\end{eqnarray}
Remarkably, this model can be partially UV completed (in the sense of having softer UV amplitudes, but still not leading to an asymptotically free theory) without violating the above soft theorems for the purely Goldstone boson amplitudes. The minimal variant of such a UV completion contains on top of the Goldstone bosons two additional massive scalar and one massive vector resonances
and can be described as follows. 

Let us assume two linear $U(N) $ multiplets, namely one complex scalar $N-$plet $\Phi $ and an additional complex singlet scalar $\psi$.
These transform under $U\in $ $U(N) $ as
\begin{eqnarray}
\Phi ^{\prime } =U\cdot \Phi  ,\,\,\,\,\,
\psi ^{\prime }=\det U~\psi. 
\end{eqnarray}
The last ingredient is the $U(1)$ gauge field $A_{\mu }$ associated with the $U(N)$ generator 
\begin{equation}
T=\left( 
\begin{array}{cc}
\boldsymbol{1}_{N-1} & 0 \\ 
0 & 1%
\end{array}%
\right). 
\end{equation}
The renormalizable Lagrangian describing the UV completion of the above sigma model is then 
\begin{equation}
\mathcal{L} = D\Phi ^{+}\cdot D\Phi +D\psi ^{\ast }\cdot D\psi -\frac{\alpha }{4}( \Phi ^{+}\cdot \Phi -v_{1}^2) ^{2}-\frac{\beta }{4}( \psi ^{\ast }\psi -v_{2}^2) ^{2}-\frac{\gamma }{2}( \Phi ^{+}\cdot \Phi -v_{1}^2)( \psi^{\ast }\psi -v_{2}^2),
\label{UV_complete_Lagrangian}
\end{equation}
where the covariant derivatives introduce two gauge couplings $g_{1,2}$
\begin{equation}
D_{\mu }\Phi =\partial _{\mu }\Phi +{\rm{i}}g_{1}A_{\mu }\Phi ,~~~~~~D_{\mu }\psi
=\partial _{\mu }\psi +{\rm{i}}g_{2}A_{\mu }\psi .  
\end{equation}
The classical vacua can be chosen as 
\begin{equation*}
\langle \Phi \rangle =\left( 
\begin{array}{c}
0_{N-1} \\ 
v_{1}%
\end{array}%
\right) ,~~~~\langle \psi \rangle =v_{2}.
\end{equation*}
Then the broken generators are linear combinations of the $U(N) $
generator $T$ $=\mathbf{1}_{N}$ and $SU(N)$ generators
\begin{equation*}
X_{i}=\left( 
\begin{array}{cc}
0_{N-1} & \mathbf{e}_{i} \\ 
\mathbf{e}_{i}^{T} & 0%
\end{array}%
\right) ,~~~Y_{i}=\left( 
\begin{array}{cc}
0_{N-1} & \mathrm{i}\mathbf{e}_{i} \\ 
-\mathrm{i}\mathbf{e}_{i}^{T} & 0%
\end{array}%
\right) ,~~~Z=\left( 
\begin{array}{cc}
\mathbf{1}_{N-1} & 0 \\ 
0 & -(N-1)%
\end{array}%
\right) ,
\end{equation*}
where $\mathbf{e}_{i}$ are the column vectors with components $%
e_{i}^{j}=\delta _{i}^{j}$. Appropriate parametrization of the quantum
fluctuations around the classical ground state reads
\begin{equation}
\Phi  =\left( 
\begin{array}{c}
\boldsymbol{\Phi /v}_{1} \\ 
\mathrm{e}^{\mathrm{i}\varphi /\sqrt{2}v_{1}}\sqrt{1-\frac{\boldsymbol{\Phi }%
^{+}\cdot \boldsymbol{\Phi }}{v_{1}^{2}}}%
\end{array}%
\right) \left( \frac{1}{\sqrt{2}}\sigma +v_{1}\right),\qquad
\psi  =\mathrm{e}^{\mathrm{i}\chi /\sqrt{2}v_{2}}\left( \frac{1}{\sqrt{2}}%
h+v_{2}\right) .  \label{field_parametrization}
\end{equation}
Here the fields $\varphi $, $\chi $ and $\boldsymbol{\Phi }$ play the role
of the Goldstone bosons corresponding to the broken generators ($\varphi $, $%
\chi $ are related to the linear combination of the generators $T$ and $Z$,
while $\boldsymbol{\Phi }$ and $\boldsymbol{\Phi }^{+}$ are associated with
the broken generators $Z_{i}^{\pm }=X_{i}\pm \mathrm{i}Y_{i}$). According to
the Higgs mechanism, since the gauged generator is broken, one linear
combination of the Goldstone bosons is eaten to generate the mass of the gauge
field $A_{\mu }$ so we end up with $(N-1)-$plet of charged Goldstone bosons $%
\boldsymbol{\Phi }$ and with one neutral Goldstone boson $\phi $, which represents a linear combination of the above fields $\varphi $ and $\chi $ with mixing angle $\vartheta$
\begin{equation}
\phi =\chi \cos \vartheta -\varphi \sin \vartheta .
\end{equation}
The orthogonal combination
\begin{equation}
\eta =\varphi \cos \vartheta +\chi \sin \vartheta 
\end{equation}
is eaten by the Higgs mechanism. The fields $\sigma $ and $h$ form two
massive Higgs scalars with masses $m_{1,2}$ which correspond to the linear
combinations of $\sigma $ and $h$ with mixing angle $\theta $
\begin{equation*}
H_{1,2}=\left\{ 
\begin{array}{c}
\sigma \cos \theta -h\sin \theta  \\ 
\sigma \sin \theta +h\cos \theta 
\end{array}%
\right. .
\end{equation*}
The last ingredient of the particle spectrum is the massive vector boson $A_{\mu }$ with mass $M$. The original parameters of the Lagrangian (\ref{UV_complete_Lagrangian}) can be expressed in terms of the masses and mixing angles as follows
\begin{eqnarray}
\alpha  &=&\frac{1}{v_{1}^{2}}\left( m_{1}^{2}\cos ^{2}\theta +m_{2}^{2}\sin
^{2}\theta \right),\quad
\beta =\frac{1}{v_{2}^{2}}\left( m_{1}^{2}\sin ^{2}\theta +m_{2}^{2}\cos
^{2}\theta \right)  \\
\gamma  &=&-\frac{1}{2v_{1}v_{2}}\left( m_{1}^{2}-m_{2}^{2}\right) \sin
2\theta,\quad
g_{1} =\frac{M}{\sqrt{2}v_{1}}\cos \vartheta, \quad g_{2}=\frac{M}{\sqrt{2}%
v_{2}}\sin \vartheta.
\end{eqnarray}
Inserting the parameterization (\ref{field_parametrization}) into the
Lagrangian (\ref{UV_complete_Lagrangian}), and passing to the unitary gauge $\eta=0$, we can integrate out the massive particles to the  second order in the derivative expansion. At tree level, it means to use the classical equation of motion for the heavy fields and expand the solutions to the desired order in derivatives. In practice this means to set $\sigma ,\eta \rightarrow 0$ and to substitute for the field $A_{\mu }$  the leading order term
\begin{equation}
A_{\mu }=-\frac{\cos\vartheta}{M}\left[ \mathrm{i}\frac{1}{\sqrt{2}v_1}\left( \partial \boldsymbol{%
\Phi }^{+}\cdot \boldsymbol{\Phi -\Phi }^{+}\cdot \partial \boldsymbol{\Phi }%
\right) +\sin\vartheta \frac{\boldsymbol{\Phi }^{+}\cdot 
\boldsymbol{\Phi }}{v_{1}^{2}}\partial \phi \right] +O\left( \partial
^{2}\right) .
\end{equation}
As a result, we obtain as the leading order  effective theory just the $U(1) $ fibred $\mathbb{C}P^{N-1}$ sigma model with the Lagrangian (\ref{fibrated_CPN}), where we identify 
\begin{equation}
F=v_{1},~~~F_{0}=-\sqrt{2}v_{1}\sin \vartheta .
\label{F_F_0_identification}
\end{equation}

The above UV completion of the $\mathbb{C}P^{1}$ sigma model provides us with a nontrivial example of renormalizable theory with Goldstone bosons violating  the Adler zero and with well defined celestial amplitudes.
Here the soft theorems  have the general form of (\ref{model_soft_theorem}). For instance, take $N=2$ and
let us discuss   the amplitude $A_{5}\left(
1,2^{-},3^{-},4^{+},5^{+}\right)$.
Here we use condensed notation identifying $p_{i}\equiv i
$ and, as above, the superscript denotes the charge. According to (\ref{CPN_soft_theorem}), we get e.g.
\begin{eqnarray}
\lim_{p_{5}\rightarrow 0}A_{5}\left( 1,2^{-},3^{-},4^{+},5^{+}\right) =-%
\mathrm{i}\frac{\sin \vartheta }{\sqrt{2}v_{1}}&&\left[ A_{4}\left(
1^{+},2^{-},3^{-},4^{+}\right)\right.\notag\\
&&\left.-A_{4}\left( 1,2,3^{-},4^{+}\right)
-A_{4}\left( 1,2^{-},3,4^{+}\right) \right]   \label{soft_theorem_CPN}
\end{eqnarray}
and observe that the leading order term of the momentum space soft theorem on the right hand side is constant in the energy of the soft (fifth) particle, thus the limit $\Delta_5\to0$ probes the leading term of the conformal soft theorem.
According to the general discussion, we expect the conformal soft theorem in the form (\ref{general_F_5_soft})\footnote{The superscript denotes the charges of the particles involved.}
\begin{eqnarray}
\lim_{t_{4}\rightarrow \overline{t_{4}}}\mathcal{F}_{5}^{\left( 0--++\right)
}\left( \Delta ,t_{4},t_{5}\right) =-\mathrm{i}\frac{\sin \vartheta }{\sqrt{2%
}v_{1}}&&\left[ \mathcal{F}_{4}^{\left( +--+\right) }\left( \Delta
,t_{4}\right)\right.\notag\\ &&\left.-\mathcal{F}_{4}^{\left( 00-+\right) }\left( \Delta
,t_{4}\right) -\mathcal{F}_{4}^{\left( 0-0+\right) }\left( \Delta
,t_{4}\right) \right].  \label{conformal_soft_theorem}
\end{eqnarray}
All the functions ${\cal{F}}_k$ have nonempty fundamental strip, e.g.
\begin{align}
\mathcal{F}_{4}^{\left( 00-+\right) }\left( \Delta ,t_{4}\right)  &=\frac{1%
}{2v_{1}}\frac{\pi }{\sin \left( \pi \Delta \right) }\left[ m_{1}^{2}\cos
\theta \left( \frac{\cos \theta \sin ^{2}\vartheta }{v_{1}}-\frac{\sin
\theta \cos ^{2}\vartheta }{v_{2}}\right) \left( -\frac{m_{1}^{2}}{\Sigma
_{12}^{(4) }}\right) ^{\Delta }\right.   \notag \\
&  \hspace{2cm}\left. +m_{2}^{2}\sin \theta \left( \frac{\sin \theta \sin ^{2}\vartheta }{%
v_{1}}+\frac{\cos \theta \cos ^{2}\vartheta }{v_{2}}\right) \left( -\frac{%
m_{2}^{2}}{\Sigma _{12}^{(4) }}\right) ^{\Delta }\right]
\end{align}
has a fundamental strip $\langle -1,0\rangle$, which can be compared with the same function within the effective theory
\begin{equation}
\mathcal{F}_{4}^{\left( 00-+\right) }\left( \Delta ,t_{4}\right) ^{eff}
=2\pi \delta \left( \Delta +1\right) \frac{F_{0}^{2}}{4F^{4}}\Sigma
_{12}^{(4) } ,  
\end{equation}
where the fundamental strip is empty\footnote{Note that $\mathcal{F}_{4}^{\left( 00-+\right) }\left( \Delta ,t_{4}\right)$ again differ from $\mathcal{F}_{4}^{\left( 00-+\right) }\left( \Delta ,t_{4}\right) ^{eff}$ by the presence of the delta function $\delta(\Delta+1)$ which makes the latter well-defined only in the distributional sense.}.
Here we denoted
\begin{equation}
\Sigma _{ij}^{\left(4\right) }=2\varepsilon _{i}\varepsilon _{j}\sigma
_{\ast i}\sigma _{\ast j}\left( q_{i}\cdot q_{j}\right) =\varepsilon
_{i}\varepsilon _{j}\sigma _{\ast i}\sigma _{\ast j}\left\vert
z_{ij}\right\vert ^{2}.
\end{equation}
The conformal soft theorem (\ref{conformal_soft_theorem}) is supposed to hold in the intersection of the fundamental strips of all ${\cal{F}}_k$'s. In Appendix \ref{app:u1_fibred} we present explicit results for the 4pt and 5pt amplitudes and explicitly verify both the above soft theorems (\ref{soft_theorem_CPN}) and (\ref{conformal_soft_theorem}).

%%%%%%%%%%%%%%%%%%%%%%%%%%%%%%%%%%%%%%%%%%%%%%%%%%%%%%%%%%%%%%%%
\subsubsection*{Conformal soft theorem as a Ward identity}\label{app:conformal}
%%%%%%%%%%%%%%%%%%%%%%%%%%%%%%%%%%%%%%%%%%%%%%%%%%%%%%%%%%%%%%

Usually, soft theorems in momentum space associated with 4D bulk are encoded in the boundary 2D CCFT theory in terms of the OPE~\footnote{OPE, i.e. the limit of two approaching points on the celestial sphere, probes collinear limits in momentum space. However, when a momentum approaches zero (soft limit) it simultaneously becomes collinear with any other momentum.} of the soft particle operator, interpreted as a current on the celestial sphere with other primary (hard) fields. However, this is not a general case, and the interpretation of the soft Goldstone particle as a local current need not be possible. 
We will illustrate this phenomenon using the U(1) fibered model. 

Let us start with the general representation of the celestial amplitude in
the form (\ref{general_npt}), i.e.\footnote{Here and in what follows, we tacitly assume the integrands to be calculated at the reference point $\left( 1,0,\varepsilon
^{-1},t_{4}^{-1},\ldots ,t_{n}^{-1}\right)$.}%
\begin{eqnarray}
&&\left\langle O_{\Delta _{1}}^{\varepsilon _{1}}\left( z_{1},\overline{z}%
_{1}\right) \ldots O_{\Delta _{n},}^{\varepsilon _{n}}\left( z_{n},\overline{%
z}_{n}\right) \right\rangle =C_{n}\left( \left\{ \Delta _{i},z_{i}\right\}
_{i=1}^{n}\right) \mathcal{H}_{n}\left( \left\{ \Delta _{i}\right\}
_{i=1}^{n},t_{4},\ldots ,t_{n}\right)   \notag \\
&=&C_{n}\left( \left\{ \Delta _{i},z_{i}\right\} \right) \lim_{\varepsilon
\rightarrow 0}\int \prod\limits_{i=6}^{n}\mathrm{d}\sigma _{i}\sigma
_{i}^{\Delta _{i}-1}\frac{\varepsilon ^{-2\Delta
_{3}}\prod\limits_{i=1}^{5}\sigma _{\ast i}^{\Delta _{i}-1}\chi _{\langle
0,1\rangle }\left( \sigma _{\ast i}\right) }{2\left\vert \det \mathbf{A}%
_{5}\right\vert }  \notag \\
&&\times  \frac{1}{2}\int_{0}^{\infty }\mathrm{d}u~u^{\Delta
-1}A_{n}\left( \sqrt{u}\varepsilon _{1}\sigma _{\ast 1}q\left( 1,1\right)
,\ldots ,\sqrt{u}\varepsilon _{n}\sigma _{n}q\left( t_{n}^{-1},\overline{%
t_{n}}^{-1}\right) \right)
\end{eqnarray}%
and take formally the limit $\Delta _{n}\rightarrow 0$.
Using (\ref{C_n_to_C_n-1}) and (\ref{soft_delta_function}), we
get 
\begin{eqnarray}
&&\lim_{\Delta _{n}\rightarrow 0}\Delta _{n}\left\langle O_{\Delta
_{1}}^{\varepsilon _{1}}\left( z_{1},\overline{z}_{1}\right) \ldots
O_{\Delta _{n},}^{\varepsilon _{n}}\left( z_{n},\overline{z}_{n}\right)
\right\rangle   \notag \\
&=&C_{n-1}\left( \left\{ \Delta _{i},z_{i}\right\} _{i=1}^{n-1}\right)
\lim_{\varepsilon \rightarrow 0}\int \prod\limits_{i=6}^{n-1}\mathrm{d}%
\sigma _{i}\lim_{\sigma _{n}\rightarrow 0}\frac{\varepsilon ^{-2\Delta
_{3}}\prod\limits_{i=1}^{5}\sigma _{\ast i}^{\Delta _{i}-1}\chi _{\langle
0,1\rangle }\left( \sigma _{\ast i}\right) }{2\left\vert \det \mathbf{A}%
_{5}\right\vert }  \notag \\
&&\times  \frac{1}{2}\int_{0}^{\infty }\mathrm{d}u~u^{\Delta
-1}A_{n}\left( \sqrt{u}\varepsilon _{1}\sigma _{\ast 1}q\left( 1,1\right)
,\ldots ,\sqrt{u}\varepsilon _{n}\sigma _{n}q\left( t_{n}^{-1},\overline{%
t_{n}}^{-1}\right) \right) 
\end{eqnarray}
and therefore, the residue of the celestial correlator for $\Delta
_{n}\rightarrow 0$ probes the momentum space amplitude in the integrand in
the limit when the $n-$th particle becomes soft. Taking into account (\ref{sigma_n_to_sigma_n-1}) and the momentum space soft theorem, we can relate the integrand in the above formula to (the linear combination of) the analogous
integrands for $(n-1) -$point amplitudes. Namely, using (\ref{CPN_soft_theorem}), we
obtain the conformal soft theorem in the UV completion of the $U\left(
1\right)$ fibered $\mathbb{C}P^{1}$ model as follows\footnote{%
Here the second superscript denotes the charge of the corresponding field.
The combination $\varepsilon \pm $ corresponds then to a particle with charge $%
\pm $, provided $\varepsilon =1$ and its antiparticle with charge $\mp $
when $\varepsilon =-1$. Double superscript $\varepsilon 0$ corresponds to a
neutral particle for both values of $\varepsilon $.}%
\begin{equation}
\lim_{\Delta _{n}\rightarrow 0}\Delta _{n}\left\langle O_{\Delta
_{1}}^{\varepsilon _{1}Q_{1}}\ldots O_{\Delta _{n-1}}^{\varepsilon
_{n-1}Q_{n-1}}O_{\Delta _{n}}^{\varepsilon _{n}\pm }X\right\rangle
=\sum_{i=1}^{n-1}\left\langle O_{\Delta _{1}}^{\varepsilon _{1}Q_{1}}\ldots
\delta ^{\varepsilon \pm }O_{\Delta _{i}}^{\varepsilon _{i}Q_{i}}\ldots
O_{\Delta _{n-1}}^{\varepsilon _{n-1}Q_{n-1}}X\right\rangle .\label{conformal_soft_theorem2}
\end{equation}%
Here we denoted by $O_{\Delta
_{i}}^{\varepsilon _{i}Q_{i}}$ the primary field corresponding to a Goldstone boson with charge $Q_i$ and inserted at the point $(z_i,\bar{z_i})$ on the celestial sphere, $X$ stays symbolically for an additional string of primary operators
corresponding to the massive non-Goldstone particles, and we introduced a shorthand notation for the charge-flipped particles on the right-hand side of the soft theorem as~\footnote{In words, neutral Goldstones get flipped to positive/negative charge depending on whether they are incoming/outgoing, while charged Goldstones get flipped to neutral ones (and their causality is of course always preserved). The soft symmetry does not act on other massive fields $X$.}
\begin{equation}
\delta ^{\varepsilon \pm }O_{\Delta }^{\pm 0}\left( z,\overline{z}\right)
\equiv\pm \mathrm{i}\varepsilon \sin ^{2}\vartheta O_{\Delta }^{\pm \mp }\left( z,%
\overline{z}\right) ,~~~~\delta ^{\varepsilon \pm }O_{\Delta }^{\pm \mp
}\left( z,\overline{z}\right) \equiv\mp \mathrm{i}\varepsilon \sin ^{2}\vartheta
O^{\pm 0}_{\Delta}\left( z,\overline{z}\right) .
\label{delta_O}
\end{equation}%
Note that the right
hand side of the above identity (\ref{conformal_soft_theorem2}) does not depend on $\left( z_{n},\overline{%
z_{n}}\right) $ at all. This suggests that the single insertion%
\begin{equation}
Q^{\varepsilon \pm }=\lim_{\Delta \rightarrow 0}\Delta O_{\Delta
}^{\varepsilon \pm }\left( z,\overline{z}\right)
\end{equation}%
formally behaves as a global operator.

Let us remind that in the general case global Ward identities corresponding to a global
symmetry transformation acting on operators $O_{i}\left( z,\overline{z}%
\right) $ as 
\begin{equation}
O_{i}^{^{\prime }}\left( z,\overline{z}\right) =O_{i}\left( z,\overline{z}%
\right) +\delta O_{i}\left( z,\overline{z}\right) ,
\label{abstract_symmetry}
\end{equation}
have the following form
\begin{equation}
\sum\limits_{i=1}^{n}\left\langle O_{1}\left( z_{1},\overline{z_{1}}\right)
\ldots \delta O_{i}\left( z_{i},\overline{z_{i}}\right) \ldots O_{n}\left(
z_{n},\overline{z_{n}}\right) \right\rangle =\int \mathrm{d}%
^{2}z\left\langle \mathcal{A}\left( z,\overline{z}\right) O_{1}\left( z_{1},%
\overline{z_{1}}\right) \ldots O_{n}\left( z_{n},\overline{z_{n}}\right)
\right\rangle,
\label{Ward_identity_global}
\end{equation}%
where $\mathcal{A}\left( z,\overline{z}\right) $ stands for a local operator insertion
expressing explicit or anomalous violation of the symmetry (\ref%
{abstract_symmetry}). Comparing this with the soft theorem (\ref{conformal_soft_theorem2}), it seems reasonable \cite{Kapec:2022axw,Kapec:2022hih,Kapec:2021eug} to suppose that the operator $Q^{\varepsilon \pm }$ is in fact a shadow transformation of some local primary operator $\mathcal{A}%
_{\Delta ^{\prime }}^{\varepsilon \pm }\left( z,\overline{z}\right) $ with
scaling dimension $\Delta ^{\prime }=2$, i.e.%
\begin{equation}
Q^{\varepsilon \pm }=\int \mathrm{d}^{2}w\mathcal{A}_{2}^{\varepsilon \pm
}\left( w,\overline{w}\right) .
\end{equation}%
The conformal soft theorem (\ref{conformal_soft_theorem2}) might be then interpreted as the global Ward
identity corresponding to the explicitly or anomalously broken symmetry
which acts on the primaries as 
\begin{equation}
\delta ^{\varepsilon \pm }O_{\Delta }^{\pm 0}\left( z,\overline{z}\right)
=\pm \mathrm{i}\varepsilon \sin ^{2}\vartheta O_{\Delta }^{\pm \mp }\left( z,%
\overline{z}\right) ,~~~~\delta ^{\varepsilon \pm }O_{\Delta }^{\pm \mp
}\left( z,\overline{z}\right) =\mp \mathrm{i}\varepsilon \sin ^{2}\vartheta
O_{\Delta }^{\pm 0}\left( z,\overline{z}\right) ,~~~~\delta ^{\varepsilon
\pm }X=0.
\end{equation}%
Here $X$ stays for conformal primaries corresponding to the massive
non-Goldstone particles.

%%%%%%%%%%%%%%%%%%%%%%%%%%%%%%%%%%%%%%%%%%%%
\subsection{The pion-dilaton model}
%%%%%%%%%%%%%%%%%%%%%%%%%%%%%%%%%%%%%%%%%%%%

In this section, we introduce a simple renormalizable model with spontaneous
symmetry breaking, which shows another feature of Goldstone bosons on
the celestial sphere, namely the conformal soft theorems of unusually complex form. The Lagrangian is
\begin{equation}
\mathcal{L}=\frac{1}{2}\left( \partial \phi \right) ^{2}+\frac{1}{2}\left(
\partial \psi \right) ^{2}+\frac{1}{2}\left( \partial \chi \right) ^{2}-%
\frac{\lambda }{16}\left( \phi ^{2}+\psi ^{2}-\chi ^{2}\right) ^{2} ,
\end{equation}
where $\phi $, $\psi $ and $\chi $ are real scalar fields. The symmetries of
the Lagrangian are
\begin{enumerate}
\item the $SO(2)$ rotation in the plane $\left( \phi ,\psi \right)$: $\delta \phi =\psi ,~~\delta \psi =-\phi, ~~\delta\chi=0$,
\item the scale transformation: $\delta \phi _{i}=-\left( 1+x\cdot \partial \right) \phi _{i},~~\phi
_{i}=\phi ,\psi ,\chi$
\end{enumerate}
The classical ground state satisfies the constraint
\begin{equation}
\phi ^{2}+\psi ^{2}-\chi ^{2}=0
\end{equation}
and we choose the broken phase with the vacuum expectation values as follows
\begin{equation}
\langle \phi \rangle =0,~~~\langle \psi \rangle =\langle \chi \rangle =v,
\end{equation}
for which both the above symmetries are broken. Appropriate parametrization
of the fluctuations around the above classical ground state reads
\begin{equation}
\phi \equiv \pi ,~~~\psi =\frac{1}{\sqrt{2}}\left( a+\sigma \right)
+v,~~~\chi =\frac{1}{\sqrt{2}}\left( a-\sigma \right) +v.
\end{equation}
The new fields transform with respect to the rotations as 
\begin{equation}
\delta \pi  =\frac{1}{\sqrt{2}}\left( a+\sigma \right) +v,\qquad 
\delta a =-\frac{1}{\sqrt{2}}\pi,\qquad
\delta \sigma =-\frac{1}{\sqrt{2}}\pi 
\end{equation}
and with respect to the scale transformations as 
\begin{equation}
\delta \pi  =-\left( 1+x\cdot \partial \right) \pi,\qquad
\delta a =-\left( 1+x\cdot \partial \right) \pi -\sqrt{2}v,\qquad
\delta \sigma =-\left( 1+x\cdot \partial \right) \sigma \,.
\end{equation}
Thus the ``pion'' $\pi $ is a Goldstone
boson of $SO(2)$ breaking while the ``dilaton'' $a$ corresponds to the breaking of the scale invariance. In these variables, the Lagrangian becomes
\begin{eqnarray}
\mathcal{L} &=&\frac{1}{2}\left( \partial \pi \right) ^{2}+\frac{1}{2}\left(
\partial a\right) ^{2}+\frac{1}{2}\left( \partial \sigma \right) ^{2}-\frac{1%
}{2}m^{2}\sigma ^{2}  \notag \\
&&\hspace{1cm} -\frac{\lambda ^{1/2}m}{2\sqrt{2}}\pi ^{2}\sigma -\frac{\lambda ^{1/2}m}{\sqrt{2}}a\sigma ^{2}-\frac{\lambda }{16}\pi ^{4}-\frac{\lambda }{4}\pi
^{2}a\sigma -\frac{\lambda }{4}a^{2}\sigma ^{2},
\end{eqnarray}
where we expressed the vacuum expectation value $v$ in terms of the mass $m^{2}=\lambda v^{2}$ of the $\sigma $ particle. The tree-level amplitudes are the subject of the following soft theorems with non-trivial right-hand sides given by linear combinations of lower point amplitudes (in the same theory):\footnote{In the following formulas, the momentum conservation $\delta$-function is not included into the amplitude $A$.}

\begin{enumerate}
\item The soft pion theorem, which reflects the spontaneous breaking of the $SO(2)$ symmetry of the form
\begin{eqnarray}
&&A\left( tp^{\left( \pi \right) },p_{1}^{\left( \pi
\right) },\ldots ,p_{m}^{\left( \pi \right) },q_{1}^{\left( a\right)
},\ldots ,q_{n}^{\left( a\right) },k_{1}^{\left( \sigma \right) },\ldots
,k_{k}^{\left( \sigma \right) }\right)   \notag \\
&\stackrel{t\rightarrow 0}{=}&\frac{1 }{m}\sqrt{\frac{\lambda }{2}}\sum\limits_{i=1}^{m}A\left(
p_{1}^{\left( \pi \right) },\ldots ,p_{i}^{\left( a\right) },\ldots
,p_{m}^{\left( \pi \right) },q_{1}^{\left( a\right) },\ldots ,q_{n}^{\left(
a\right) },k_{1}^{\left( \sigma \right) },\ldots ,k_{k}^{\left( \sigma
\right) }\right)   \notag \\
&&-\frac{1 }{m}\sqrt{\frac{\lambda }{2}}\sum\limits_{i=1}^{n}A\left(
p_{1}^{\left( \pi \right) },\ldots ,p_{m}^{\left( \pi \right)
},q_{1}^{\left( a\right) },\ldots ,,q_{i}^{\left( \pi \right) },\ldots
q_{n}^{\left( a\right) },k_{1}^{\left( \sigma \right) },\ldots
,k_{k}^{\left( \sigma \right) }\right)+O(t) \notag
\label{soft_pion_theorem_for_pion_dilaton_model}
\end{eqnarray}
The leading soft term on the right hand side is constant in the energy of the soft pion and thus the leading conformal soft theorem will be probed by the limit $\Delta^{(\pi)}\to 0$.

\item The soft dilaton theorem, which is a consequence of the nonlinearly realized dilation symmetry, and which reads
\begin{align}\label{soft_dilaton_theorem}
&A_{n+1}\left( tp^{\left( a\right) },p_{1}^{\left( f_{1}\right) },\ldots
,p_{n}^{\left( f_{n}\right) }\right) \stackrel{t\rightarrow 0}{=}\frac{1}{t}\sqrt{\frac{\lambda }{2}}
\sum\limits_{i=1}^{n}\left(\delta_{f_{i}\sigma }\frac{m}{p\cdot p_{i}}\right)A_{n}\left( p_{1}^{\left( f_{1}\right) },\ldots
,p_{n}^{\left( f_{n}\right) }\right)  \\
&+\sqrt{\frac{\lambda }{2}}\left[ \sum\limits_{i=1}^{n}\delta
_{f_{i}\sigma }\frac{m}{p\cdot p_{i}}p\cdot \partial _{p_{i}} + \frac{4}{m} - \frac{1}{m} \sum\limits_{i=1}^{n}%
\left( 1+p_{i}\cdot \partial _{p_{i}}\right)\right] A_n\left( p_{1}^{\left( f_{1}\right) },\ldots ,p_{n}^{\left(
f_{n}\right) }\right) +O\left( t\right)\notag
\end{align}
Here the superscript $f_i=\pi,\,a,\,\sigma$ denotes the flavor of the corresponding external particles.
The leading soft term on the first line is proportional to the inverse energy of the soft dilaton, while the subleading one on the last two lines is constant. This corresponds via Mellin transform to $\Delta^{(a)}\to1$ for the leading conformal soft theorem and $\Delta^{(a)}\to0$ for the subleading one. Taking into account the homogeneity of the amplitudes as functions of the momenta and the mass parameter $m$, namely
\begin{equation}
    A_n(\{tp_i\}_{i=1}^{n};tm)=t^{4-n}A_n(\{p_i\}_{i=1}^{n};m),
\end{equation}
we can rewrite the right-hand side using the Euler theorem in the form
\begin{align}
&A_{n+1}\left( tp^{\left( a\right) },p_{1}^{\left( f_{1}\right) },\ldots
,p_{n}^{\left( f_{n}\right) }\right)\overset{t\rightarrow 0}{=}\frac{1}{t}\sqrt{\frac{\lambda }{2}}%
\sum\limits_{i=1}^{n}\delta _{f_{i}\sigma }\frac{m}{p\cdot p_{i}}A_{n}\left(
p_{1}^{\left( f_{1}\right) },\ldots ,p_{n}^{\left( f_{n}\right) }\right)  \\
&\hspace{3cm}+\sqrt{\frac{\lambda }{2}}\left[ \frac{\partial }{\partial m}%
+\sum\limits_{i=1}^{n}\delta _{f_{i}\sigma }\frac{m}{p\cdot p_{i}}p\cdot
\partial _{p_{i}}\right] A\left( p_{1}^{\left( f_{1}\right) },\ldots
,p_{n}^{\left( f_{n}\right) }\right) +O\left( t\right)\notag.
\label{soft_dilaton_theorem}
\end{align}
Here the partial derivative with respect to $m$ takes into account only the explicit dependence of the vertices and propagators (i.e. not the implicit dependence stemming from the massive momenta).
\end{enumerate}

\noindent Let us demonstrate the validity of these theorems on the following simple example of a tree-level 4pt amplitude corresponding to the scattering $\pi\pi\to a\sigma$. There are just two tree-level Feynman diagrams leading to 
\begin{equation}
A_{4}\left( p_{1}^{\left( \pi \right) },p_{2}^{\left( \pi \right)
},q^{\left( a\right) },k^{\left( \sigma \right) }\right) =-\frac{\lambda
m^{2}}{2\left( q\cdot k\right) }-\frac{\lambda }{2}.
\end{equation}
The 3pt amplitudes entering the right hand side of the leading soft pion theorem~\eqref{soft_pion_theorem_for_pion_dilaton_model} take the form 
\begin{equation}
A_{3}\left( p_{2}^{\left( a\right) },q^{\left( a\right) },k^{\left( \sigma
\right) }\right) =0,~~~~A_{3}\left( p_{2}^{\left( \pi \right) },q^{\left(
\pi \right) },k^{\left( \sigma \right) }\right) =-\sqrt{\frac{\lambda }{2}}m,
\label{3pt_amplitudes}
\end{equation}
and as a consequence of the 3pt kinematics $\lim_{p_{1}\rightarrow 0}2\left( q\cdot k\right) =-m^{2}$. Thus the soft pion theorem is manifest
\begin{equation}
\lim_{p_{1}\rightarrow 0}A_{4}\left( p_{1}^{\left( \pi \right)
},p_{2}^{\left( \pi \right) },q^{\left( a\right) },k^{\left( \sigma \right)
}\right) =\frac{\lambda }{2}=-\frac{1 }{m}\sqrt{\frac{\lambda }{2}}A_{3}\left(
p_{2}^{\left( \pi \right) },q^{\left( \pi \right) },k^{\left( \sigma \right)
}\right) .
\end{equation}
Similarly, from (\ref{3pt_amplitudes}) and since
\begin{eqnarray}
\sqrt{\frac{\lambda }{2}}\frac{\partial }{\partial m}A_{3}\left(
p_{1}^{\left( \pi \right) },p_{2}^{\left( \pi \right) },k^{\left( \sigma
\right) }\right) =-\frac{\lambda }{2} ,\,\, \sqrt{\frac{\lambda }{2}}\frac{m}{ q\cdot k }A_{3}\left(
p_{1}^{\left( \pi \right) },p_{2}^{\left( \pi \right) },k^{\left( \sigma
\right) }\right)  =-\frac{\lambda m^{2}}{2 q\cdot k }\,,
\end{eqnarray}
we have the exact identity 
\begin{equation}
A_{4}\left( p_{1}^{\left( \pi \right) },p_{2}^{\left( \pi \right)
},q^{\left( a\right) },k^{\left( \sigma \right) }\right) =\sqrt{\frac{\lambda }{2}}\frac{m}{ q\cdot k }A_{3}\left(
p_{1}^{\left( \pi \right) },p_{2}^{\left( \pi \right) },k^{\left( \sigma
\right) }\right) +%
\sqrt{\frac{\lambda }{2}}\frac{\partial }{\partial m}A_{3}\left(
p_{1}^{\left( \pi \right) },p_{2}^{\left( \pi \right) },k^{\left( \sigma
\right) }\right) ,
\end{equation}
which verifies the validity of the soft dilaton theorem. Let us now discuss the conformal soft theorems for the celestial amplitudes.

%============================================
\subsubsection*{Conformal soft theorems}
%============================================

The conformal soft pion theorem is similar to the one discussed in section~\ref{app:conformal}. It can be obtained by means of a transformation of
both sides of (\ref{soft_pion_theorem_for_pion_dilaton_model}) to the
celestial sphere and taking the residue in one of the conformal weights of
the pions at $\Delta ^{\left( \pi \right) }=0$. As a result we obtain
\begin{equation}
\lim_{\Delta \rightarrow 0}\Delta \left\langle O_{\Delta }^{\varepsilon \left( \pi
\right) }O_{\Delta _{1}^{\left( f_{1}\right) }}^{\varepsilon _{1}\left(
f_{1}\right) }\ldots O_{\Delta _{i}^{\left( f_{i}\right) }}^{\varepsilon
_{i}\left( f_{i}\right) }\ldots O_{\Delta _{n}^{\left( f_{n}\right)
}}^{\varepsilon _{n}\left( f_{n}\right) }\right\rangle
=\sum\limits_{i=1}^{n}\left\langle O_{\Delta _{1}^{\left( f_{1}\right)
}}^{\varepsilon _{1}\left( f_{1}\right) }\ldots \delta ^{\left( \pi \right)
}O_{\Delta _{i}^{\left( f_{i}\right) }}^{\varepsilon _{i}\left( f_{i}\right)
}\ldots O_{\Delta _{n}^{\left( f_{n}\right) }}^{\varepsilon _{n}\left(
f_{n}\right) }\right\rangle\,, \label{pionsoft}
\end{equation}
where $f_{i}=\pi $, $a$, $\sigma $ stays for the flavor of the particles and
we denoted
\begin{equation}
\delta ^{\left( \pi \right) }O_{\Delta }^{\varepsilon \left( \pi \right) }=%
\frac{\lambda ^{1/2}}{\sqrt{2}m}O_{\Delta }^{\varepsilon \left( a\right)
},~~~~\delta ^{\left( \pi \right) }O_{\Delta }^{\varepsilon \left( a\right)
}=-\frac{\lambda ^{1/2}}{\sqrt{2}m}O_{\Delta }^{\varepsilon \left( \pi
\right) },~~~~\delta ^{\left( \pi \right) }O_{\Delta }^{\varepsilon \left(
\sigma \right) }=0.
\end{equation}
The form of the soft theorem can be interpreted as a global Ward identity~(\ref{Ward_identity_global}), with the soft insertion defined as an integrated operator of an exactly marginal shadow dual local field.

The leading conformal soft dilaton theorem corresponds to the residue in
the dilaton conformal weight at $\Delta ^{\left( a\right) }=1$. Direct but rather long calculation  shows that this conformal soft theorem takes a rather complicated form, namely
\begin{eqnarray}
&&\hspace{-2cm}\lim_{\Delta \rightarrow 1}\left( \Delta -1\right) \left\langle O_{\Delta
}^{\varepsilon \left( a \right) }\left( w,\overline{w}\right) O_{\Delta
_{1}}^{\varepsilon _{1}\left( f_{1}\right) }\left( z_{1},\overline{z}%
_{1}\right) \ldots O_{\Delta _{n}}^{\varepsilon _{n}\left( f_{n}\right)
}\left( z_{n},\overline{z}_{n}\right) \right\rangle   \notag \\
&=&4\sqrt{\frac{\lambda }{2}}\sum\limits_{i=1}^{n}\delta _{f_{i}\sigma }\int 
\mathrm{d}\nu \mu \left( \nu \right) \mathrm{d}^{2}x\frac{C\left( \Delta ,1+%
\mathrm{i}\nu ,1\right) }{\left\vert w-x\right\vert ^{2-\Delta _{i}+\mathrm{i%
}\nu }\left\vert z_{i}-w\right\vert ^{\Delta _{i}-\mathrm{i}\nu }\left\vert
z_{i}-x\right\vert ^{\Delta _{i}+\mathrm{i}\nu }}  \notag \\
&&\hspace{2cm} \times \left\langle O_{\Delta _{1}}^{\varepsilon _{1}\left( f_{1}\right)
}\left( z_{1},\overline{z}_{1}\right) \ldots O_{1-\mathrm{i}\nu
}^{\varepsilon _{i}\left( f_{i}\right) }\left( x,\overline{x}\right) \ldots
O_{\Delta _{n}}^{\varepsilon _{n}\left( f_{n}\right) }\left( z_{n},\overline{%
z}_{n}\right) \right\rangle .  \label{leading_soft_dilaton_CS}
\end{eqnarray}
All the details are discussed in Appendix~\ref{app:soft} including special limits. The form of (\ref{leading_soft_dilaton_CS}) suggests that the dilaton soft theorem can not be easily interpreted as a Ward identity. In fact, the right hand side of (\ref{leading_soft_dilaton_CS}) is not just a sum of celestial correlators like in {\ref{pionsoft}), but it is further integrated with a certain integration kernel. 

Subleading conformal soft dilaton theorem corresponds to the limit $\Delta\rightarrow 0$, the result now reads,
\begin{eqnarray}
&&\hspace{-1cm}\lim_{\Delta \rightarrow 0}\Delta \left\langle O_{\Delta }^{\varepsilon
\left( \sigma \right) }\left( \mathbf{z}\right) O_{\Delta _{1}}^{\varepsilon
_{1}\left( f_{1}\right) }\left( \mathbf{z}_{1}\right) \ldots O_{\Delta
_{n}}^{\varepsilon _{n}\left( f_{n}\right) }\left( \mathbf{z}_{n}\right)
\right\rangle  \notag \\
&& =-\frac{1}{m}\sqrt{\frac{\lambda }{2}}\left[ n-\sum\limits_{i=1}^{n}%
\Delta _{i}\right] \left\langle O_{\Delta _{1}}^{\varepsilon _{1}\left(
f_{1}\right) }\left( \mathbf{z}_{1}\right) \ldots O_{\Delta
_{n}}^{\varepsilon _{n}\left( f_{n}\right) }\left( \mathbf{z}_{n}\right)
\right\rangle  \notag \\
&&\hspace{0.5cm} -\frac{1}{m\pi ^{3}}\sqrt{\frac{\lambda }{2}}\sum\limits_{i=1}^{n}\delta
_{f_{i}\sigma }\int \mathrm{d}\nu \mathrm{d}^{2}\mathbf{x}\frac{\nu
^{2}\left( 1+\mathrm{i}\nu \right) C\left( \Delta _{i},2+\mathrm{i}\nu
,1\right) }{\left\vert \mathbf{x}-\mathbf{z}\right\vert ^{1-\Delta _{i}+%
\mathrm{i}\nu }\left\vert \mathbf{z}_{i}-\mathbf{x}\right\vert ^{1+\Delta
_{i}+\mathrm{i}\nu }\left\vert \mathbf{z}_{i}-\mathbf{z}\right\vert
^{-1+\Delta _{i}-\mathrm{i}\nu }}  \notag \\
&&\hspace{5cm} \times \left\langle O_{\Delta _{1}}^{\varepsilon _{1}\left( f_{1}\right)
}\left( \mathbf{z}_{1}\right) \ldots O_{1-\mathrm{i}\nu }^{\varepsilon
_{i}\left( f_{i}\right) }\left( \mathbf{x}\right) \ldots O_{\Delta
_{n}}^{\varepsilon _{n}\left( f_{n}\right) }\left( \mathbf{z}_{n}\right)
\right\rangle.
\end{eqnarray}
More details can be again found in Appendix~\ref{app:soft}. 

%=====================================================
\subsection{Z-theory: UV completion of non-linear sigma model}
%=======================================================

Quite recently, an interesting stringy UV completion of the nonlinear sigma
model relevant for the dynamics of mesons was found, namely the so-called Abelian $Z-$theory~\cite{Carrasco:2016ldy,Carrasco:2016ygv}. It is related to
the $Z-$theory disc integrals, which originally depend on two orderings $
\sigma ,\rho \in S_{n}$%
\begin{equation}
Z_{\rho }(\sigma (1) ,\ldots ,\sigma (n) )=\alpha
^{\prime n-3}\int_{D\left( \rho \right) }\frac{\prod\limits_{i=1}^{n}%
\mathrm{d}z_{i}}{\mathrm{Vol}\left( SL\left( 2,%
\mathbb{R}
\right) \right) }\frac{\prod\limits_{j<k}z_{jk}^{\alpha ^{\prime }s_{jk}}}{%
z_{\sigma (1) \sigma (2) }z_{\sigma (2)
\sigma \left( 3\right) }\ldots z_{\sigma (n) \sigma \left(
1\right) }}.
\end{equation}
Here the integration domain is
\begin{equation}
D\left( \rho \right) =\left\{ \left( z_{1},z_{2},\ldots ,z_{n}\right) \in 
\mathbb{R}
^{n},-\infty <z_{\rho (1) }<z_{\rho (2) }<\ldots
<z_{\rho (n) }<\infty \right\} ,
\end{equation}
and 
\begin{equation}
s_{i_{1}\ldots i_{k}}=\frac{1}{2}\left(
\sum\limits_{j=1}^{k}p_{i_{j}}\right) ^{2},~~~p_{i}^{2}=0.
\end{equation}
The abelian amplitudes (abelian disc integral) are then defined as%
\begin{equation}
Z_{\times }\left( \sigma (1) ,\ldots ,\sigma (n)
\right) =\sum\limits_{\rho \in S_{n}/%
\mathbb{Z}
_{n}}Z_{\rho }\left( \sigma (1) ,\ldots ,\sigma (n)
\right) .
\end{equation}
In \cite{Carrasco:2016ldy} it was claimed that the field theory limit of the amplitudes $Z_{\times }$ corresponds to the stripped amplitudes of the nonlinear sigma model including higher order corrections with specific values of the higher
order low energy constants. Here we will concentrate on the four-point
amplitude, which is known explicitly and which reads (in what follows, we use
the units in which $\alpha ^{\prime }=1$ for simplicity) 
\begin{equation}
Z_{\times }\left( 1,2,3,4\right) =2\left[ B\left( -s,1-t\right) +B\left(
-u,1-t\right) -B\left( -s,-u\right) \right] ,
\label{Z_x}
\end{equation}
where
\begin{equation}
B\left( a,b\right) =\frac{\Gamma \left( a\right) \Gamma \left( b\right) }{%
\Gamma \left( a+b\right) }
\end{equation}
is the Euler beta function and $s,t$ and $u$ are the usual Mandelstam
variables\footnote{The $\alpha ^{\prime }$ dependence can be easily restored by the substitution $s\rightarrow \alpha ^{\prime }s$, $t\rightarrow \alpha
^{\prime }t$, $u\rightarrow \alpha ^{\prime }u$.} (\ref{Mandelstam_variables}).

The transformation of this amplitude to the celestial sphere is, according
to (\ref{G_4}) and (\ref{A_4_simplified}), determined by the non-universal conformal invariant function $\mathcal{G}_{4}\left( \Delta ,t_{4}\right) $, where $t_{4}=z_{1234}$ is the cross-ratio and $\Delta $ is given by~\eqref{eq:def_delta}. Explicitly 
\begin{eqnarray}\label{eq:Z_mellin}
&&\mathcal{G}_{4}\left( \Delta ,t_{4}\right) 
\equiv \int_{0}^{\infty }\mathrm{d}s~s^{\Delta -1}Z_{4}\left(
s,t_{4}\right)
\phantom{B\left( -s,1+\frac{s}{t_{4}}\right) +B\left( \left( 1-%
\frac{1}{t_{4}}\right) s,1+\frac{s}{t_{4}}\right) }\notag\\
&&=2\int_{0}^{\infty }\mathrm{d}%
s~s^{\Delta -1}\left[ B\left( -s,1+\frac{s}{t_{4}}\right) +B\left( \left( 1-%
\frac{1}{t_{4}}\right) s,1+\frac{s}{t_{4}}\right) -B\left( -s,\left( 1-\frac{%
1}{t_{4}}\right)s\right) \right]  
,\notag\\
\end{eqnarray}
where, according to our general prescription, the poles on the real axis
should be approached from the upper complex half-plane (see Fig.~\ref{fig:hank_cont}).

Let us first discuss the general properties of $\mathcal{G}_{4}\left( \Delta
,t_{4}\right) $. It is given by the above Mellin transform and the corresponding
fundamental strip is determined by the asymptotics of $Z_{4}\left(
s,t_4\right) $ for $s\rightarrow 0$ and $s\rightarrow \infty $. The low
energy behavior is fixed by the nonlinear sigma model amplitude (see section~\ref{eft_amps} for the discussion of NLSM amplitudes),
\begin{eqnarray}
&&\hspace{-2.2cm} Z_{4}\left( s,t_{4}\right) \overset{s\rightarrow 0}{=}A_{4}^{NLSM}\left(
s,-\frac{s}{t_{4}};\left( 1-\frac{1}{t_{4}}\right) s\right) +O\left(
s^{3}\right)   \notag \\
&=&-\pi ^{2}\frac{s}{t_{4}}-\frac{\pi ^{4}}{12}\frac{(1-t_{4}+t_{4}^{2})}{%
t_{4}^{3}}s^{3}+\frac{\pi ^{2}}{2}\psi ^{\prime \prime }(1) 
\frac{t_{4}-1}{t_{4}^{3}}s^{4}+O\left( s^{5}\right) .
\label{Z_4_IR_asymptotics}
\end{eqnarray}
This means that the Mellin amplitude has poles at $\Delta=-1,-3,-4,\dots$ where $\Delta=-1$ corresponds to the leading soft theorem. The high $s$ asymptotics follows from the general formula
\begin{equation}
\Gamma \left( z\right) \overset{z\rightarrow \infty }{=}\sqrt{2\pi }\mathrm{e%
}^{\left( z-1/2\right) \ln z-z}\left( 1+O\left( \frac{1}{z}\right) \right) ,
\label{Gamma_asymptotic}
\end{equation}
which is valid for\footnote{
Here we take $\arg \left( z\right) \in \left( -\pi ,\pi \right) $.} $
\left\vert \arg \left( z\right) \right\vert <\pi -\delta $, for any $0<\delta <\pi $. Let us remind that the kinematics constraints give $t_{4}>1$. Suppose first $s$ to be complex $s\rightarrow z=t\mathrm{e}^{\mathrm{i}%
\varepsilon }$ and let $t\rightarrow \infty $. Then for  $0<\delta
<\left|\varepsilon\right| <\pi-\delta $ and for $t$ large enough, the arguments of $-z$, $1+z/t_{4}$ and $\left( 1-t_{4}^{-1}\right) z$ satisfy the applicability
conditions of the above asymptotic relation (\ref{Gamma_asymptotic}). We get
then
\begin{eqnarray}
&&\hspace{-1.5cm} Z_{4 }\left( z,t_4\right) |_{ z=t\mathrm{e}^{\mathrm{i}%
\varepsilon }}\overset{t\rightarrow \infty }{=}-2\sqrt{\frac{2\pi }{z\left(
t_{4}-1\right) }}\exp \left[ -z~f\left( t_{4}\right) +O\left( \frac{1}{z}%
\right) \right]   \notag \\
&&\times \left[ 1+ \exp \left( \mathrm{i}\pi \eta(z)\left(1+ \frac{z}{t_{4}}\right)\right) -\exp
\left( \mathrm{i}\pi\eta(z) z\left( 1-\frac{1}{t_{4}}\right) \right) \right] \left[
1+O\left( \frac{1}{z}\right) \right] \,,
\end{eqnarray}
where
\begin{align}
f\left( t_{4}\right) &=\frac{\ln t_{4}}{t_{4}}-\left( 1-\frac{1}{t_{4}}%
\right) \ln \left( 1-\frac{1}{t_{4}}\right), \\
\eta(z)&=\mathrm{sign}\left(\mathrm{Im}(z)\right).
\end{align}
The function $f\left( t_{4}\right) $ is positive for $t_{4}>1$, and
therefore the amplitude $%
Z_{4}\left( s,t_{4}\right) $ is exponentially suppressed along any path $s=t%
\mathrm{e}^{\mathrm{i}\varepsilon }$ for $t\rightarrow \infty $, with $%
\varepsilon $ fixed in the range $0<\delta<\left|\varepsilon\right|<\pi/2$. 

For calculation of the Mellin transform, we  define the amplitude for real $s$  to be a boundary value of the function $Z_{4}\left( s,t_{4}\right) $ which is analytic in the  upper complex half plane. 
This defines a prescription for avoiding the poles of the Mellin integrand on the real axis. They are being bypassed by deformation of the integration contour in the upper half plane.
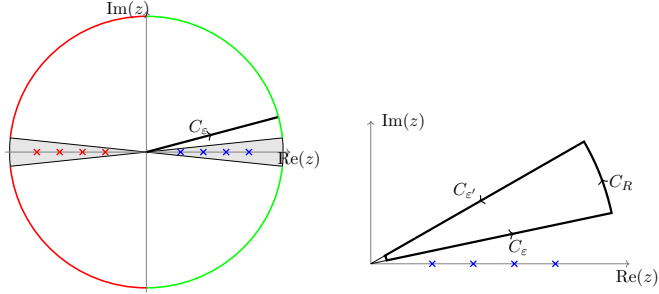
\begin{figure}[ht]
\centering
\scalebox{.6}{
\begin{tikzpicture}[scale=.5]
\tikzset{cross/.style={cross out, draw=black, minimum size=2*(#1-\pgflinewidth), inner sep=0pt, outer sep=0pt},
%default radius will be 1pt. 
cross/.default={2.7pt}}

% Axes
\draw[fill=gray!20,line width=0.3pt] (0,0) -- (5.96713, -0.627171) arc (-6:6:6) -- cycle;
\draw[fill=gray!20,line width=0.3pt] (0,0) -- (-5.96713, 0.627171) arc (174:186:6) -- cycle;

\draw [help lines,->] (-6.2, 0) -- (6.3,0);
\draw [help lines,->] (0, -6.2) -- (0, 6.2);

\draw[line width=1.4pt,   decoration={ markings,
  mark=at position 0.5 with {\arrow[line width=.7pt]{>}}},
  postaction={decorate}]	(0,0) -- (5.79555, 1.55291);
\draw[green,line width=1pt]	(5.96713, 0.627171) arc (6:90:6);
\draw[green,line width=1pt]	(5.96713, -0.627171) arc (-6:-90:6);
\draw[red,line width=1pt]	(-5.96713, 0.627171) arc (174:90:6);
\draw[red,line width=1pt]	(-5.96713, -0.627171) arc (-174:-90:6);
% The labels
	\node at (6.7,-0.35){$\mathrm{Re}(z)$};
	\node at (-0.79,6.3) {$\mathrm{Im}(z)$};
\node at (2.3,1.1) {$C_\varepsilon$};

\draw (1.5,0) node[cross,blue,thick] {};
\draw (2.5,0) node[cross,blue,thick] {};
\draw (3.5,0) node[cross,blue,thick] {};
\draw (4.5,0) node[cross,blue,thick] {};
\draw (-1.8,0) node[cross,red,thick] {};
\draw (-2.8,0) node[cross,red,thick] {};
\draw (-3.8,0) node[cross,red,thick] {};
\draw (-4.8,0) node[cross,red,thick] {};
\end{tikzpicture}
\qquad
\begin{tikzpicture}[scale=0.9]
\tikzset{cross/.style={cross out, draw=black, minimum size=2*(#1-\pgflinewidth), inner sep=0pt, outer sep=0pt},
%default radius will be 1pt. 
cross/.default={2.7pt}}

% Axes
\draw [help lines,->] (0, 0) -- (6.3,0);
\draw [help lines,->] (0, 0) -- (0, 3.5);

\draw[line width=1.4pt,   decoration={ markings,
  mark=at position 0.25 with {\arrow[line width=.7pt]{>}},
  mark=at position 0.5 with {\arrow[line width=.7pt]{>}},
  mark=at position 0.8 with {\arrow[line width=.7pt]{>}}},
  postaction={decorate}]	(0.34641, 0.2) arc (30:12:0.4) -- (5.86889, 1.24747) arc(12:30:6) -- cycle;
\draw[line width=.4pt] (0,0) -- (0.34641, 0.2);
\draw[line width=.4pt] (0,0) -- (0.391259, 0.0831647);
% The labels
	\node at (6.5,-0.35){$\mathrm{Re}(z)$};
	\node at (.8,3.5) {$\mathrm{Im}(z)$};
\node at (3.6,.4) {$C_\varepsilon$};
\node at (6.1,2) {$C_R$};
\node at (2.3,1.8) {$C_{\varepsilon^\prime}$};

\draw (1.5,0) node[cross,blue,thick] {};
\draw (2.5,0) node[cross,blue,thick] {};
\draw (3.5,0) node[cross,blue,thick] {};
\draw (4.5,0) node[cross,blue,thick] {};
\end{tikzpicture}}
\caption{\emph{Left:} Asymptotic behavior of the integrand in~\eqref{eq:Z_mellin} is shown. It has poles both on the positive and negative real axis, thus infinitesimal wedges (shaded gray) where the asymptotics cannot be controlled have to be excised. On the green arcs at infinity the integrand is exponentially suppressed, while on the red ones it is large. The Mellin transform is defined by an integral along the contour $C_\varepsilon$, such that it lies outside the excised wedge and asymptotes to a region where the behavior of the integrand can be controlled. Both the contour $C_\varepsilon$ and the excised wedge will be pushed on the real axis by a limiting process, but their relative position is fixed as depicted. \emph{Right:} Cauchy contour argument for the limit in~\eqref{Z_theory_Mellin}. } \label{fig:wedge}
\end{figure}

\noindent
The Mellin integral can be then defined as 
\begin{equation}
\mathcal{G}_{4}\left( \Delta ,t_{4}\right) =\lim_{\varepsilon\to 0^+}\int_{C_\varepsilon}dz z^{\Delta-1}Z_4(z,t_4)=\lim_{\varepsilon\to 0^+}\mathrm{e}^{\mathrm{i}%
\varepsilon \Delta }\int_{0}^{\infty }\mathrm{d}t~t^{\Delta -1}Z_{4}\left( t%
\mathrm{e}^{\mathrm{i}\varepsilon },t_{4}\right).  \label{Z_theory_Mellin}
\end{equation}
Applying the Cauchy theorem to the closed contour depicted in the right panel of Fig.~\ref{fig:wedge} and taking into account the exponential
suppression of the integrand along the arc $z=R\mathrm{e}^{\mathrm{i}\phi }$%
, $\phi \in \left( \varepsilon,\varepsilon^\prime \right) $ for $R\to\infty$, we find that the integral before taking the limit $\varepsilon\to 0^+$ in fact does not depend on $\varepsilon$ and the limit therefore need not to be taken.
Using the above asymptotics for $Z_{4}\left(t
\mathrm{e}^{\mathrm{i}\varepsilon },t_{4}\right)$, we can conclude that the fundamental strip of the Mellin transform of the function $Z_{4}\left( t\mathrm{e}^{\mathrm{i}\varepsilon },t_{4}\right) $ is $\left\langle -1,\infty
\right\rangle$. 

Unfortunately, we are not able to calculate the Mellin transform explicitly but we can still derive some properties. In particular, inside the fundamental strip, $\mathcal{G}_{4}\left( \Delta
,t_{4}\right) $ is holomorphic, and in the rest of the complex plane it is
meromorphic with simple poles $\Delta =-1,-2,\ldots $, whose residues are in one-to-one correspondence with the coefficients of the low energy
expansion of the amplitude $Z_{4}\left( t\mathrm{e}^{\mathrm{i}\varepsilon
},t_{4}\right) $. Namely, as a consequence of (\ref{Z_4_IR_asymptotics}), the
singular expansion\footnote{Here we mean symbolic expression, which is a formal sum of all of the singular elements of the function (i.e. the singular parts of the Laurent expansion at all of the poles).} of $\mathcal{G}_{4}\left( \Delta ,t_{4}\right)$ reads
\begin{equation}
\mathcal{G}_{4}\left( \Delta ,t_{4}\right) \asymp -\pi ^{2}\frac{1}{t_{4}}%
\frac{1}{\Delta +1}-\frac{\pi ^{4}}{12}\frac{(1-t_{4}+t_{4}^{2})}{t_{4}^{3}}%
\frac{1}{\Delta +3}+\frac{\pi ^{2}}{2}\psi ^{\prime \prime }(1) 
\frac{t_{4}-1}{t_{4}^{3}}\frac{1}{\Delta +4}+\ldots . 
\label{G_4_singular_expansion}
\end{equation}
This is a formal sum as it is not convergent. The simple poles of (\ref{G_4_singular_expansion}) are the only singularities of the function $\mathcal{G}_{4}\left( \Delta ,t_{4}\right) $. This can be compared with the case of the leading order nonlinear sigma model amplitude, where the
celestial amplitude has a $\delta -$function singularity
\begin{equation}
\mathcal{G}_{4}^{NLSM}\left( \Delta ,t_{4}\right) =-\frac{2\pi ^{3}}{t_{4}}%
\delta \left( \Delta +1\right) .
\end{equation}
Interestingly, this result can be derived formally directly from the stringy function $\mathcal{G}_{4}\left( \Delta ,t_{4}\right)$ as follows (cf. \cite{Arkani-Hamed:2020gyp}). 
Note that the field theory limit of the amplitude $Z_{4}\left( s,t_{4}\right) $ can be obtained as the $\alpha ^{\prime }$ expansion for $\alpha ^{\prime}\rightarrow 0$. Restoring the $\alpha ^{\prime }$ dependence setting $s\rightarrow \alpha ^{\prime }s$, we get
\begin{equation}
Z_{4}\left( \alpha ^{\prime }s,t_{4}\right) =-\alpha ^{\prime }\pi ^{2}\frac{%
s}{t_{4}}+O\left( \alpha ^{\prime 3}\right) ,
\end{equation}
and thus 
\begin{equation}
A_{4}^{NLSM}\left( s,-\frac{s}{t_{4}};\left( 1-\frac{1}{t_{4}}\right)
s\right) =\lim_{\alpha ^{\prime }\rightarrow 0}\frac{Z_{4}\left( \alpha
^{\prime }s,t_{4}\right) }{\alpha ^{\prime }}.
\end{equation}
On the other hand, the $\alpha^{\prime}$ dependence of the Mellin transform $\mathcal{G}_{4}$ is simple, namely
\begin{eqnarray}
\mathcal{G}_{4}\left( \Delta ,t_{4},\alpha ^{\prime }\right) &\equiv& \mathrm{e%
}^{\mathrm{i}\varepsilon \Delta }\int_{0}^{\infty }\mathrm{d}t~t^{\Delta
-1}Z_{4}\left( \alpha ^{\prime }t\mathrm{e}^{\mathrm{i}\varepsilon
},t_{4}\right) \notag\\
&=&\frac{1}{\alpha ^{\prime \Delta }}\mathrm{e}^{\mathrm{i}%
\varepsilon \Delta }\int_{0}^{\infty }\mathrm{d}t~t^{\Delta -1}Z_{4}\left( t%
\mathrm{e}^{\mathrm{i}\varepsilon },t_{4}\right) =\frac{\mathcal{G}%
_{4}\left( \Delta ,t_{4}\right) }{\alpha ^{\prime \Delta }},
\end{eqnarray}
therefore, interchanging formally the limit and integration and using the formula
(\ref{delta_representation_limit}) and the singular expansion (\ref{G_4_singular_expansion}), we get
\begin{eqnarray}
\lim_{\alpha ^{\prime }\rightarrow 0}\frac{\mathcal{G}_{4}\left( \Delta
,t_{4},\alpha ^{\prime }\right) }{\alpha ^{\prime }}&=&\lim_{\alpha ^{\prime
}\rightarrow 0}\frac{\mathcal{G}_{4}\left( \Delta ,t_{4}\right) }{\alpha
^{\prime \Delta +1}}=2\pi \delta \left( \Delta +1\right) \lim_{\Delta
\rightarrow -1}\left( \Delta +1\right) \mathcal{G}_{4}\left( \Delta
,t_{4}\right)\notag\\
&=&-\frac{2\pi ^{3}}{t_{4}}\delta \left( \Delta +1\right)=\mathcal{G}_{4}^{NLSM}\left( \Delta ,t_{4}\right). 
\end{eqnarray}
Note again that the celestial amplitude for the $Z$-theory is a well-defined function, the counterpart in the effective theory is only defined in the distributional sense as evident from the presence of the delta function $\delta(\Delta+1)$ in ${\cal G}_4^{NLSM}(\Delta,t_4)$. 

In Appendix~\ref{Z_trunc}, we comment on a truncated version of the Z-theory $4$-pt amplitude for which the Mellin transform can be performed explicitly.

%%%%%%%%%%%%%%%%%%%%%%%%%%%%%%%%%%%%%%%%%%%
\section{Towards celestial soft bootstrap}\label{sec:7}
%%%%%%%%%%%%%%%%%%%%%%%%%%%%%%%%%%%%%%%%%%%

In this section we present several ideas how to implement methods in momentum space to calculate celestial amplitudes. We first discuss recursion relations and their soft extensions, leading to recursive formulas that include products of celestial amplitudes with one on-shell massive leg. 

%=========================================
\subsection*{Towards celestial recursion relations}
%=========================================

As we have seen in~\eqref{mellin_transform_complete}, the Mellin transform of the massless amplitude is proportional to the Mellin integral
\begin{equation}
\int_{0}^{\infty }\mathrm{d}u~u^{\Delta -1}A_{n}\left( \sqrt{u}\varepsilon
_{1}\sigma _{1}q\left( z_{1},\overline{z}_{1}\right) ,\ldots ,\sqrt{u}%
\varepsilon _{n}\sigma _{n}q\left( z_{n},\overline{z}_{n}\right) \right) ,
\end{equation}
which can be calculated by the residue theorem using the Hankel contour as described in Appendix \ref{app:A} (cf. 
formula (\ref{residue_formula_for_A_n}). Note that the configuration
of the null momenta
\begin{equation}
p_{i}\left( u\right) =\sqrt{u}\sigma _{i}q\left( z_{i},\overline{z}%
_{i}\right)
\end{equation}
is constrained by the $\delta -$function in (\ref{mellin_transform_complete}) to satisfy the momentum conservation and thus represents a valid all-line
BCFW-like shift. The residues at the poles 
\begin{equation}
u\to u_{\mathcal{F}}=\frac{M_{\mathcal{F}}^{2}}{Q_{\mathcal{F}}^{2}}=\frac{M_{%
\mathcal{F}}^{2}}{\sum\limits_{\left( i<j\right) \in \mathcal{F}}\varepsilon
_{i}\varepsilon _{j}\sigma _{i}\sigma _{j}\left\vert z_{ij}\right\vert ^{2}}
\label{Mellin_poles}
\end{equation}
in formula (\ref{integral_mellin}) correspond
then to the massive factorization channels of the amplitude, and are therefore
determined by the one-particle unitarity, namely (see Fig. \ref{fig:fact_channel})
\begin{eqnarray}
\mathrm{res}\left( A_{n}\left( u\right) ,u_{\mathcal{F}}\right) &=&-\frac{1}{%
Q_{\mathcal{F}}^{2}}\sum\limits_{I}A_{\mathcal{F}}\left( \left\{ \sqrt{u_{%
\mathcal{F}}}\varepsilon _{i}\sigma _{i}q\left( z_{i},\overline{z}%
_{i}\right) \right\} _{i\in \mathcal{F}},-\sqrt{u_{\mathcal{F}}}Q_{\mathcal{F%
}},I\right)  \notag \\
&&\hspace{1cm} \times A_{\mathcal{F}^{c}}\left( \left\{ \sqrt{u_{\mathcal{F}}}\varepsilon
_{j}\sigma _{j}q\left( z_{j},\overline{z}_{j}\right) \right\} _{j\in 
\mathcal{F}^{c}},\sqrt{u_{\mathcal{F}}}Q_{\mathcal{F}},I\right) ,
\label{residue}
\end{eqnarray}
where $I$ stays for the collection of the quantum numbers of the
one-particle intermediate states with momentum $\sqrt{u_{\mathcal{F}}}Q_{%
\mathcal{F}}$ and mass $M_{{\cal{F}}}$ and $\mathcal{F}^{c}$ is the complementary set of momenta with
respect to $\mathcal{F}$. According to the sign of $Q_{\mathcal{F}}^{2}$,
the $\sqrt{u_{\mathcal{F}}}$ is either real or purely imaginary, therefore $%
\sqrt{u_{\mathcal{F}}}Q_{\mathcal{F}}$ is always time-like and on shell (recall~(\ref{Mellin_poles})). The
amplitudes $A_{\mathcal{F}}$ and $A_{\mathcal{F}^{c}}$ on the right hand
side of (\ref{residue}) are then in general analytically continued on-shell
amplitudes to complex momenta. The formula (\ref{residue}) suggests a
possibility to calculate the Mellin transform (\ref{correlator_definition})
recursively~\footnote{Celestial BCFW recursion was treated e.g. in~\cite{Pasterski:2017ylz,Guevara:2019ypd,Hu:2022bpa}.}. The computation splits into two distinct branches, according to the causality of $Q_{\mathcal{F}}$, which have to be dealt with separately. A detailed derivation can be found in Appendix~\ref{celest_recurs_detail}, here we present the final result
\begin{eqnarray}
&&\int_{0}^{\infty }\mathrm{d}u~u^{\Delta -1}A_{n}\left( \sqrt{u}\varepsilon
_{1}\sigma _{1}q\left( z_{1},\overline{z}_{1}\right) ,\ldots ,\sqrt{u}%
\varepsilon _{n}\sigma _{n}q\left( z_{n},\overline{z}_{n}\right) \right)  
\notag \\
&=&2\frac{\pi }{\sin \pi \Delta }\int \mathrm{d}\nu \mu
\left( \nu \right) \mathrm{d}^{2}z\sum\limits_{\mathcal{F}}M_{\mathcal{F}}^{2}\left[ \theta
\left( Q_{\mathcal{F}}^{2}\right) \sum\limits_{I}\mathrm{e}^{\pi \mathrm{i}%
\Delta }\left( \widetilde{\mathcal{A}}_{\mathcal{F}}^{-,1-\mathrm{i}\nu ,I}%
\widetilde{\mathcal{A}}_{\mathcal{F}^{c}}^{+,1+\mathrm{i}\nu ,I}+\widetilde{%
\mathcal{A}}_{\mathcal{F}}^{+,1-\mathrm{i}\nu ,I}\widetilde{\mathcal{A}}_{%
\mathcal{F}^{c}}^{-,1+\mathrm{i}\nu ,I}\right) \right.   \notag \\
&&\hspace{7cm} \left. +\theta \left( -Q_{\mathcal{F}}^{2}\right) \sum\limits_{I}%
\widetilde{\mathcal{A}}_{\mathcal{F}}^{C,-,1-\mathrm{i}\nu ,I}\widetilde{%
\mathcal{A}}_{\mathcal{F}^{c}}^{C,+,1+\mathrm{i}\nu ,I}\right],
\label{recursion_resulting_formula}
\end{eqnarray}
where we abbreviated the celestial amplitudes with just one massive external state $I$ as%
\begin{eqnarray}
\widetilde{\mathcal{A}}_{\mathcal{F}}^{\varepsilon ,\Delta }
&\equiv &\widetilde{\mathcal{A}}_{\mathcal{F}}\left( \left\{ \varepsilon
_{i},\Delta _{i},z_{i},\overline{z}_{i}\right\} ,\left\{ \varepsilon ,\left(
\Delta \right) ,z,\overline{z},I\right\} \right),   
\end{eqnarray}
and similarly for the generalized celestial amplitude (you can find the full definition in the Appendix~\ref{celest_recurs_detail})
\begin{eqnarray}
\widetilde{\mathcal{A}}_{\mathcal{F}}^{C,\varepsilon ,\Delta ,I} &\equiv &%
\widetilde{\mathcal{A}}^C_{\mathcal{F}}\left( \left\{ \varepsilon
_{i},\Delta _{i},z_{i},\overline{z}_{i}\right\} _{i\in \mathcal{F}},\left\{
\varepsilon ,\Delta ,z,\overline{z},I\right\} \right)\,.
\end{eqnarray}%
In the formula (\ref{recursion_resulting_formula}), the integrals are over parameters of the internal massive leg. Importantly, the individual terms in the sum are products of (generalized) celestial   amplitudes where one of the on-shell legs is massive. Therefore, we can not continue the recursion beyond the first step unless we consider a more general framework with both massive and massless legs for the celestial amplitudes. This is of course possible to do but it goes beyond the scope of this paper.

%==========================================
\subsection*{Celestial six-point functions in soft-reconstructible
theories}
%==========================================

We can apply the same logic for soft-reconstructible theories. These are theories which fail the BCFW criterion for vanishing at infinity, but this deficiency is compensated by improved soft behavior. See section~\ref{sec:2} for review. We will concentrate on amplitudes in these theories at 6pt. Let us remind the general form of the 6pt function of massless scalar
particles
\begin{equation}
\widetilde{\mathcal{A}}_{6}\left( \left\{ \Delta _{i},z_{i}\right\} \right)
=C_{6}\left( \left\{ \Delta _{i},z_{i}\right\} \right) \mathcal{H}_{6}\left(
\left\{ \Delta _{i}\right\} ,t_{4},t_{5},t_{6}\right) \,,
\end{equation}%
where $C_{6}\left( \left\{ \Delta _{i},z_{i}\right\} \right) $ is given by
the formula (\ref{conformal_factor_n}) and where
\begin{eqnarray}
&&\hspace{-0.5cm}\mathcal{H}_{6}\left( \left\{ \Delta _{i}\right\}
,t_{4},t_{5},t_{6}\right) =\lim_{\varepsilon \rightarrow 0}\int_{0}^{1}%
\mathrm{d}\sigma _{6}\sigma _{6}^{\Delta _{6}-1}\frac{\varepsilon ^{-2\Delta
_{3}}\prod\limits_{i=1}^{5}\sigma _{\ast i}^{\Delta _{i}-1}\chi _{\langle
0,1\rangle }\left( \sigma _{\ast i}\right) }{2\left\vert \det \mathbf{A}%
_{5}\right\vert }  \\
&&\times \left. \frac{1}{2}\int_{0}^{\infty }\mathrm{d}u~u^{\Delta
-1}A_{6}\left( \sqrt{u}\varepsilon _{1}\sigma _{\ast 1}q\left( 1,1\right)
,\ldots ,\sqrt{u}\varepsilon _{6}\sigma _{6}q\left( t_{6}^{-1},\overline{%
t_{6}}^{-1}\right) \right) \right\vert _{\left( 1,0,\varepsilon
^{-1},t_{4}^{-1},\ldots ,t_{6}^{-1}\right) }\,.\notag
\end{eqnarray}
For soft-reconstructible scalar theories~\cite{Cheung:2016drk} we can proceed as follows. Suppose that the power-counting parameter $\rho $, which is defined as the ratio
\begin{equation}
\rho =\frac{D_{V}-2}{N_{V}-2}\,,
\end{equation}
where $D_{V}$ is number of derivatives and $N_{V}$ number of external legs
of the vertex $V$, is fixed, i.e. it does not depend on $V$. Then we have
\begin{equation}
A_{6}\left( \sqrt{u}\varepsilon _{1}\sigma _{\ast 1}q\left( 1,1\right)
,\ldots ,\right) =\left( \sqrt{u}\right) ^{2+4\rho }A_{6}\left( \varepsilon
_{1}\sigma _{\ast 1}q\left( 1,1\right) ,\ldots \right) ,
\end{equation}
and thus
\begin{eqnarray}
&&\hspace{-1cm}\mathcal{H}_{6}\left( \left\{ \Delta _{i}\right\}
,t_{4},t_{5},t_{6}\right) =\lim_{\varepsilon \rightarrow 0}\int_{0}^{1}%
\mathrm{d}\sigma _{6}\sigma _{6}^{\Delta _{6}-1}\frac{\varepsilon ^{-2\Delta
_{3}}\prod\limits_{i=1}^{5}\sigma _{\ast i}^{\Delta _{i}-1}\chi _{\langle
0,1\rangle }\left( \sigma _{\ast i}\right) }{2\left\vert \det \mathbf{A}%
_{5}\right\vert }  \notag \\
&&\times \left. \pi \delta \left( \Delta +2\rho +1\right) A_{6}\left(
\varepsilon _{1}\sigma _{\ast 1}q\left( 1,1\right) ,\ldots ,\varepsilon
_{6}\sigma _{6}q\left( t_{6}^{-1},\overline{t_{6}}^{-1}\right) \right)
\right\vert _{\left( 1,0,\varepsilon ^{-1},t_{4}^{-1},\ldots
,t_{6}^{-1}\right) }.  \label{6pt_correlator}
\end{eqnarray}
The limit $\varepsilon \rightarrow 0$ can be taken by formal exchange
\begin{equation}
\det \mathbf{A}_{5} \rightarrow \det \widetilde{\mathbf{A}_{5}}\equiv
\lim_{\varepsilon \rightarrow 0}\varepsilon ^{2}\det \mathbf{A}_{5},\quad
\sigma _{\ast i} \rightarrow \widetilde{\sigma }_{\ast i}\equiv \left\{ 
\begin{array}{c}
\sigma _{\ast i}\text{,~~~for}~~~i\neq 3 \\ 
\lim_{\varepsilon \rightarrow 0}\frac{\sigma _{\ast 3}}{\varepsilon ^{2}}%
\end{array}%
\right. ,
\end{equation}
and
\begin{equation}
\sigma _{\ast 3}q\left( \frac{1}{\varepsilon },\frac{1}{\varepsilon }\right)
\rightarrow \widetilde{\sigma }_{\ast 3}n,~~~~n=\frac{1}{2}\left(
1,0,0,1\right) ,
\end{equation}
in the formula (\ref{6pt_correlator}). The $\sigma _{\ast i}$'s have the
general form $\widetilde{\sigma }_{\ast i}\left( \sigma _{6}\right) =\widetilde{\sigma }%
_{\ast i}^{\left( 5\right) }+\widetilde{K}_{i}\sigma _{6}$ where $\sigma _{\ast i}^{\left( 5\right) }$ are the solutions of equations
(\ref{sigma_equations}) with 5pt kinematics (see Appendix \ref{5pt_sigmas}) and $\widetilde{K}_{i}$'s are listed in Appendix \ref{sigma_6pt}. At the tree level, the function 
\begin{equation}
\widehat{A}_{6}\left( \sigma _{6}\right) \equiv A_{6}\left( \varepsilon _{1}%
\widetilde{\sigma }_{\ast 1}q\left( 1,1\right) ,\ldots ,\widetilde{\sigma }%
_{\ast 3}n,\ldots ,\varepsilon _{6}\sigma _{6}q\left( t_{6}^{-1},\overline{%
t_{6}}^{-1}\right) \right) 
\end{equation}
is therefore a meromorphic function with simple poles $z_{\mathcal{F}}$
corresponding to the factorization channels $\mathcal{F}$ of the 6pt
amplitude. These are defined as the solutions of the equations\footnote{Here we tacitly assume $q_{3}=n$ as above$.$}
\begin{equation}
P_{\mathcal{F}}^{2}(z)\equiv \left( \sum\limits_{i\in \mathcal{F}%
}\varepsilon _{i}\widetilde{\sigma }_{\ast i}\left( z_{\mathcal{F}}\right)
q_{i}\right) ^{2}=0.  \label{pole_factorization}
\end{equation}
Provided the theory satisfies the enhanced Adler zero condition with soft
index $\sigma $, i.e.
\begin{equation}
A_{6}\left( p_{1},\ldots ,p_{6}\right) \overset{p_{i}\rightarrow 0}{=}%
O\left( p_{i}^{\sigma }\right) 
\end{equation}
and is reconstructible, i.e. $\sigma\ge\rho \geq \sigma -1$, the function $\widehat{A}_{6}\left( \sigma _{6}\right) $ can be
reconstructed from its poles and zeroes. Namely, applying the residue
theorem to the function
\begin{equation}
f\left( z\right) \equiv \frac{\widehat{A}_{6}\left( z\right) }{F\left(
z\right) \left( z-\sigma _{6}\right) } \quad\mbox{where}\quad
F\left( z\right) \equiv \left[ z\prod\limits_{i=1}^{5}\widetilde{\sigma }%
_{\ast i}\left( z\right) \right] ^{\sigma }  \label{F(z)_definition}
\end{equation}
and taking into account that there is neither residue for $\sigma
_{6}\rightarrow \infty $ nor for $\widetilde{\sigma }_{\ast i}\left( \sigma
_{6}\right) \rightarrow 0$, we get
\begin{equation}
\widehat{A}_{6}\left( \sigma _{6}\right)  =-F\left( \sigma _{6}\right)
\sum\limits_{\mathcal{F}}\mathrm{res}\left( \frac{\widehat{A}_{6}\left(
z\right) }{F\left( z\right) \left( z-\sigma _{6}\right) },z_{\mathcal{F}%
}\right)=\sum\limits_{\mathcal{F}}\frac{\mathrm{res}\left( \widehat{A}_{6}\left(
z\right) ,z_{\mathcal{F}}\right) }{F\left( z_{\mathcal{F}}\right) }\frac{%
F\left( \sigma _{6}\right) }{\sigma _{6}-z_{\mathcal{F}}}\,.
\end{equation}
The residue $\mathrm{res}\left( \widehat{A}_{6}\left( z\right) ,z_{%
\mathcal{F}}\right) $ can be calculated using the factorization at the pole (\ref{pole_factorization}) as
\begin{equation}
\mathrm{res}\left( \widehat{A}_{6}\left( z\right) ,z_{\mathcal{F}}\right)
=-A_{L}^{\mathcal{F}}\left( z_{\mathcal{F}}\right) A_{R}^{\mathcal{F}%
^{c}}\left( z_{\mathcal{F}}\right) \mathrm{res}\left( \frac{1}{P_{\mathcal{F}%
}^{2}(z)},z_{\mathcal{F}}\right) \,,
\end{equation}
where $A_{L,R}^{\mathcal{F}}\left( z_{\mathcal{F}}\right) $ are the
lower-point amplitudes with momenta $\left\{ \varepsilon _{i}\widetilde{%
\sigma }_{\ast i}\left( z_{\mathcal{F}}\right) q_{i}\right\} $ with $i\in 
\mathcal{F},\mathcal{F}^{c}$. Finally, using (\ref{F(z)_definition}) we can write
\begin{eqnarray}
\mathcal{H}_{6}\left( \left\{ \Delta _{i}\right\} ,t_{4},t_{5},t_{6}\right) 
&=&-\pi \delta \left( \Delta +2\rho +1\right) \sum\limits_{\mathcal{F}%
}\frac{A_{L}^{\mathcal{F}}\left( z_{\mathcal{F}}\right) A_{R}^{\mathcal{F}%
^{c}}\left( z_{\mathcal{F}}\right)}{F(z_{\cal F})} \label{H6rec}\\
&&\hspace{1cm}\times \mathrm{res}\left( \frac{1}{P_{\mathcal{F}}^{2}(z)},z_{\mathcal{F}%
}\right) \mathcal{K}\left( \left\{ \Delta _{i}+\sigma \right\} ,\left\{
\varepsilon _{i}\right\} ,\left\{ t_{i}\right\} |z_{\mathcal{F}}\right)  \,,
\notag
\end{eqnarray}
where we introduced the theory-independent universal function, depending
only on the 6pt kinematics, by the formula
\begin{align}
\mathcal{K}\left( \left\{ \Delta _{i}\right\} ,\left\{ \varepsilon
_{i}\right\} ,\left\{ t_{i}\right\} |z\right) &=\int \mathrm{d}\sigma
_{6}\sigma _{6}^{\Delta _{6}-1}\chi _{\langle 0,1\rangle }\left( \sigma
_{6}\right) \frac{\prod\limits_{i=1}^{5}\widetilde{\sigma }_{\ast i}^{\Delta
_{i}-1}\chi _{\langle 0,1\rangle }\left( \sigma _{\ast i}\right) }{%
2\left\vert \det \widetilde{\mathbf{A}}_{5}\right\vert  }%
\frac{1}{\sigma _{6}-z} \notag\\
&\hspace{-1cm}=\int_{\,0}^{1}\prod\limits_{i=1}^{6}\mathrm{d}\sigma _{i}\sigma
_{i}^{\Delta _{i}-1}\delta \left( 1-\sum\limits_{j=1}^{6}\sigma _{i}\right)
\delta ^{(4) }\left( \sum\limits_{k=1}^{6}\varepsilon
_{k}\sigma _{k}q_{k}\right) \frac{1}{\sigma _{6}-z}\,,
\end{align}
where the last line is written in a more symmetric form. It would be interesting to extend this procedure to a general $n$-pt amplitude. 

%===============================
\subsection*{Soft bootstrap on the celestial sphere?}
%===============================

It is a natural idea to explore the possibility of bootstrap approach on the celestial sphere. In this approach, we would like to expand the celestial amplitude $\widetilde{\cal A}_n$ in terms of some well defined building and carefully chosen blocks ${\cal B}_k$ with arbitrary coefficients,
\begin{equation}
\widetilde{\cal A}_n = \sum_k c_k {\cal B}_k
\end{equation}
impose a set of certain conditions and fix the coefficients $c_k$ uniquely. These conditions should include the celestial avatars of factorization and soft limit properties (and any others like double soft limits and collinear limits). In a trivial sense, this could be done by writing ${\cal B}_k$ using the Mellin transforms of the momentum space building blocks $B_k(p_j)$
and impose the conditions directly on $B_k(p_j)$. This just reduces the problem back to the momentum space problem and we have not really learned anything. Ideally, we would like to impose constraints directly on ${\cal B}_k$. While the constraints of locality and unitarity in this basis are not well understood, the celestial basis conjecturally makes manifest different properties of amplitudes not visible in the momentum space representation. However, our current understanding of the celestial amplitudes is not deep enough to formulate and efficiently run the bootstrap methods. We leave that for future work in the years to come.

%%%%%%%%%%%%%%%%%%%%%%%%%%%%%%%%%%%%%%
\section{Summary}\label{sec:8}
%%%%%%%%%%%%%%%%%%%%%%%%%%%%%%%%%%%%%%

In this paper, we continue the effort (initiated e.g. in~\cite{Hamada:2017atr,Garcia-Sepulveda:2022lga,Kapec:2022axw,Kapec:2022hih}) of analyzing celestial amplitudes of Goldstone bosons and their conformal soft theorems.
Our intention was to focus on the form of soft theorems for effective theories of Goldstone bosons in a celestial basis. In particular, we have collected a representative selection of soft behaviors of scalar particles, transformed it to the celestial basis and thus obtained a  database of conformally soft theorems for scalar Goldstone bosons.

One takeaway message is that for celestial amplitudes of massless particles to be well defined, their fundamental strip associated with the Mellin transform has to be non-empty. It is governed by low and high energy properties of the amplitude and there is tension between them. An amplitude with enhanced soft behavior but bad high-energy behavior can have a well-defined celestial dual. The opposite is also true, a very soft UV behavior (like in string theories) does not require particularly nice IR properties for the celestial amplitude to be well-defined. In this work we emphasize the IR properties, consider the soft behavior as fixed and described by a given effective theory of Goldstone bosons (a non-linear sigma model). Then, in order to obtain well-defined celestial amplitudes we first have to construct partial UV completions of these effective theories -- renormalizable theories known as linear sigma models. 

In particular, our examples encompass two linear sigma models (the so-called $U(1)$ and $O(N)$ sigma models) including also massive particles in addition to the Goldstone bosons. 
Integrating out these massive states in the leading order of low energy expansion we obtain the corresponding nonlinear sigma models, the celestial amplitudes of  which have empty fundamental strip and exist only in the sense of distributions.
In the momentum space, the amplitudes within the linear and nonlinear version can be related by appropriate limit sending the masses of the massive states to infinity.
Taking the corresponding limit for the celestial dual of these amplitudes in the distributive sense, we might naively expect to obtain the same interrelation.
However, this naive correspondence is rather subtle: the result depends on the way how the infinity mass limit is performed in terms of the original parameters of the linearized theory, namely the dimensionless coupling and the vacuum expectation value, which sets the scale of the spontaneous symmetry breaking.
We have discussed in detail two possible scenarios of this procedure, namely either sending the dimensionless coupling to infinity while keeping the scale of spontaneous symmetry breaking fixed or sending just the  latter scale to infinity when keeping the coupling fixed.
On the Lagrangian level, the first scenario leads to the NLSM while within the second one we end up with a free theory of decoupled Goldstone bosons.
Though on the level of the momentum space amplitudes, the result of both scenarios might be the same (as established for the $U(1)$ case), this is not true for their celestial duals.
Moreover, while the first scenario interrelates the celestial amplitudes of linearized and nonlinear sigma models correctly, the second one does not and instead of yielding the zero celestial amplitude we get nonzero result reflecting the UV properties of the linearized amplitudes in the momentum space.
The latter case is a manifestation of the anti-Wilsonian paradigm for amplitudes on the celestial sphere.

The next entry on the list of theories we present is the $U(1)$ fibered $\mathbb{C}P^{N-1}$ non-linear sigma model. It has non-zero odd amplitudes and thus allows for soft limits with non-trivial right-hand sides. In accord with the strategy outlined above, we then constructed its partial UV completion and computed celestial amplitudes within this linear sigma model. 
We have found, that the corresponding conformal soft theorem on the celestial sphere can be formally interpreted as a global Ward identity associated with an explicitly or anomalously broken global symmetry acting on the primaries as a linear transformation.

The final example within the framework of spontaneously broken QFT is the pion-dilaton model. It is presented as last since it features the most complex soft behavior. In particular, its soft theorems in momentum space have nonzero right-hand side and the corresponding soft factors depend on the momenta of massive particles. This has far-reaching consequences for the associated conformally soft theorems: on the celestial sphere, the right-hand side  of the conformal soft theorem is highly non-local  and causes difficulties in their interpretation as Ward identities. For this reason it requires further study, especially along the lines of~\cite{Guevara:2021tvr,Kapec:2021eug}.  

The closing pair of theories we discuss is of a slightly different nature in that the UV completion is not just a partial one. As the effective theory we consider the $\mathrm{SU}(N)$ non-linear sigma model relevant for low energy QCD. Its conjecturally full UV completion of stringy nature (therefore UV finite) is known as Z-theory~\cite{Carrasco:2016ldy}. Despite the fact that we did not manage to compute the $4$-pt celestial amplitude in a closed form, we were still able to comment on its properties and derive a truncation that approximates it to some degree.
Also for this example we compared celestial amplitudes for this stringy UV completion with those in the low energy effective theory and commented on recovering them in the QFT limit.

We closed the paper by investigating a BCFW-like celestial recursion. At this point it does not have direct applicability as it requires celestial amplitudes with one massive leg as input. Those are technically hard to compute.

\acknowledgments

We thank Paolo di Vecchia for turning our attention to the pion-dilaton model. This work is supported by GA\v{C}R 21-26574S, MEYS LUAUS23126, DOE grant No. SC0009999 and the funds of the University of California.

%%%%%%%%%%%%%%%%%%%%%%%%%%%%%%%%%%%%%%%%%%%%%%%%%%%%%%%%%%%%%%%%%%%%%%%%%
\appendix 

%%%%%%%%%%%%%%%%%%%%%%%%%
\section{Useful formulas}
%%%%%%%%%%%%%%%%%%%%%%%%%

%===============================
\subsection*{Notation and conventions}\label{app:A}
%===============================

We use the mostly minus metric 
\begin{equation*}
\eta _{\mu \nu }=\mathrm{diag}\left( 1,-1,-1,-1\right) .
\end{equation*}%
The spinors corresponding to the light-like momentum $p^{2}=0$, $p^{0}>0$,
are given as follows%
\begin{eqnarray}
|p\rangle &=&\sqrt{\omega }\left( 
\begin{array}{c}
z \\ 
1%
\end{array}%
\right) ,~~~~|p]=\sqrt{\omega }\left( 
\begin{array}{c}
1 \\ 
-\overline{z}%
\end{array}%
\right)  \notag \\
\langle p| &=&\sqrt{\omega }\left( 1,-z\right) \text{,~~~}[p|=\sqrt{\omega }%
\left( \overline{z},1\right),
\end{eqnarray}%
where $|p\rangle $ is right-handed and $|p]$ left-handed, and where 
\begin{equation*}
\omega =p^{-},~~~z=\frac{p_{\perp }^{\ast }}{p^{-}},~~~\overline{z}=\frac{%
p_{\perp }}{p^{-}}
\end{equation*}%
with%
\begin{equation*}
~p^{\pm }=p^{0}\pm p^{3},~~~p_{\perp }=p^{1}+\mathrm{i}p^{2},~~~p_{\perp
}^{\ast }=p^{1}-\mathrm{i}p^{2}.
\end{equation*}%
Then, defining $\sigma ^{\mu }=\left( 1,\sigma ^{i}\right) $ and $\overline{%
\sigma }^{\mu }=\left( 1,-\sigma ^{i}\right) $ we have 
\begin{equation*}
p_{A\overset{.}{A}}=p_{\mu }\overline{\sigma }_{A\overset{.}{A}}^{\mu
}=|p\rangle _{A}[p|_{\overset{.}{A}}=\left( 
\begin{array}{cc}
p^{+} & p_{\perp }^{\ast } \\ 
p_{\perp } & p^{-}%
\end{array}%
\right) =\omega \left( 
\begin{array}{cc}
z\overline{z} & z \\ 
\overline{z} & 1%
\end{array}%
\right)
\end{equation*}%
and similarly%
\begin{equation*}
p^{\overset{.}{A}A}=p^{\mu }\sigma _{\mu }^{\overset{.}{A}A}=\left( 
\begin{array}{cc}
p^{-} & p_{\perp } \\ 
p_{\perp }^{\ast } & p^{+}%
\end{array}%
\right) =\omega \left( 
\begin{array}{cc}
1 & \overline{z} \\ 
z & z\overline{z}%
\end{array}%
\right),
\end{equation*}%
which results in~\footnote{Note that in the embedding formalism $q(z,\overline{z})$ corresponds to the $q^-=1$ section of the Minkowski light-cone. Most authors in the celestial literature use either $q^+=2$ or $q^+=1$ sections.}
\begin{equation*}
p^{\mu }=\frac{1}{2}\omega \left( 1+z\overline{z},z+\overline{z},\mathrm{i}%
\left( z-\overline{z}\right) ,z\overline{z}-1\right) \equiv \omega q\left(
z,\overline{z}\right) .
\end{equation*}%
For the spinor products we get%
\begin{equation*}
\langle i,j\rangle =\sqrt{\omega _{i}\omega _{j}}\left( z_{j}-z_{i}\right)
,~~~~\left[ i,j\right] =\sqrt{\omega _{i}\omega _{j}}\left( \overline{z}_{i}-%
\overline{z}_{j}\right)
\end{equation*}%
therefore scalar products read%
\begin{equation*}
2p_{i}\cdot p_{j}=\omega _{i}\omega _{j}\left\vert z_{i}-z_{j}\right\vert
^{2}\equiv \omega _{i}\omega _{j}\left\vert z_{ij}\right\vert ^{2}.
\end{equation*}%
Under Lorentz transformations $\Lambda $ the spinors transform according to%
\begin{eqnarray*}
|\Lambda p\rangle &=&U_{R}\left( \Lambda \right) |p\rangle e^{-\frac{i}{2}%
\theta \left( \Lambda ,p\right) } \\
|\Lambda p] &=&U_{L}\left( \Lambda \right) |p]e^{\frac{i}{2}\theta \left(
\Lambda ,p\right) },
\end{eqnarray*}%
where $U_{R}\left( \Lambda \right) $ and $U_{L}\left( \Lambda \right) $ are $SL(2,\mathbb{C})$ matrices 
\begin{equation*}
U_{R}\left( \Lambda \right) =\left( 
\begin{array}{cc}
a & b \\ 
c & d%
\end{array}%
\right) ,~~~~U_{L}\left( \Lambda \right) =\left( 
\begin{array}{cc}
d^{\ast } & -c^{\ast } \\ 
-b^{\ast } & a^{\ast }%
\end{array}%
\right) ,
\end{equation*}%
and 
\begin{equation*}
\theta \left( \Lambda ,p\right) =2\arg \left( cz+d\right) .
\end{equation*}%
Then the transformed spinors take the form%
\begin{equation*}
|\Lambda p\rangle =\sqrt{\omega ^{\prime }}\left( 
\begin{array}{c}
z^{\prime } \\ 
1%
\end{array}%
\right) ,~~~|\Lambda p]=\sqrt{\omega ^{\prime }}\left( 
\begin{array}{c}
1 \\ 
-\overline{z}^{\prime }%
\end{array}%
\right),
\end{equation*}%
where%
\begin{equation*}
z^{\prime }=\frac{az+b}{cz+d},~~\overline{z}^{\prime }=\frac{a^{\ast }%
\overline{z}+b^{\ast }}{c^{\ast }\overline{z}+d^{\ast }},~~~\omega ^{\prime
}=\left\vert cz+d\right\vert ^{2}\omega.
\end{equation*}%
The invariant measure on the forward light cone is 
\begin{equation*}
\widetilde{\mathrm{d}}p=\int_{p^{0}}\mathrm{d}^{4}p\theta \left(
p^{0}\right) \delta ^{(4) }\left( p^{2}\right) =\frac{1}{2}%
\omega \mathrm{d}\omega \mathrm{d}^{2}z,
\end{equation*}%
where%
\begin{equation*}
\mathrm{d}^{2}z=\mathrm{d}{\rm{Re}}z~\mathrm{d}{\rm{Im}}z.
\end{equation*}%
For massive particles with mass $m$, an appropriate parametrization of the
on-shell momenta reads%
\begin{eqnarray*}
p &=&m\widehat{p}\left( w,\overline{w},y\right) =\frac{m}{2y}\left( 1+w%
\overline{w}+y^{2},w+\overline{w},i\left( w-\overline{w}\right) ,w\overline{w%
}+y^{2}-1\right) \\
&=&\frac{m}{y}\left[ q\left( w,\overline{w}\right) +{y^{2}}n\right],
\end{eqnarray*}%
where
$n=(1,0,0,1)/2$.
The invariant measure on the two-sheeted mass hyperboloid $\widehat{p}^{2}=1$ then takes the form%
\begin{equation*}
\widetilde{\mathrm{d}}\widehat{p}=\frac{1}{2y^{3}}\mathrm{d}y\mathrm{d}^{2}w,
\end{equation*}
where the upper sheet is covered by $y>0$, while the lower one by $y<0$.

%%%%%%%%%%%%%%%%%%%%%%%%%%%%%%%%%%%%
\subsection*{Some useful integrals \label{Appendix_Integrals}}
%%%%%%%%%%%%%%%%%%%%%%%%%%%%%%%%%%%%

Here we give a survey of the integrals 
\begin{equation}
I_{n}\left( \left\{ \Delta _{i},\mathbf{z}_{i}\right\} _{i=1}^{n}\right)
=\int_{H_{+}}\widetilde{\mathrm{d}}\widehat{p}\prod\limits_{i=1}^{n}\left( 
\frac{1}{2\widehat{p}\cdot q\left( \mathbf{z}_{i}\right) }\right) ^{\Delta
_{i}}=\int \frac{\mathrm{d}y}{y^{3}}\mathrm{d}^{2}\mathbf{w}%
\prod\limits_{i=1}^{n}\left( \frac{y}{y^{2}+\left\vert \mathbf{w}-\mathbf{z}%
_{i}\right\vert ^{2}}\right) ^{\Delta _{i}}
\end{equation}%
for $n=2,3$, which are used in the main text. These are a special case of contact (Euclidean) AdS Witten diagrams and a much more complete analysis of their structure can be found e.g. in~\cite{Liu:1998ty,Freedman:1998bj,DHoker:1998ecp,Mueck:1998wkz,DHoker:1999mqo,Hijano:2015zsa}. Using Schwinger
parametrization we have%
\begin{equation}
I_{n}\left( \left\{ \Delta _{i},\mathbf{z}_{i}\right\} _{i=1}^{n}\right)
=\int_{0}^{\infty }\prod\limits_{i=1}^{n}\mathrm{d}\alpha _{i}\frac{\alpha
_{i}^{\Delta _{i}-1}}{\Gamma \left( \Delta _{i}\right) }\int_{H_{+}}%
\widetilde{\mathrm{d}}\widehat{p}~\exp \left( -2\widehat{p}\cdot Q\right) ,
\end{equation}%
where the following relations hold%
\begin{equation*}
Q=\sum\limits_{i=1}^{n}\alpha _{i}q\left( \mathbf{z}_{i}\right)
,~~~Q^{2}=\sum\limits_{i<j=1}^{n}\alpha _{i}\alpha _{j}\left\vert \mathbf{z}%
_{i}-\mathbf{z}_{j}\right\vert ^{2}>0.
\end{equation*}%
Due to the Lorentz invariance of $I_{n}\left( \left\{ \Delta _{i},\mathbf{z}%
_{i}\right\} _{i=1}^{n}\right) $, we can use the frame where $Q=\left( Q^{0},%
\mathbf{0}\right) $ with $Q^{0}=\sqrt{Q^{2}}$and then%
\begin{equation*}
2\widehat{p}\cdot Q=\frac{Q^{0}}{y}\left( 1+\left\vert \mathbf{w}\right\vert
^{2}+y^{2}\right) .
\end{equation*}%
The integration over $\mathbf{w}$ is elementary with the result%
\begin{eqnarray}
I_{n}\left( \left\{ \Delta _{i},\mathbf{z}_{i}\right\} _{i=1}^{n}\right)
&=&\pi \int_{0}^{\infty }\prod\limits_{i=1}^{n}\mathrm{d}\alpha _{i}\frac{%
\alpha _{i}^{\Delta _{i}-1}}{\Gamma \left( \Delta _{i}\right) }%
\int_{0}^{\infty }\frac{\mathrm{d}y}{y^{3}}\frac{y}{Q^{0}}\exp \left(
-yQ^{0}-\frac{Q^{0}}{y}\right)  \notag \\
&=&\pi \int_{0}^{\infty }\prod\limits_{i=1}^{n}\mathrm{d}\alpha _{i}\frac{%
\alpha _{i}^{\Delta _{i}-1}}{\Gamma \left( \Delta _{i}\right) }%
\int_{0}^{\infty }\frac{\mathrm{d}t}{t^{2}}\exp \left( -t-\frac{Q^{2}}{t}%
\right),\label{I_n_modified}
\end{eqnarray}%
where we denoted $t=yQ^{0}$.

For $n=2$ we have $Q^{2}=\alpha _{1}\alpha _{2}\left\vert \mathbf{z}_{1}-%
\mathbf{z}_{2}\right\vert ^{2}$. Let us integrate over $\alpha _{1}$ first
and then over $t$. We get%
\begin{eqnarray}
I_{2}\left( \left\{ \Delta _{i},\mathbf{z}_{i}\right\} _{i=1}^{2}\right) &=&%
\frac{\pi }{\Gamma \left( \Delta _{2}\right) }\int_{0}^{\infty }\frac{%
\mathrm{d}t}{t^{2}}\mathrm{e}^{-t}\mathrm{d}\alpha _{2}\alpha _{2}^{\Delta
_{2}-1}\left( \frac{t}{\alpha _{2}\left\vert \mathbf{z}_{1}-\mathbf{z}%
_{2}\right\vert ^{2}}\right) ^{\Delta _{1}}  \notag \\
&=&\pi \frac{\Gamma \left( \Delta _{1}-1\right) }{\Gamma \left( \Delta
_{2}\right) }\frac{1}{\left\vert \mathbf{z}_{1}-\mathbf{z}_{2}\right\vert
^{2\Delta _{1}}}\int_{0}^{\infty }\mathrm{d}\alpha _{2}\alpha _{2}^{\Delta
_{2}-\Delta _{1}-1}.
\end{eqnarray}
Using now
\begin{equation}
\int_{0}^{\infty }\mathrm{d}\alpha _{2}\alpha _{2}^{\Delta _{2}-\Delta
_{1}-1}=2\pi \delta \left( \Delta _{1}-\Delta _{2}\right)\,,
\end{equation}
we get the final result for $n=2$
\begin{equation}
I_{2}\left( \left\{ \Delta _{i},\mathbf{z}_{i}\right\} _{i=1}^{2}\right) =%
\frac{2\pi ^{2}}{\Delta _{1}-1}\delta \left( \Delta _{1}-\Delta _{2}\right) 
\frac{1}{\left\vert \mathbf{z}_{1}-\mathbf{z}_{2}\right\vert ^{2\Delta _{1}}}\,.
\end{equation}
For $n=3$ we can write
\begin{equation}
Q^{2}=\alpha _{1}\alpha _{2}\left\vert \mathbf{z}_{1}-\mathbf{z}%
_{2}\right\vert ^{2}+\alpha _{1}\alpha _{3}\left\vert \mathbf{z}_{1}-\mathbf{%
z}_{3}\right\vert ^{2}+\alpha _{2}\alpha _{3}\left\vert \mathbf{z}_{2}-%
\mathbf{z}_{3}\right\vert ^{2}.
\end{equation}%
Let us substitute%
\begin{equation}
\frac{\alpha _{i}\alpha _{j}}{t}=x_{k},~~~i,j,k,=1,2,3,~~~i\neq j\neq k
\end{equation}%
with the inverse%
\begin{equation}
\alpha _{i}=\sqrt{\frac{tx_{j}x_{k}}{x_{i}}}i,j,k,=1,2,3,~~~i\neq j\neq k
\end{equation}%
and with the Jacobian%
\begin{equation}
\frac{\partial \left( \alpha _{1},\alpha _{2},\alpha _{3}\right) }{\partial
\left( x_{1},x_{2},x_{3}\right) }=\frac{t^{3/2}}{2\sqrt{x_{1},x_{2},x_{3}}}.
\end{equation}%
Then the integral (\ref{I_n_modified}) factorizes
\begin{eqnarray}
I_{3}\left( \left\{ \Delta _{i},\mathbf{z}_{i}\right\} _{i=1}^{3}\right) &=&%
\frac{\pi }{2}\int_{0}^{\infty }\mathrm{d}t~t^{-2+\frac{3}{2}+\frac{1}{2}%
\sum\limits_{i=1}^{3}(\Delta _{i}-1)}\mathrm{e}^{-t}  \notag \\
&&\times \prod\limits_{i=1,i\neq j\neq k}^{3}\int_{0}^{\infty }\frac{%
\mathrm{d}x_{i}}{\Gamma \left( \Delta _{i}\right) }x_{i}^{-\frac{1}{2}-\frac{%
\Delta _{i}-1}{2}+\frac{\Delta _{j}-1}{2}+\frac{\Delta _{k}-1}{2}}\exp
\left( -x_{i}\left\vert \mathbf{z}_{j}-\mathbf{z}_{k}\right\vert ^{2}\right)
.\notag\\
\end{eqnarray}%
As a result of the above integration one obtains
\begin{eqnarray}
I_{3}\left( \left\{ \Delta _{i},\mathbf{z}_{i}\right\} _{i=1}^{3}\right) &=&%
\frac{\pi }{2}\Gamma \left( \frac{1}{2}\sum\limits_{i=1}^{3}\Delta
_{i}-1\right) \prod\limits_{i=1,i\neq j\neq k}^{3}\frac{\Gamma \left( \frac{%
\Delta _{j}}{2}+\frac{\Delta _{k}}{2}-\frac{\Delta _{i}}{2}\right) }{\Gamma
\left( \Delta _{i}\right) }\frac{1}{\left\vert \mathbf{z}_{j}-\mathbf{z}%
_{k}\right\vert ^{\Delta _{j}+\Delta _{k}-\Delta _{i}}}  \notag \\
&\equiv &\frac{C\left( \Delta _{1},\Delta _{2},\Delta _{3}\right) }{%
\left\vert \mathbf{z}_{1}-\mathbf{z}_{2}\right\vert ^{\Delta _{1}+\Delta
_{2}-\Delta _{3}}\left\vert \mathbf{z}_{1}-\mathbf{z}_{3}\right\vert
^{\Delta _{1}+\Delta _{3}-\Delta _{2}}\left\vert \mathbf{z}_{2}-\mathbf{z}%
_{3}\right\vert ^{\Delta _{2}+\Delta _{3}-\Delta _{1}}}.
\label{I_3}
\end{eqnarray}%

%========================
\subsection*{Hankel contour \label{App:Hankel}}
%========================

The Mellin integral from section~\ref{Non-homogeneous amplitudes and UV completion} can be calculated using the complex Hankel contour $C_{H}$ consisting of the two straight
lines $C_{\pm }$ infinitesimally above and below the positive real axes
and an infinite radius circle $C_{R}$ with center in the origin (see Fig.~\ref{fig:hank_cont}), i.e.
\begin{eqnarray*}
I_{C_{H}} &=&I_{C_{+}}+I_{C_{-}}+I_{C_{R}} \\
I_{C} &\equiv &\int_{C}\mathrm{d}u~u^{\Delta -1}A_{n}\left( u\right) 
\end{eqnarray*}
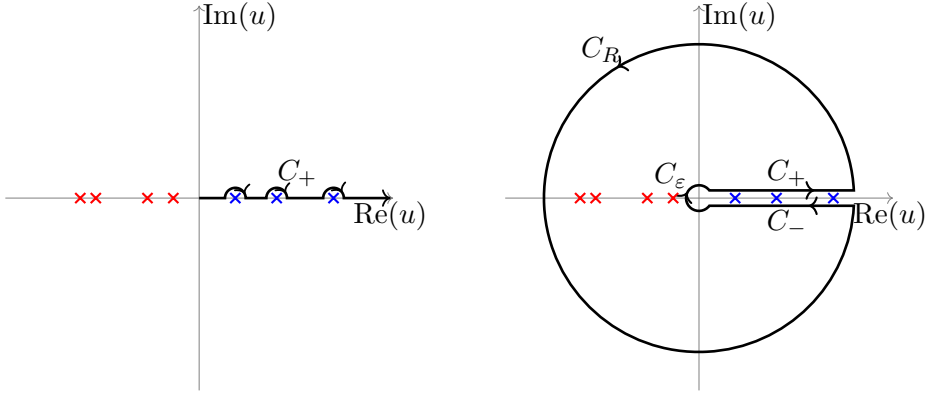
\begin{figure}[ht]
\centering
\begin{tikzpicture}[scale=.68]
\tikzset{cross/.style={cross out, draw=black, minimum size=2*(#1-\pgflinewidth), inner sep=0pt, outer sep=0pt},
%default radius will be 1pt. 
cross/.default={2.7pt}}
% configurable parameters
\def\bigradius{3}
\def\littleradius{0.25}

% Axes
\draw [help lines,->] (-1.25*\bigradius, 0) -- (1.25*\bigradius,0);
\draw [help lines,->] (0, -1.25*\bigradius) -- (0, 1.25*\bigradius);

\draw[line width=1pt,   decoration={ markings,
  mark=at position 0.24 with {\arrow[line width=.7pt]{>}},
  mark=at position 0.47 with {\arrow[line width=.7pt]{>}},
  mark=at position 0.77 with {\arrow[line width=.7pt]{>}},
  mark=at position 1.0 with {\arrow[line width=.7pt]{>}}},
  postaction={decorate}]	(0,0) -- (0.5,0) arc (180:0:.2) -- (1.3,0) arc (180:0:.2) -- (2.4,0) arc (180:0:.2) -- (3.7,0);   
% The labels
	\node at (3.7,-0.35){$\mathrm{Re}(u)$};
	\node at (0.79,3.53) {$\mathrm{Im}(u)$};
\node at (1.9,0.49) {$C_+$};

\draw (.7,0) node[cross,blue,thick] {};
\draw (1.5,0) node[cross,blue,thick] {};
\draw (2.6,0) node[cross,blue,thick] {};
\draw (-.5,0) node[cross,red,thick] {};
\draw (-1,0) node[cross,red,thick] {};
\draw (-2,0) node[cross,red,thick] {};
\draw (-2.3,0) node[cross,red,thick] {};
\end{tikzpicture}
\qquad
\begin{tikzpicture}[scale=.68]
\tikzset{cross/.style={cross out, draw=black, minimum size=2*(#1-\pgflinewidth), inner sep=0pt, outer sep=0pt},
%default radius will be 1pt. 
cross/.default={2.7pt}}
% configurable parameters
\def\gap{0.3}
\def\bigradius{3}
\def\littleradius{0.25}
% Axes
\draw [help lines,->] (-1.25*\bigradius, 0) -- (1.25*\bigradius,0);
\draw [help lines,->] (0, -1.25*\bigradius) -- (0, 1.25*\bigradius);
% Red path
\draw[line width=1pt,   decoration={ markings,
  mark=at position 0.2455 with {\arrow[line width=.8pt]{>}},
  mark=at position 0.765 with {\arrow[line width=.8pt]{>}},
  mark=at position 0.87 with {\arrow[line width=.8pt]{>}},
  mark=at position 0.97 with {\arrow[line width=.8pt]{>}}},
  postaction={decorate}]
  let
     \n1 = {asin(\gap/2/\bigradius)},
     \n2 = {asin(\gap/2/\littleradius)}
  in (\n1:\bigradius) arc (\n1:360-\n1:\bigradius)
  -- (-\n2:\littleradius) arc (-\n2:-360+\n2:\littleradius)
  -- cycle;

% The labels
	\node at (3.7,-0.35){$\mathrm{Re}(u)$};
	\node at (0.79,3.53) {$\mathrm{Im}(u)$};
\node at (-0.55,0.44) {$C_{\varepsilon}$};
\node at (-1.9,2.9) {$C_{R}$};
\node at (1.7,0.49) {$C_+$};
\node at (1.7,-0.49) {$C_-$};

\draw (.7,0) node[cross,blue,thick] {};
\draw (1.5,0) node[cross,blue,thick] {};
\draw (2.6,0) node[cross,blue,thick] {};
\draw (-.5,0) node[cross,red,thick] {};
\draw (-1,0) node[cross,red,thick] {};
\draw (-2,0) node[cross,red,thick] {};
\draw (-2.3,0) node[cross,red,thick] {};
\end{tikzpicture}
\caption{Contours relevant for computing the Mellin transform in $u$.(\emph{Left}): The physical amplitude is defined by approaching the real axis from above, which results in a prescription that poles are avoided on upper semi-circles. (\emph{Right}): By contour deformation, the Mellin integral can be evaluated by employing the Hankel contour. Poles with both $u_{\mathcal{F}}>0$ (blue) and $u_{\mathcal{F}}<0$ (red) contribute. The value of the integral on $C_+$ and $C_-$ differs by a phase, while $C_R$ and $C_{\varepsilon}$ can be dropped.} \label{fig:hank_cont}
\end{figure}

For tree-level amplitudes and for $\Delta $ from the above region, the only singularities of $u^{\Delta -1}A_n\left( u\right)$ are the branching point for $u=0$ and simple poles $u_{\mathcal{F}}$ corresponding to the massive factorization channels $\mathcal{F}$ (see Fig.~\ref{fig:fact_channel}) which are determined by the
conditions 
\begin{equation}
uQ_{\mathcal{F}}^{2}-M_{\mathcal{F}}^{2}=0,
\end{equation}%
where%
\begin{equation}
Q_{\mathcal{F}}=\sum\limits_{i\in \mathcal{F}}\varepsilon _{i}\sigma
_{i}q\left( z_{i},\overline{z}_{i}\right) .
\end{equation}%
This leads to the location of the simple pole corresponding to a given factorization channel $\mathcal{F}$
\begin{equation}
u_{\mathcal{F}}=\frac{M_{\mathcal{F}}^{2}}{Q_{\mathcal{F}}^{2}}=\frac{M_{%
\mathcal{F}}^{2}}{\sum\limits_{\left( i<j\right) \in \mathcal{F}}\varepsilon
_{i}\varepsilon _{j}\sigma _{i}\sigma _{j}\left\vert z_{ij}\right\vert ^{2}}.
\end{equation}%

Due to the asymptotics (\ref{asymptotics}), the integral $I_{R}$ around the infinite circle vanishes, while the integrals on the half-lines above and below the positive real axis give (second terms on r.h.s. come from infinitesimal semicircles around blue poles in Fig.~\ref{fig:hank_cont}, while the overall phase in $I_{C_-}$ from looping around the $u=0$ branch point on $C_\varepsilon$)
\begin{eqnarray}
I_{C_{+}} &=&I_{VP}-\mathrm{i}\pi \sum\limits_{\mathcal{F}}\theta \left( u_{%
\mathcal{F}}\right) u_{\mathcal{F}}^{\Delta -1}\mathrm{res}\left(
A_{n}\left( u\right) ,u_{\mathcal{F}}\right) , \\
I_{C_{-}} &=&-\mathrm{e}^{2\pi \mathrm{i}\Delta }\left( I_{VP}+\mathrm{i}\pi
\sum\limits_{\mathcal{F}}\theta \left( u_{\mathcal{F}}\right) u_{\mathcal{F}%
}^{\Delta -1}\mathrm{res}\left( A_{n}\left( u\right) ,u_{\mathcal{F}}\right)
\right) ,
\end{eqnarray}%
respectively, where%
\begin{equation}
I_{VP}=VP\int_{0}^{\infty }\mathrm{d}u~u^{\Delta -1}A_{n}\left( u\right) 
\end{equation}%
is the corresponding principal value. On the other hand, according to the
residue theorem, the integral over the Hankel contour is given by the sum of
residues inside $C_{H}$ (red poles in Fig.~\ref{fig:hank_cont})
\begin{equation}
\int_{C_{H}}\mathrm{d}u~u^{\Delta -1}A_{n}\left( u\right) =2\pi \mathrm{i}%
\sum\limits_{\mathcal{F}}\theta \left( -u_{\mathcal{F}}\right) u_{\mathcal{F}%
}^{\Delta -1}\mathrm{res}\left( A_{n}\left( u\right) ,u_{\mathcal{F}}\right)
.
\end{equation}%
Supposing the usual Feynman $\mathrm{i}\varepsilon $ prescription for the propagators (i.e. the $u$ Mellin integral is defined by the contour in the left panel of Fig.~\ref{fig:hank_cont}), we get\footnote{This prescription might differ from those used by other authors to compute celestial amplitudes. Recently, it was observed in~\cite{Sleight:2023ojm} that the original definition of the conformal primary wavefunction is equivalent to an integral transform of a bulk Wightman function, which does not take properly into account causality. It was suggested to replace the Wightman function with a Feynman propagator. It would be interesting to further investigate if their proposed definition of celestial amplitudes agrees with ours.} 
\begin{eqnarray}
\int_{0}^{\infty }\mathrm{d}u~u^{\Delta -1}A_{n}\left( u\right) 
&=&I_{C_{+}}=-\frac{\pi }{\sin \pi \Delta }\mathrm{e}^{\pi \mathrm{i}\Delta
}\sum\limits_{\mathcal{F}}\theta \left( u_{\mathcal{F}}\right) u_{\mathcal{F}%
}^{\Delta -1}\mathrm{res}\left( A_{n}\left( u\right) ,u_{\mathcal{F}}\right) 
\notag \\
&&-\frac{\pi }{\sin \pi \Delta }\mathrm{e}^{-\pi \mathrm{i}\Delta
}\sum\limits_{\mathcal{F}}\theta \left( -u_{\mathcal{F}}\right) u_{\mathcal{F%
}}^{\Delta -1}\mathrm{res}\left( A_{n}\left( u\right) ,u_{\mathcal{F}%
}\right) .  \label{integral_mellin}
\end{eqnarray}
which can be combined into
\begin{equation}
\int_{0}^{\infty }\mathrm{d}u~u^{\Delta -1}A_{n}\left( u\right) 
= -\frac{\pi }{\sin(\pi\Delta)} \sum\limits_{\mathcal{F}}\mathrm{e}^{\pi \mathrm{i}\Delta \cdot \mathrm{
sign}\left( u_{\mathcal{F}}\right) }u_{\mathcal{F}}^{\Delta -1}\mathrm{res}
\left( A_{n}\left( u\right) ,u_{\mathcal{F}}\right).
\end{equation}
%
%===========================
\section{Explicit formulas for $\protect\sigma_{\ast i}$}\label{appendix_sigmas}}
%================================

Here we summarize explicit solutions~\eqref{sigma_i*_solution} of delta function constraints~\eqref{sigma_equations} (four from momentum conservation and one from scaling) needed to construct lower point (4, 5, 6-pt) celestial amplitudes.
\newcommand{\nocontentsline}[3]{}
\newcommand{\tocless}[2]{\bgroup\let\addcontentsline=\nocontentsline#1{#2}\egroup}
%=============================
\tocless\subsection{The 5pt amplitude}\label{5pt_sigmas}
%=============================

We assume the process $1+2\rightarrow 3+4+5$, i.e. $\varepsilon
_{1}=\varepsilon _{2}=-\varepsilon _{3}=-\varepsilon _{4}=-\varepsilon
_{5}=-1$. In the 5pt reference point $\left( z_{1},\ldots ,z_{5}\right)
\rightarrow \left( 1,0,\varepsilon ^{-1},t_{4}^{-1},t_{5}^{-1}\right) $ we
get
\begin{eqnarray}
\det \mathbf{A}_{5} &=&\frac{\mathrm{i}}{2}\left( \frac{1}{t_{4}}-\frac{1}{%
\overline{t_{4}}}-\frac{1}{t_{5}}+\frac{1}{\overline{t_{5}}}\right)
\varepsilon ^{-2}+O\left( \varepsilon ^{-1}\right) \notag \\
&\equiv &\frac{\mathrm{i}}{2}A\varepsilon ^{-2}+O\left( \frac{1}{\varepsilon 
}\right) 
\label{detA5}
\end{eqnarray}
and
\begin{eqnarray*}
\sigma _{\ast 1} &=&\frac{1}{2A}\left( \frac{1}{t_{4}\overline{t_{5}}}-\frac{%
1}{\overline{t_{4}}t_{5}}\right) +O\left( \varepsilon \right)  
\overset{\overline{t_{4}}\rightarrow t_{4}}{\rightarrow }\frac{1}{2t_{4}}%
+O\left( \varepsilon \right)  \\
\sigma _{\ast 2} &=&\frac{1}{2}\left[ 1-\frac{1}{A}\left( \frac{1}{t_{4}%
\overline{t_{5}}}-\frac{1}{\overline{t_{4}}t_{5}}\right) \right] +O\left(
\varepsilon \right)  
\overset{\overline{t_{4}}\rightarrow t_{4}}{\rightarrow }\frac{1}{2}\left(
1-\frac{1}{t_{4}}\right) +O\left( \varepsilon \right)  \\
\sigma _{\ast 3} &=&\frac{\varepsilon ^{2}}{2A}\left[ \frac{1}{\left\vert
t_{4}\right\vert ^{2}}\left( \frac{1}{t_{5}}-\frac{1}{\overline{t_{5}}}%
\right) -\frac{1}{\left\vert t_{5}\right\vert ^{2}}\left( \frac{1}{t_{4}}-%
\frac{1}{\overline{t_{4}}}\right) +\left( \frac{1}{t_{4}\overline{t_{5}}}-%
\frac{1}{\overline{t_{4}}t_{5}}\right) \right] +O\left( \varepsilon
^{3}\right)  \\
&&\overset{\overline{t_{4}}\rightarrow t_{4}}{\rightarrow }\frac{\varepsilon
^{2}}{2}\frac{1}{t_{4}}\left( 1-\frac{1}{t_{4}}\right) +O\left( \varepsilon
^{3}\right)  \\
\sigma _{\ast 4} &=&\frac{1}{2A\left\vert t_{5}\right\vert ^{2}}\left( t_{5}-%
\overline{t_{5}}\right) +O\left( \varepsilon \right) \overset{\overline{t_{4}%
}\rightarrow t_{4}}{\rightarrow }\frac{1}{2}+O\left( \varepsilon \right)  \\
\sigma _{\ast 5} &=&-\frac{1}{2A\left\vert t_{4}\right\vert ^{2}}\left(
t_{4}-\overline{t_{4}}\right) +O\left( \varepsilon \right) \overset{%
\overline{t_{4}}\rightarrow t_{4}}{\rightarrow }O\left( \varepsilon \right) \,.
\end{eqnarray*}
Note that since $A^*=-A$, the solutions $\sigma_{*i}$ are real.

%=============================
\tocless\subsection{The 4pt amplitude \label{4pt_sigmas}}
%=============================

Here we assume the process $1+2\rightarrow 3+4$, i.e. $\varepsilon
_{1}=\varepsilon _{2}=-\varepsilon _{3}=-\varepsilon _{4}=-1$. We get on the
support of the $\delta -$function $t_{4}=\overline{t_{4}}$%
\begin{eqnarray*}
\sigma _{\ast 1} &=&\frac{1}{2t_{4}}+O\left( \varepsilon \right)  \\
\sigma _{\ast 2} &=&\frac{1}{2}\left( 1-\frac{1}{t_{4}}\right) +O\left(
\varepsilon \right)  \\
\sigma _{\ast 3} &=&\frac{\varepsilon ^{2}}{2}\frac{1}{t_{4}}\left( 1-\frac{1%
}{t_{4}}\right) +O\left( \varepsilon ^{3}\right)  \\
\sigma _{\ast 4} &=&\frac{1}{2}+O\left( \varepsilon \right) \,,
\end{eqnarray*}%
which coincides with the $\overline{t_{4}}\rightarrow t_{4}$ limit of the
5pt case.

%=============================
\tocless\subsection{The 6pt amplitude \label{sigma_6pt}}
%=============================

Here we assume the process $1+2+3\to 4+5+6$. In the 6pt reference point $\left( z_{1},\ldots ,z_{6}\right)
\rightarrow \left( 1,0,\varepsilon ^{-1},t_{4}^{-1},t_{5}^{-1},t_6^{-1}\right) $ we have the general formula%
\begin{equation*}
\sigma _{\ast i}=\sigma _{\ast i}^{\left( 5\right) }+K_{i}\sigma _{6}\,,
\end{equation*}
where $\sigma _{\ast i}^{\left( 5\right) }$ are the solutions for 5pt
kinematics listed above and 
\begin{eqnarray*}
K_{1} &=&\frac{1}{A}\left( \frac{1}{t_{5}\overline{t_{4}}}-\frac{1}{t_{4}%
\overline{t_{5}}}+\frac{1}{t_{4}\overline{t_{6}}}-\frac{1}{t_{6}\overline{%
t_{4}}}+\frac{1}{t_{5}\overline{t_{6}}}-\frac{1}{t_{6}\overline{t_{5}}}%
\right) +O\left( \varepsilon \right)  \\
K_{2} &=&\frac{1}{A}\left( -\frac{1}{t_{5}\overline{t_{4}}}+\frac{1}{t_{4}%
\overline{t_{5}}}-\frac{1}{t_{4}\overline{t_{6}}}+\frac{1}{t_{6}\overline{%
t_{4}}}+\frac{1}{t_{5}\overline{t_{6}}}-\frac{1}{t_{6}\overline{t_{5}}}%
\right) +O\left( \varepsilon \right)  \\
K_{3} &=&\frac{\varepsilon ^{2}}{A}\left[ \left( \frac{1}{t_{5}\overline{%
t_{4}}}-\frac{1}{t_{4}\overline{t_{5}}}+\frac{1}{t_{4}\overline{t_{6}}}-%
\frac{1}{t_{6}\overline{t_{4}}}-\frac{1}{t_{5}\overline{t_{6}}}+\frac{1}{%
t_{6}\overline{t_{5}}}\right) \right.  \\
&&\left. +\frac{1}{\left\vert t_{4}\right\vert ^{2}}\left( -\frac{1}{t_{5}}+%
\frac{1}{\overline{t_{5}}}-\frac{1}{t_{6}}+\frac{1}{\overline{t_{6}}}\right)
\right.  \\
&&\left. +\frac{1}{\left\vert t_{5}\right\vert ^{2}}\left( \frac{1}{t_{4}}-%
\frac{1}{\overline{t_{4}}}-\frac{1}{t_{6}}+\frac{1}{\overline{t_{6}}}\right)
\right.  \\
&&\left. +\frac{1}{\left\vert t_{6}\right\vert ^{2}}\left( \frac{1}{t_{4}}-%
\frac{1}{\overline{t_{5}}}+\frac{1}{t_{5}}-\frac{1}{\overline{t_{5}}}\right) %
\right] +O\left( \varepsilon ^{3}\right)  \\
K_{4} &=&\frac{1}{A}\left( \frac{1}{t_{5}}-\frac{1}{\overline{t_{5}}}-\frac{1%
}{t_{6}}+\frac{1}{\overline{t_{6}}}\right) +O\left( \varepsilon \right)  \\
K_{5} &=&\frac{1}{A}\left( \frac{1}{t_{6}}-\frac{1}{\overline{t_{6}}}-\frac{1%
}{t_{4}}+\frac{1}{\overline{t_{4}}}\right) +O\left( \varepsilon \right) 
\end{eqnarray*}%
with $A$ given by equation (\ref{detA5}).

%%%%%%%%%%%%%%%%%%%%%%%%%%%%%%%%%%%%%%%%%%%%%%%%%%%%%%%%%%%%%%%
\section{Amplitudes and soft theorems for the $U(1)$ fibered model}\label{app:u1_fibred}
%%%%%%%%%%%%%%%%%%%%%%%%%%%%%%%%%%%%%%%%%%%%%%%%%%%%%%%%%%%%%%%

Let us discuss in details 5pt amplitudes in the model from section~\ref{sec:fibred_model}. We consider the amplitude $A_{5}\left(
1,2^{-},3^{-},4^{+},5^{+}\right)$ in the UV completion of the $%
U(1) $ fibered $\mathbb{C}P^{1}$ sigma model. Here we use condensed notation identifying $p_{i}\equiv i
$ and, as above, the superscript denotes the charge. According to (\ref{CPN_soft_theorem}), we get
\begin{eqnarray}
\lim_{p_{5}\rightarrow 0}A_{5}\left( 1,2^{-},3^{-},4^{+},5^{+}\right) =-%
\mathrm{i}\frac{\sin \vartheta }{\sqrt{2}v_{1}}&&\left[ A_{4}\left(
1^{+},2^{-},3^{-},4^{+}\right)\right.\notag\\
&&\left.-A_{4}\left( 1,2,3^{-},4^{+}\right)
-A_{4}\left( 1,2^{-},3,4^{+}\right) \right]   
\end{eqnarray}
and observe that the leading order term of the momentum space soft theorem on the right hand side is constant in the energy of the soft (fifth) particle, thus the limit $\Delta_5\to0$ probes the leading term of the conformally soft theorem.
We, therefore, expect the conformal soft theorem in the form
\begin{eqnarray}
\lim_{t_{4}\rightarrow \overline{t_{4}}}\mathcal{F}_{5}^{\left( 0--++\right)
}\left( \Delta ,t_{4},t_{5}\right) =-\mathrm{i}\frac{\sin \vartheta }{\sqrt{2%
}v_{1}}&&\left[ \mathcal{F}_{4}^{\left( +--+\right) }\left( \Delta
,t_{4}\right)\right.\notag\\ &&\left.-\mathcal{F}_{4}^{\left( 00-+\right) }\left( \Delta
,t_{4}\right) -\mathcal{F}_{4}^{\left( 0-0+\right) }\left( \Delta
,t_{4}\right) \right] ,  
\end{eqnarray}
which is supposed to hold in the intersection of the fundamental strips.
This can be confirmed explicitly. The tree-level momentum space 5pt
amplitude reads
\begin{eqnarray}
A_{5}\left( 1,2^{-},3^{-},4^{+},5^{+}\right)  &=&\mathrm{i}\cos \vartheta
\sin 2\vartheta \frac{M^{2}}{4\sqrt{2}v_{1}^{2}}\left\{ -\frac{1}{v_{1}}%
\left[ \frac{s_{14}-s_{12}}{s_{24}-M^{2}}+\frac{s_{14}-s_{13}}{s_{34}-M^{2}}%
\right. \right.   \notag \\
&&\left. \left. +\frac{s_{15}-s_{12}}{s_{25}-M^{2}}+\frac{s_{15}-s_{13}}{%
s_{35}-M^{2}}\right] \right.   \notag \\
&&\left. +\cos \theta \left( \frac{\cos \theta }{v_{1}}+\frac{\sin \theta }{%
v_{2}}\right) \left[ \frac{s_{14}-s_{12}}{s_{24}-M^{2}}\frac{s_{35}}{%
s_{35}-m_{1}^{2}}+\frac{s_{14}-s_{13}}{s_{34}-M^{2}}\frac{s_{25}}{%
s_{25}-m_{1}^{2}}\right. \right.   \notag \\
&&\left. \left. +\frac{s_{15}-s_{12}}{s_{25}-M^{2}}\frac{s_{34}}{%
s_{34}-m_{1}^{2}}+\frac{s_{15}-s_{13}}{s_{35}-M^{2}}\frac{s_{24}}{%
s_{24}-m_{1}^{2}}\right] \right.   \notag \\
&&\left. +\sin \theta \left( \frac{\sin \theta }{v_{1}}-\frac{\cos \theta }{%
v_{2}}\right) \left[ \frac{s_{14}-s_{12}}{s_{24}-M^{2}}\frac{s_{35}}{%
s_{35}-m_{2}^{2}}+\frac{s_{14}-s_{13}}{s_{34}-M^{2}}\frac{s_{25}}{%
s_{25}-m_{2}^{2}}\right. \right.   \notag \\
&&\left. \left. +\frac{s_{15}-s_{12}}{s_{25}-M^{2}}\frac{s_{34}}{%
s_{34}-m_{2}^{2}}+\frac{s_{15}-s_{13}}{s_{35}-M^{2}}\frac{s_{24}}{%
s_{24}-m_{2}^{2}}\right] \right\} \label{A_5_full},
\end{eqnarray}%
where, as usual, $s_{ij}=\left( p_{i}+p_{j}\right) ^{2}$ and all momenta are
treated as outgoing. Clearly, the soft limit for the neutral Goldstone boson $p_{1}\rightarrow 0$ means $%
s_{1j}\rightarrow 0$ and we get the Adler zero according to (\ref{CPN_Adler_zero}). On the other hand, the soft limit for the charged one,  $p_{5}\rightarrow 0$ reads%
\begin{eqnarray}
A_{5}\left( 1,2^{-},3^{-},4^{+},5^{+}\right)  &=&\mathrm{i}\cos \vartheta
\sin 2\vartheta \frac{M^{2}}{4\sqrt{2}v_{1}^{2}}\left\{ -\frac{1}{v_{1}}%
\left[ \frac{s_{14}-s_{12}}{s_{24}-M^{2}}+\frac{s_{14}-s_{13}}{s_{34}-M^{2}}%
 +\frac{s_{12}}{M^{2}}+\frac{s_{13}}{M^{2}}\right] \right.  
\notag \\
&&\left. +\cos \theta \left( \frac{\cos \theta }{v_{1}}+\frac{\sin \theta }{%
v_{2}}\right) \left[ \frac{s_{12}}{M^{2}}\frac{s_{34}}{s_{34}-m_{1}^{2}}+%
\frac{s_{13}}{M^{2}}\frac{s_{24}}{s_{24}-m_{1}^{2}}\right] \right.   \notag
\\
&&\left. +\sin \theta \left( \frac{\sin \theta }{v_{1}}-\frac{\cos \theta }{%
v_{2}}\right) \left[ \frac{s_{12}}{M^{2}}\frac{s_{34}}{s_{34}-m_{2}^{2}}+%
\frac{s_{13}}{M^{2}}\frac{s_{24}}{s_{24}-m_{2}^{2}}\right] \right\} .\notag\\
\label{A_5_CPN}
\end{eqnarray}
In order to check the soft theorem in this case, we need also 4pt amplitudes%
\begin{eqnarray}
A_{4}\left( 1^{+},2^{-},3^{-},4^{+}\right)  &=&\frac{1}{2v_{1}^{2}}\left\{
s_{12}+s_{24}-\cos ^{2}\theta \left[ \frac{s_{12}^2}{s_{12}-m_{1}^{2}}+\frac{%
s_{24}^2}{s_{24}-m_{1}^{2}}\right] \right.   \notag \\
&&\left. -\sin ^{2}\theta \left[ \frac{s_{12}^2}{s_{12}-m_{2}^{2}}+\frac{s_{24}^2%
}{s_{24}-m_{2}^{2}}\right] \right.   \notag \\
&&\left. +M^{2}\cos ^{2}\vartheta \left[ \frac{s_{14}-s_{24}}{s_{12}-M^{2}}+%
\frac{s_{14}-s_{12}}{s_{24}-M^{2}}\right] \right\} 
\label{A_4_1}
\end{eqnarray}%
and%
\begin{eqnarray}
A_{4}\left( 1,2,3^{-},4^{+}\right)  &=&\sin ^{2}\vartheta \frac{1}{2v_{1}^{2}%
}s_{12}-\frac{1}{2v_{1}}\left[ \cos \theta \left( \frac{\cos \theta \sin
^{2}\vartheta }{v_{1}}-\frac{\sin \theta \cos ^{2}\vartheta }{v_{2}}\right) 
\frac{s_{12}^2}{s_{12}-m_{1}^{2}}\right.   \notag \\
&&\left. +\sin \theta \left( \frac{\sin \theta \sin ^{2}\vartheta }{v_{1}}+%
\frac{\cos \theta \cos ^{2}\vartheta }{v_{2}}\right) \frac{s_{12}^2}{%
s_{12}-m_{2}^{2}}\right]. 
\label{A_4_2}
\end{eqnarray}
It is then straightforward to verify explicitly the validity of the (momentum) soft theorem (\ref{soft_theorem_CPN}) within the UV completed linear model. 

Or we can verify explicitly the validity of the (momentum) soft theorem directly in the effective theory, i.e. within the $U(1)$ fibered $\mathbb{C}P^{1}$ nonlinear sigma model. Expanding the above amplitudes in $s_{ij}$ and keeping only leading order terms, we can easily obtain the
effective amplitudes (cf. (\ref{F_F_0_identification}) and ref. \cite{Kampf:2019mcd})
\begin{eqnarray}
A_{5}^{eff}\left( 1,2^{-},3^{-},4^{+},5^{+}\right)  &=&\mathrm{i}\frac{\cos
\vartheta \sin 2\vartheta }{2\sqrt{2}v_{1}^{3}}\left(
s_{14}+s_{15}-s_{12}-s_{13}\right)   \notag \\
&=&\mathrm{i}\frac{F_{0}}{F^{6}}\left( F^{2}-\frac{1}{2}F_{0}^{2}\right)
\left( s_{14}+s_{15}\right)   \notag \\
A_{4}^{eff}\left( 1^{+},2^{-},3^{-},4^{+}\right)  &=&\frac{1}{2v_{1}^{2}}%
\left[ s_{12}+s_{24}-\cos ^{2}\vartheta \left(
s_{14}-s_{24}+s_{14}-s_{12}\right) \right]   \notag \\
&=&\frac{1}{4F^{4}}\left( 3F_{0}^{2}-8F^{2}\right) s_{14}  \notag \\
A_{4}^{eff}\left( 1,2,3^{-},4^{+}\right)  &=&\frac{\sin ^{2}\vartheta }{%
2v_{1}^{2}}s_{12}=\frac{F_{0}^{2}}{4F^{4}}s_{12}.
\label{infrared_expansion_CPN}
\end{eqnarray}%
By inspection we immediately conclude that the (momentum) soft theorem (\ref{soft_theorem_CPN}) holds manifestly within the effective theory.

In order to check the conformal soft theorem in the general form (\ref{conformal_soft_theorem}),
we have to calculate explicitly the function $\mathcal{F}_{5}^{\left(
0--++\right) }\left( \Delta ,t_{4},t_{5}\right) $.

Let us first check the validity of the conformal soft theorem (\ref{conformal_soft_theorem}) for the
effective theory amplitudes. In this case, the fundamental strips are empty
and we have to treat the Mellin transform in the sense of distributions.
Explicitly we get%
\begin{eqnarray}
\mathcal{F}_{5}^{\left( 0--++\right) }\left( \Delta ,t_{4},t_{5}\right)
^{eff} &=&2\pi \delta \left( \Delta +1\right) \mathrm{i}\frac{F_{0}}{F^{6}}%
\left( F^{2}-\frac{1}{2}F_{0}^{2}\right) \left( \Sigma _{14}^{\left(
5\right) }+\Sigma _{15}^{\left( 5\right) }\right)   \notag \\
\mathcal{F}_{4}^{\left( +--+\right) }\left( \Delta ,t_{4}\right) ^{eff}
&=&2\pi \delta \left( \Delta +1\right) \frac{1}{4F^{4}}\left(
3F_{0}^{2}-8F^{2}\right) \Sigma _{14}^{(4) }  \notag \\
\mathcal{F}_{4}^{\left( 00-+\right) }\left( \Delta ,t_{4}\right) ^{eff}
&=&2\pi \delta \left( \Delta +1\right) \frac{F_{0}^{2}}{4F^{4}}\Sigma
_{12}^{(4) }  \notag \\
\mathcal{F}_{4}^{\left( 0-0+\right) }\left( \Delta ,t_{4}\right) ^{eff}
&=&2\pi \delta \left( \Delta +1\right) \frac{F_{0}^{2}}{4F^{4}}\Sigma
_{13}^{(4) }.
\end{eqnarray}
Here we denoted
\begin{equation}
\Sigma _{ij}^{\left( 5,4\right) }=2\varepsilon _{i}\varepsilon _{j}\sigma
_{\ast i}\sigma _{\ast j}\left( q_{i}\cdot q_{j}\right) =\varepsilon
_{i}\varepsilon _{j}\sigma _{\ast i}\sigma _{\ast j}\left\vert
z_{ij}\right\vert ^{2},
\end{equation}
where on the right-hand side we insert the reference points $\left(
z_{1},\ldots ,z_{5}\right) \rightarrow \left( 1,0,\varepsilon
^{-1},t_{4}^{-1},t_{5}^{-1}\right) $ for $\Sigma _{ij}^{\left( 5\right) }$
and $\left( z_{1},\ldots ,z_{4}\right) \rightarrow \left( 1,0,\varepsilon
^{-1},t_{4}^{-1}\right) $ for $\Sigma _{ij}^{(4) }$ and the limit $\epsilon\to 0$ is tacitly assumed. Since (cf.
Appendix \ref{appendix_sigmas} ) 
\begin{equation}
\Sigma _{15}^{\left( 5\right) }=O\left( t_{4}-\overline{t_{4}}\right) 
\end{equation}%
and%
\begin{equation}
\lim_{\overline{t_{4}}\rightarrow t_{4}}\Sigma _{14}^{\left( 5\right)
}=\Sigma _{14}^{(4) }
\end{equation}%
and since 
\begin{equation}
\Sigma _{12}^{(4) }+\Sigma _{13}^{(4) }=-\Sigma
_{14}^{(4) }
\end{equation}%
due to the momentum conservation $\delta -$function, the conformal soft
theorem for the 5pt effective amplitude is manifest algebraically in the same
way as the momentum space one.

Let us now check the validity of the conformal soft theorem (\ref{conformal_soft_theorem}) within the original UV completed theory.
The behavior of the amplitudes (\ref{A_5_CPN}), (\ref{A_4_1}) and (\ref{A_4_2}) for $s_{ij}\rightarrow \infty $ is $O\left( s_{ij}^{-1}\right)$, $O\left( s_{ij}^{0}\right)$ and $O\left( s_{ij}^{0}\right)$ respectively
as expected from their softer UV behavior in the linear theory.
These asymptotics, together with the leading order behavior (\ref{infrared_expansion_CPN}) in the infrared, determine  the fundamental strips of $
\Delta $ for  $\mathcal{F}_{5}^{\left( 0--++\right) }\left( \Delta
,t_{4},t_{5}\right) $ and for $\mathcal{F}_{4}^{\left( +--+\right) }\left(
\Delta ,t_{4}\right) $ (and $\mathcal{F}_{4}^{\left( 00-+\right) }\left(
\Delta ,t_{4}\right) $) as $\langle -1,1\rangle $ and  $\langle -1,0\rangle $
respectively.
Therefore the intersection of the fundamental strips for the left-hand side and the right-hand side of the conformal soft theorem (\ref{conformal_soft_theorem}) is nonempty.
Though the Mellin transform (\ref{F_5}) of the amplitude (\ref{A_5_full}) is less trivial, nevertheless it
can be easily obtained using the general formula (\ref{residue_formula}). As a result, we get
\begin{eqnarray}
\mathcal{F}_{5}^{\left( 0--++\right) }\left( \Delta ,t_{4},t_{5}\right)  &=&%
\mathrm{i}\cos \vartheta \sin 2\vartheta \frac{M^{2}}{4\sqrt{2}v_{1}^{2}}%
\frac{\pi }{\sin \left( \pi \Delta \right) }  \notag \\
&&\times \left\{ \sum\limits_{\left( i,j\right) \in I}\frac{\Sigma
_{1i}^{\left( 5\right) }-\Sigma _{1j}^{\left( 5\right) }}{\Sigma
_{ij}^{\left( 5\right) }}\left( -\frac{M^{2}}{\Sigma _{ij}^{\left( 5\right) }%
}\right) ^{\Delta }\right.   \notag \\
&&\times \left. \left[ -\frac{1}{v_{1}}+\cos \theta \left( \frac{\cos \theta 
}{v_{1}}+\frac{\sin \theta }{v_{2}}\right) \left( 1-\frac{\Sigma
_{ij}^{\left( 5\right) }}{\Sigma _{\left( ij\right) ^{c}}^{\left( 5\right) }}%
\frac{m_{1}^{2}}{M^{2}}\right) ^{-1}\right. \right.   \notag \\
&&\left. \left. +\sin \theta \left( \frac{\sin \theta }{v_{1}}-\frac{\cos
\theta }{v_{2}}\right) \left( 1-\frac{\Sigma _{ij}^{\left( 5\right) }}{%
\Sigma _{\left( ij\right) ^{c}}^{\left( 5\right) }}\frac{m_{2}^{2}}{M^{2}}%
\right) ^{-1}\right] \right.   \notag \\
&&\left. -\sum\limits_{\left( i,j\right) \in I}\frac{\Sigma _{1i}^{\left(
5\right) }-\Sigma _{1j}^{\left( 5\right) }}{\Sigma _{ij}^{\left( 5\right) }}%
\right.   \notag \\
&&\left. \times \left[ \cos \theta \left( \frac{\cos \theta }{v_{1}}+\frac{%
\sin \theta }{v_{2}}\right) \left( -\frac{m_{1}^{2}}{\Sigma _{\left(
ij\right) ^{c}}^{\left( 5\right) }}\right) ^{\Delta }\left( 1-\frac{\Sigma
_{\left( ij\right) ^{c}}^{\left( 5\right) }}{\Sigma _{ij}^{\left( 5\right) }}%
\frac{M^{2}}{m_{1}^{2}}\right) ^{-1}\right. \right.   \notag \\
&&\left. \left. +\sin \theta \left( \frac{\sin \theta }{v_{1}}-\frac{\cos
\theta }{v_{2}}\right) \left( -\frac{m_{2}^{2}}{\Sigma _{\left( ij\right)
^{c}}^{\left( 5\right) }}\right) ^{\Delta }\left( 1-\frac{\Sigma _{\left(
ij\right) ^{c}}^{\left( 5\right) }}{\Sigma _{ij}^{\left( 5\right) }}\frac{%
M^{2}}{m_{2}^{2}}\right) ^{-1}\right] \right\}   \label{F_5_full},\notag\\
\end{eqnarray}%
where we denoted $I=\left\{ (24),(34),(25),(35)\right\} $ and the pair of
indices $\left( ij\right) ^{c}$ means the complementary pair with respect to
the set $\left\{ 2,3,4,5\right\} $. We have also used identities like%
\begin{equation}
\exp \left( i\pi \Delta \mathrm{sign}\Sigma _{ij}^{\left( 5\right) }\right)
\left( \frac{M^{2}}{\Sigma _{ij}^{\left( 5\right) }}\right) ^{\Delta
}=\left( -\frac{M^{2}}{\Sigma _{ij}^{\left( 5\right) }}\right) ^{\Delta }.
\end{equation}
Similarly, we get for the Mellin transform of the 4pt amplitudes
\begin{eqnarray}
\mathcal{F}_{4}^{\left( +--+\right) }\left( \Delta ,t_{4}\right)  &=&\frac{1%
}{2v_{1}^{2}}\frac{\pi }{\sin \left( \pi \Delta \right) }\left\{
m_{1}^{2}\cos ^{2}\theta \left[ \left( -\frac{m_{1}^{2}}{\Sigma
_{12}^{(4) }}\right) ^{\Delta }+\left( -\frac{m_{1}^{2}}{\Sigma
_{24}^{(4) }}\right) ^{\Delta }\right] \right.   \notag \\
&&\left. m_{2}^{2}\sin ^{2}\theta \left[ \left( -\frac{m_{2}^{2}}{\Sigma
_{12}^{(4) }}\right) ^{\Delta }+\left( -\frac{m_{2}^{2}}{\Sigma
_{24}^{(4) }}\right) ^{\Delta }\right] \right.   \notag \\
&&\left. -M^{2}\cos ^{2}\vartheta \left[ \frac{\Sigma _{14}^{(4)
}-\Sigma _{24}^{(4) }}{\Sigma _{12}^{(4) }}\left( -%
\frac{M^{2}}{\Sigma _{12}^{(4) }}\right) ^{\Delta }+\frac{\Sigma
_{14}^{(4) }-\Sigma _{12}^{(4) }}{\Sigma
_{24}^{(4) }}\left( -\frac{M^{2}}{\Sigma _{24}^{(4) }%
}\right) ^{\Delta }\right] \right\},\notag\\   \label{F4+--+}
\end{eqnarray}
and
\begin{eqnarray}
\mathcal{F}_{4}^{\left( 00-+\right) }\left( \Delta ,t_{4}\right)  &=&\frac{1%
}{2v_{1}}\frac{\pi }{\sin \left( \pi \Delta \right) }\left[ m_{1}^{2}\cos
\theta \left( \frac{\cos \theta \sin ^{2}\vartheta }{v_{1}}-\frac{\sin
\theta \cos ^{2}\vartheta }{v_{2}}\right) \left( -\frac{m_{1}^{2}}{\Sigma
_{12}^{(4) }}\right) ^{\Delta }\right.   \notag \\
&&\left. +m_{2}^{2}\sin \theta \left( \frac{\sin \theta \sin ^{2}\vartheta }{%
v_{1}}+\frac{\cos \theta \cos ^{2}\vartheta }{v_{2}}\right) \left( -\frac{%
m_{2}^{2}}{\Sigma _{12}^{(4) }}\right) ^{\Delta }\right], 
\label{F_400-+}
\end{eqnarray}
while by crossing
\begin{eqnarray}
\mathcal{F}_{4}^{\left( 0-0+\right) }\left( \Delta ,t_{4}\right)  &=&\frac{1%
}{2v_{1}}\frac{\pi }{\sin \left( \pi \Delta \right) }\left[ m_{1}^{2}\cos
\theta \left( \frac{\cos \theta \sin ^{2}\vartheta }{v_{1}}-\frac{\sin
\theta \cos ^{2}\vartheta }{v_{2}}\right) \left( -\frac{m_{1}^{2}}{\Sigma
_{13}^{(4) }}\right) ^{\Delta }\right.   \notag \\
&&\left. +m_{2}^{2}\sin \theta \left( \frac{\sin \theta \sin ^{2}\vartheta }{%
v_{1}}+\frac{\cos \theta \cos ^{2}\vartheta }{v_{2}}\right) \left( -\frac{%
m_{2}^{2}}{\Sigma _{13}^{(4) }}\right) ^{\Delta }\right] .
\label{F_40-0+}
\end{eqnarray}
It can be shown by explicit calculation that for $\Delta $ in the intersection of the fundamental strip of the 5pt and 4pt amplitudes, i.e. for 
$\Delta \in \langle -1,0\rangle $ , the $\overline{t_{4}}\rightarrow t_{4}$
limit of the individual members of the first sum on the right-hand side of (%
\ref{F_5_full}) either vanishes (for $\left( i,j\right) =\left( 2,5\right) $
or $\left( 3,5\right) $) or it is finite and can be simply obtained by the
replacement $\Sigma _{ij}^{\left( 5\right) }\rightarrow \Sigma _{ij}^{\left(
4\right) }$ and
\begin{equation}
\left( 1-\frac{\Sigma _{ij}^{\left( 5\right) }}{\Sigma _{\left( ij\right)
^{c}}^{\left( 5\right) }}\frac{m_{1}^{2}}{M^{2}}\right) ^{-1}\rightarrow 0.
\end{equation}
Similarly, the individual members of the second sum on the right-hand side
of (\ref{F_5_full}) vanish in the limit $\overline{t_{4}}\rightarrow t_{4}$
for $\left( i,j\right) =\left( 2,4\right) $ or $\left( 3,4\right) $, while
for $\left( i,j\right) =\left( 2,5\right) $ or $\left( 3,5\right) $ the
limit is finite, though the result is less transparent than above. Finally we get
\begin{eqnarray}
\lim_{\overline{t_{4}}\rightarrow t_{4}}\mathcal{F}_{5}^{\left( 0--++\right)
}\left( \Delta ,t_{4},t_{5}\right)  &=&\mathrm{i}\cos \vartheta \sin
2\vartheta \frac{M^{2}}{4\sqrt{2}v_{1}^{2}}\frac{\pi }{\sin \left( \pi
\Delta \right) }  \notag \\
&&\times \left\{ -\frac{1}{v_{1}}\frac{\Sigma _{12}^{(4)
}-\Sigma _{14}^{(4) }}{\Sigma _{24}^{(4) }}\left( -%
\frac{M^{2}}{\Sigma _{24}^{(4) }}\right) ^{\Delta }-\frac{1}{%
v_{1}}\frac{\Sigma _{13}^{(4) }-\Sigma _{14}^{(4) }}{%
\Sigma _{34}^{(4) }}\left( -\frac{M^{2}}{\Sigma _{34}^{\left(
4\right) }}\right) ^{\Delta }\right.   \notag \\
&&\left. -\frac{m_{1}^{2}}{M^{2}}\cos \theta \left( \frac{\cos \theta }{v_{1}%
}+\frac{\sin \theta }{v_{2}}\right) \left[ \left( -\frac{m_{1}^{2}}{\Sigma
_{24}^{(4) }}\right) ^{\Delta }+\left( -\frac{m_{1}^{2}}{\Sigma
_{34}^{(4) }}\right) ^{\Delta }\right] \right.   \notag \\
&&\left. -\frac{m_{2}^{2}}{M^{2}}\cos \theta \left( \frac{\sin \theta }{v_{1}%
}-\frac{\cos \theta }{v_{2}}\right) \left[ \left( -\frac{m_{2}^{2}}{\Sigma
_{24}^{(4) }}\right) ^{\Delta }+\left( -\frac{m_{2}^{2}}{\Sigma
_{34}^{(4) }}\right) ^{\Delta }\right] \right\} .\notag\\
\end{eqnarray}%
Using now (\ref{F4+--+}), (\ref{F_400-+}), (\ref{F_40-0+}) and the relations 
$\Sigma _{12}^{(4) }=\Sigma _{34}^{(4) }$ and $%
\Sigma _{13}^{(4) }=\Sigma _{24}^{(4) }$ which
follow from momentum conservation, we easily confirm the validity of the
conformal soft theorem (\ref{conformal_soft_theorem}). Let us stress that constraining $\Delta$ to the intersection of the fundamental strips was crucial for its validity; beyond this region of $\Delta$, the limit $\bar{t_4}\to  t_4$ need not to be well defined and finite.

%%%%%%%%%%%%%%%%%%%%%%%%%%%%%%%%%%%%%%%%%%%%%%%%%%%
\section{Soft theorems for pion-dilaton model}\label{app:soft}
%%%%%%%%%%%%%%%%%%%%%%%%%%%%%%%%%%%%%%%%%%%

%========================================
\subsection*{Leading conformal soft dilaton}
%========================================

The leading conformal soft theorem corresponds to the residue in
the dilaton conformal weight at $\Delta ^{\left( a\right) }=1$, which can be
expressed by the formal identity%
\begin{equation}
\lim_{\Delta \rightarrow 1}\left( \Delta -1\right) \omega ^{\Delta
-1}=\omega \delta \left( \omega \right) .
\end{equation}%
When transformed to the celestial sphere, the following integral appears on
the right-hand side of the soft theorem~\eqref{soft_dilaton_theorem} as one of the terms in the sum over all massive $\sigma$ particles
\begin{equation}
I_{\Delta }\left( z,\overline{z}\right) \equiv \int \widetilde{\mathrm{d}}%
\widehat{p}\,G_{\Delta }\left( \widehat{p},q\left( z,\overline{z}\right)
\right) \frac{1}{2\widehat{p}\cdot q\left( w,\overline{w}\right) }A\left(
\ldots ,m\varepsilon \widehat{p}^{\left( \sigma \right) },\ldots \right).
\label{I_Delta(z,barz)}
\end{equation}%
Since we are dealing with a massive particle, the celestial transform~\eqref{eq:massive_transf} is implemented by the bulk-to-boundary propagator $G_{\Delta }\left( \widehat{p},q\left( z,\overline{z}\right)\right)$, where $\widehat{p}$ is the unit momentum of the massive $\sigma$ particle whose CCFT dual operator $O_{\Delta }^{\varepsilon \left( \sigma \right) }\left( z,%
\overline{z}\right) $ is inserted at the point $(z,\bar{z})$ on the celestial sphere. Finally, $\omega q(w,\bar{w})$ is the momentum of the soft dilaton ($\omega$ was stripped off as it represents the energy expansion parameter denoted by $t$ in the second line of~\eqref{soft_dilaton_theorem} and the factor of $2$ is for convenience in order to interpret $\left(2\widehat{p}\cdot q\left( w,\overline{w}\right)\right)^{-1}$ as a bulk-to-boundary operator with scaling dimension $\Delta=1$).  

Let us use the completeness
relation (\ref{completeness}) for the bulk-to-boundary propagators to
rewrite $I_{\Delta }\left( z,\overline{z}\right) $ in the form%
\begin{eqnarray}
I_{\Delta }\left( z,\overline{z}\right)  &=&\int \widetilde{\mathrm{d}}%
\widehat{p}\,G_{\Delta }\left( \widehat{p},q\left( z,\overline{z}\right)
\right) \frac{1}{2\widehat{p}\cdot q\left( w,\overline{w}\right) }  \notag \\
&\times &2\int \mathrm{d}\nu \mu \left( \nu \right) \mathrm{d}^{2}x\,G_{1+%
\mathrm{i}\nu }\left( \widehat{p},q\left( x,\overline{x}\right) \right) 
\widetilde{A}\left( \ldots ,\left\{ 1-\mathrm{i}\nu ;x,\overline{x}\right\}
,\ldots \right) ,
\end{eqnarray}%
where we denoted%
\begin{equation}
\widetilde{A}\left( \ldots ,\left\{ 1-\mathrm{i}\nu ;x,\overline{x}\right\}
,\ldots \right) =\int \widetilde{\mathrm{d}}\widehat{p}^{\prime }G_{1-%
\mathrm{i}\nu }\left( \widehat{p}^{\prime },q\left( x,\overline{x}\right)
\right) A\left( \ldots ,m\varepsilon \widehat{p}^{\prime \left( \sigma
\right) },\ldots \right) .
\end{equation}%
In the parametrization 
\begin{equation}
\widehat{p}=\frac{1}{2y}\left( 1+\left\vert u\right\vert ^{2}+y^{2},u+%
\overline{u},\mathrm{i}(u-\overline{u}),\left\vert u\right\vert
^{2}+y^{2}-1\right) ,
\end{equation}%
we get%
\begin{equation}
I_{\Delta }\left( z,\overline{z}\right) =2\int \mathrm{d}\nu \mu \left( \nu
\right) \mathrm{d}^{2}x~\widetilde{A}\left( \ldots ,\left\{ 1-\mathrm{i}\nu
;x,\overline{x}\right\} ,\ldots \right) I_{3}\left( \left\{ \Delta
,z\right\} ,\left\{ 1+i\nu ,x\right\} ,\left\{ 1,w\right\} \right) ,
\end{equation}%
where%
\begin{equation}
I_{3}\left( \left\{ \Delta _{i},\mathbf{z}_{i}\right\} _{i=1}^{3}\right)
=\int_{H_{+}}\widetilde{\mathrm{d}}\widehat{p}\prod_{i=1}^{3}\left( 
\frac{1}{2\widehat{p}\cdot q\left( \mathbf{z}_{i}\right) }\right) ^{\Delta
_{i}}=\int \frac{\mathrm{d}y}{y^{3}}\mathrm{d}^{2}\mathbf{w}%
\prod_{i=1}^{3}\left( \frac{y}{y^{2}+\left\vert \mathbf{w}-\mathbf{z}%
_{i}\right\vert ^{2}}\right) ^{\Delta _{i}}.
\end{equation}%
The latter integral can be formally related to a tree-level contact 3pt Witten diagram in $EAdS_{3}$ dual to a 3pt correlator in a $\mathrm{CFT}_2$ living on the boundary of $EAdS_{3}$. The result
of the integration is standard~\cite{DHoker:1999mqo} (for completeness, the computation is summarized in Appendix \ref{Appendix_Integrals})%
\begin{equation}
I_{3}\left( \left\{ \Delta ,z\right\} ,\left\{ 1+\mathrm{i}\nu ,x\right\}
,\left\{ 1,w\right\} \right) =\frac{C\left( \Delta ,1+\mathrm{i}\nu
,1\right) }{\left\vert w-x\right\vert ^{2-\Delta +\mathrm{i}\nu }\left\vert
z-x\right\vert ^{\Delta +\mathrm{i}\nu }\left\vert z-w\right\vert ^{\Delta -%
\mathrm{i}\nu }}
\end{equation}%
with (cf. \ref{I_3})%
 \begin{equation}
C\left( \Delta ,1+\mathrm{i}\nu ,1\right) =\frac{{\pi }}{2}\frac{\Gamma
\left( 1-\frac{\Delta }{2}+\mathrm{i}\frac{\nu }{2}\right) \Gamma \left( 
\frac{\Delta }{2}+\mathrm{i}\frac{\nu }{2}\right) ^{2}\Gamma \left( \frac{%
\Delta }{2}-\mathrm{i}\frac{\nu }{2}\right) }{\Gamma \left( \Delta \right)
\Gamma \left( 1+\mathrm{i}\nu \right) }.  \label{C_Delta_nu}
\end{equation}

Putting these ingredients together, we get the leading order conformal soft
dilaton theorem in the form%
\begin{eqnarray}
&&\lim_{\Delta \rightarrow 1}\left( \Delta -1\right) \left\langle O_{\Delta
}^{\varepsilon \left( a \right) }\left( w,\overline{w}\right) O_{\Delta
_{1}}^{\varepsilon _{1}\left( f_{1}\right) }\left( z_{1},\overline{z}%
_{1}\right) \ldots O_{\Delta _{n}}^{\varepsilon _{n}\left( f_{n}\right)
}\left( z_{n},\overline{z}_{n}\right) \right\rangle   \notag \\
&=&4\sqrt{\frac{\lambda }{2}}\sum\limits_{i=1}^{n}\delta _{f_{i}\sigma }\int 
\mathrm{d}\nu \mu \left( \nu \right) \mathrm{d}^{2}x\frac{C\left( \Delta ,1+%
\mathrm{i}\nu ,1\right) }{\left\vert w-x\right\vert ^{2-\Delta _{i}+\mathrm{i%
}\nu }\left\vert z_{i}-w\right\vert ^{\Delta _{i}-\mathrm{i}\nu }\left\vert
z_{i}-x\right\vert ^{\Delta _{i}+\mathrm{i}\nu }}  \notag \\
&&\times \left\langle O_{\Delta _{1}}^{\varepsilon _{1}\left( f_{1}\right)
}\left( z_{1},\overline{z}_{1}\right) \ldots O_{1-\mathrm{i}\nu
}^{\varepsilon _{i}\left( f_{i}\right) }\left( x,\overline{x}\right) \ldots
O_{\Delta _{n}}^{\varepsilon _{n}\left( f_{n}\right) }\left( z_{n},\overline{%
z}_{n}\right) \right\rangle .  \label{leading_soft_dilaton_CS2}
\end{eqnarray}%
Let us also calculate the limit $w\rightarrow z_{i}$ when the momentum of the soft dilaton becomes collinear with that of the $i$-th $\sigma$ particle. Starting with (\ref%
{I_Delta(z,barz)}), we get immediately%
\begin{eqnarray}
&&\lim_{w\rightarrow z_{i}}\lim_{\Delta \rightarrow 1}\left( \Delta
-1\right) \left\langle O_{\Delta }^{\varepsilon \left( a \right)
}\left( w,\overline{w}\right) O_{\Delta _{1}}^{\varepsilon _{1}\left(
f_{1}\right) }\left( z_{1},\overline{z}_{1}\right) \ldots O_{\Delta
_{n}}^{\varepsilon _{n}\left( f_{n}\right) }\left( z_{n},\overline{z}%
_{n}\right) \right\rangle   \notag \\
&=&2\sqrt{\frac{\lambda }{2}}\delta _{f_{i}\sigma }\left\langle O_{\Delta
_{1}}^{\varepsilon _{1}\left( f_{1}\right) }\left( z_{1},\overline{z}%
_{1}\right) \ldots O_{\Delta _{i}+1}^{\varepsilon _{i}\left( f_{i}\right)
}\left( z_{i},\overline{z_{i}}\right) \ldots O_{\Delta _{n}}^{\varepsilon
_{n}\left( f_{n}\right) }\left( z_{n},\overline{z}_{n}\right) \right\rangle 
\notag \\
&&+4\sqrt{\frac{\lambda }{2}}\sum\limits_{j=1,j\neq i}^{n}\delta
_{f_{j}\sigma }\int \mathrm{d}\nu \mu \left( \nu \right) \mathrm{d}^{2}x%
\frac{C\left( \Delta ,1+\mathrm{i}\nu ,1\right) }{\left\vert
z_{i}-x\right\vert ^{2-\Delta _{j}+\mathrm{i}\nu }\left\vert
z_{i}-z_{j}\right\vert ^{\Delta _{j}-\mathrm{i}\nu }\left\vert
z_{j}-x\right\vert ^{\Delta _{j}+\mathrm{i}\nu }}  \notag \\
&&\times \left\langle O_{\Delta _{1}}^{\varepsilon _{1}\left( f_{1}\right)
}\left( z_{1},\overline{z}_{1}\right) \ldots O_{1-\mathrm{i}\nu
}^{\varepsilon _{j}\left( f_{j}\right) }\left( x,\overline{x}\right) \ldots
O_{\Delta _{n}}^{\varepsilon _{n}\left( f_{n}\right) }\left( z_{n},\overline{%
z}_{n}\right) \right\rangle .
\end{eqnarray}%
Note that this coincidence limit is regular and that the scaling dimension of the $i$-th sigma particle got shifted by one unit in the first term on the right hand side. 

This result can be also obtained directly from (\ref{leading_soft_dilaton_CS2}). Indeed,
in the limit $w\rightarrow z_{i}$, we get in the $i$-th term on the right hand side
\begin{eqnarray}
&&4\sqrt{\frac{\lambda }{2}}\int \mathrm{d}\nu \mu \left( \nu \right) 
\mathrm{d}^{2}x\frac{C\left( \Delta ,1+\mathrm{i}\nu ,1\right) }{\left\vert
z_{i}-w\right\vert ^{\Delta _{i}-\mathrm{i}\nu }\left\vert w-x\right\vert
^{2-\Delta _{i}+\mathrm{i}\nu }\left\vert z_{i}-x\right\vert ^{\Delta _{i}+%
\mathrm{i}\nu }}\left\langle \ldots O_{1-\mathrm{i}\nu }^{\varepsilon
_{i}\left( f_{i}\right) }\left( x,\overline{x}\right) \ldots \right\rangle 
\notag\\
&&\overset{w\rightarrow z_{i}}{\rightarrow }4\sqrt{\frac{\lambda }{2}}\int 
\mathrm{d}\nu \mu \left( \nu \right) 4\pi \delta \left( \Delta _{i}-\mathrm{i%
}\nu \right) \frac{C\left( \Delta ,1+\mathrm{i}\nu ,1\right) }{\Gamma \left( 
\frac{\Delta }{2}-\mathrm{i}\frac{\nu }{2}\right) }  \notag \\
&&\times \int \mathrm{d}^{2}x\frac{1}{\left\vert z_{i}-x\right\vert
^{2\left( 1+\mathrm{i}\nu \right) }}\left\langle \ldots O_{1-\mathrm{i}\nu
}^{\varepsilon _{i}\left( f_{i}\right) }\left( x,\overline{x}\right) \ldots
\right\rangle \,,
\end{eqnarray}%
where we have used (cf. (\ref{delta_representation_limit}))%
\begin{equation}
\lim_{w\rightarrow z_{i}}\frac{1}{\left\vert z_{i}-w\right\vert ^{\Delta
_{i}-\mathrm{i}\nu }}=\frac{4\pi }{\Gamma \left( \frac{\Delta }{2}-\mathrm{i}%
\frac{\nu }{2}\right) }\delta \left( \Delta _{i}-\mathrm{i}\nu \right) .
\end{equation}%
The $x-$integration can be recognized as the shadow transformation of the
operator $O_{1-\mathrm{i}\nu }^{\varepsilon _{i}\left( f_{i}\right) }\left(
x,\overline{x}\right) $. This transformation satisfies the identity%
\begin{equation}
\int \mathrm{d}^{2}x\frac{1}{\left\vert z-x\right\vert ^{2\left( 2-\Delta
\right) }}O_{\Delta }\left( x,\overline{x}\right) =\frac{\pi }{\Delta -1}%
O_{2-\Delta }\left( z,\overline{z}\right) ,
\end{equation}%
which is valid for the conformal primaries corresponding to the massive
scalar particles. Therefore%
\begin{eqnarray}
&&4\sqrt{\frac{\lambda }{2}}\int \mathrm{d}\nu \mu \left( \nu \right) 
\mathrm{d}^{2}x\frac{C\left( \Delta ,1+\mathrm{i}\nu ,1\right) }{\left\vert
z_{i}-w\right\vert ^{\Delta _{i}-\mathrm{i}\nu }\left\vert w-x\right\vert
^{2-\Delta _{i}+\mathrm{i}\nu }\left\vert z_{i}-x\right\vert ^{\Delta _{i}+%
\mathrm{i}\nu }}\left\langle \ldots O_{1-\mathrm{i}\nu }^{\varepsilon
_{i}\left( f_{i}\right) }\left( x,\overline{x}\right) \ldots \right\rangle  
\notag \\
&&\overset{w\rightarrow z_{i}}{\rightarrow }16\pi \sqrt{\frac{\lambda }{2}}%
\left. \left( \frac{\nu ^{2}}{4\pi ^{3}}\right) \frac{C\left( \Delta ,1+%
\mathrm{i}\nu ,1\right) }{\Gamma \left( \frac{\Delta }{2}-\mathrm{i}\frac{%
\nu }{2}\right) }\left( \frac{\pi }{-\mathrm{i}\nu }\right) \left\langle
\ldots O_{1+\mathrm{i}\nu }^{\varepsilon _{i}\left( f_{i}\right) }\left(
z_{i},\overline{z_{i}}\right) \ldots \right\rangle \right\vert _{\nu
\rightarrow -\mathrm{i}\Delta _{i}}  \notag \\
&&=2\sqrt{\frac{\lambda }{2}}\left\langle \ldots O_{1+\Delta _{i}}^{\varepsilon _{i}\left(
f_{i}\right) }\left( z_{i},\overline{z_{i}}\right) \ldots \right\rangle \,,
\end{eqnarray}%
since 
\begin{equation}
16\pi \sqrt{\frac{\lambda }{2}}\left. \left( \frac{\nu ^{2}}{4\pi ^{3}}%
\right) \frac{C\left( \Delta ,1+\mathrm{i}\nu ,1\right) }{\Gamma \left( 
\frac{\Delta }{2}-\mathrm{i}\frac{\nu }{2}\right) }\left( \frac{\pi }{-%
\mathrm{i}\nu }\right) \right\vert _{\nu \rightarrow -\mathrm{i}\Delta _{i}}=2\sqrt{\frac{\lambda }{2}}.
\end{equation}

%%%%%%%%%%%%%%%%%%%%%%%%%%%%%%%%%%%%%%%%%%%%%%%%%%%%%
\subsection*{Subleading conformal soft dilaton}
%%%%%%%%%%%%%%%%%%%%%%%%%%%%%%%%%%%%%%%%%%%%%%%%%%%%%

Using the usual parameterization of the massless and massive momenta
\begin{equation}
p_{i}^{\left( a ,\pi \right) }=\omega _{i}q\left( \mathbf{z}_{i}\right)
,~~~~p_{j}^{\left( \sigma \right) }=\frac{m}{y_{j}}\left( q\left( \mathbf{w}%
_{i}\right) +y_{j}^{2}n\right) ,
\end{equation}%
where 
\begin{equation}
n=\frac{1}{2}\left( 1,0,0,1\right) ,
\end{equation}%
we can rewrite the subleading soft theorem stated in the last row of~\eqref{soft_dilaton_theorem} in the form (see the last section of this appendix for
details)%
\begin{eqnarray}
\lim_{\omega \rightarrow 0}\left( 1+\omega \frac{\partial }{\partial \omega }%
\right) \mathcal{A}_{n+1}\left( \omega q\left( \mathbf{z}\right)
,\varepsilon _{1}p_{1}^{\left( f_{1}\right) },\ldots ,\varepsilon
_{n}p_{n}^{\left( f_{n}\right) }\right)\notag\\ =-\frac{1}{m}\sqrt{\frac{\lambda }{2}%
}\sum\limits_{i=1}^{n}\left( D_{i}^{\left( f_{i}\right) }+1\right) \mathcal{%
A}_{n}\left( \varepsilon _{1}p_{1}^{\left( f_{1}\right) },\ldots
,\varepsilon _{n}p_{n}^{\left( f_{n}\right) }\right) \label{dil_sub_fin},
\end{eqnarray}%
where we denoted by $D_{i}^{\left( f\right) }$ the following differential
operators acting on the amplitude $\mathcal{A}_{n}$%
\begin{eqnarray}
D_{i}^{\left( a,\pi \right) } &=&\omega _{i}\frac{\partial }{\partial \omega
_{i}}, \\
D_{i}^{\left( \sigma \right) } &=&-y_{i}\frac{\partial }{\partial y_{i}}+%
\frac{2y_{i}^{2}}{y_{i}^{2}+\left\vert \mathbf{w}_{i}-\mathbf{z}\right\vert
^{2}}\left( y_{i}\frac{\partial }{\partial y_{i}}+\left( \mathbf{w}_{i}-%
\mathbf{z}\right) \cdot \boldsymbol{\nabla }_{\mathbf{w}_{i}}\right).
\end{eqnarray}%
The left-hand side of the subleading soft theorem can be identified as the
residue at $\Delta =0$ of the Mellin transformed amplitude, i.e. symbolically%
\begin{eqnarray}
\lim_{\omega \rightarrow 0}\left( 1+\omega \frac{\partial }{\partial \omega }%
\right) \mathcal{A}_{n+1}\left( \omega q\left( \mathbf{z}\right)
,\varepsilon _{1}p_{1}^{\left( f_{1}\right) },\ldots ,\varepsilon
_{n}p_{n}^{\left( f_{n}\right) }\right)\notag\\ =\lim_{\Delta \rightarrow 0}\Delta 
\widetilde{\mathcal{A}}_{n+1}\left( \left\{ \Delta ,\mathbf{z}\right\}
,\varepsilon _{1}p_{1}^{\left( f_{1}\right) },\ldots ,\varepsilon
_{n}p_{n}^{\left( f_{n}\right) }\right) ,
\end{eqnarray}%
where%
\begin{equation}
\widetilde{\mathcal{A}}_{n+1}\left( \left\{ \Delta ,\mathbf{z}\right\}
,\varepsilon _{1}p_{1}^{\left( f_{1}\right) },\ldots ,\varepsilon
_{n}p_{n}^{\left( f_{n}\right) }\right) =\int \mathrm{d}\omega \omega
^{\Delta -1}\mathcal{A}_{n+1}\left( \omega q\left( \mathbf{z}\right)
,\varepsilon _{1}p_{1}^{\left( f_{1}\right) },\ldots ,\varepsilon
_{n}p_{n}^{\left( f_{n}\right) }\right) .
\end{equation}%
Similarly, on the right-hand side of the theorem, we can either use the
formula%
\begin{eqnarray}
&&\int \mathrm{d}\omega _{i}\omega _{i}^{\Delta _{i}-1}D_{i}^{\left( a,\pi
\right) }\mathcal{A}_{n}\left( \varepsilon _{1}p_{1}^{\left( f_{1}\right)
},\ldots ,\varepsilon _{i}p_{i}^{\left( a,\pi \right) }\left( \omega _{i},%
\mathbf{z}_{i}\right) ,\ldots ,\varepsilon _{n}p_{n}^{\left( f_{n}\right)
}\right)  \notag \\
&=&-\Delta _{i}\widetilde{\mathcal{A}}_{n}\left( \varepsilon
_{1}p_{1}^{\left( f_{1}\right) },\ldots ,\left\{ \varepsilon _{i},\Delta
_{i},\mathbf{z}_{i}\right\} ,\ldots ,\varepsilon _{n}p_{n}^{\left(
f_{n}\right) }\right) ,
\end{eqnarray}%
where%
\begin{eqnarray}
&&\widetilde{\mathcal{A}}_{n}\left( \varepsilon _{1}p_{1}^{\left(
f_{1}\right) },\ldots ,\left\{ \varepsilon _{i},\Delta _{i},\mathbf{z}%
_{i}\right\} ,\ldots ,\varepsilon _{n}p_{n}^{\left( f_{n}\right) }\right) 
\notag \\
&=&\int \mathrm{d}\omega _{i}\omega _{i}^{\Delta _{i}-1}\mathcal{A}%
_{n}\left( \varepsilon _{1}p_{1}^{\left( f_{1}\right) },\ldots ,\varepsilon
_{i}p_{i}^{\left( a,\pi \right) }\left( \omega _{i},\mathbf{z}_{i}\right)
,\ldots ,\varepsilon _{n}p_{n}^{\left( f_{n}\right) }\right)\,,
\end{eqnarray}%
or insert the completeness relation for the bulk-to-boundary propagators
and write%
\begin{eqnarray}
&&\mathcal{A}_{n}\left( \varepsilon _{1}p_{1}^{\left( f_{1}\right) },\ldots
,\varepsilon _{i}p_{i}^{\left( \sigma \right) }\left( y_{i},\mathbf{w}%
_{i}\right) ,\ldots ,\varepsilon _{n}p_{n}^{\left( f_{n}\right) }\right) 
\notag \\
&=&2\int \mathrm{d}\nu \mu \left( \nu \right) \mathrm{d}^{2}\mathbf{x}\left( 
\frac{y_{i}}{y_{i}^{2}+\left\vert \mathbf{w}_{i}-\mathbf{x}\right\vert ^{2}}%
\right) ^{1+\mathrm{i}\nu }\widetilde{\mathcal{A}}_{n}\left( \varepsilon
_{1}p_{1}^{\left( f_{1}\right) },\ldots ,\left\{ \varepsilon _{i},1-\mathrm{i%
}\nu ,\mathbf{x}\right\} ,\ldots ,\varepsilon _{n}p_{n}^{\left( f_{n}\right)
}\right) ,\notag\\
\end{eqnarray}%
where%
\begin{eqnarray}
&&\widetilde{\mathcal{A}}_{n}\left( \varepsilon _{1}p_{1}^{\left(
f_{1}\right) },\ldots ,\left\{ \varepsilon _{i},1-\mathrm{i}\nu ,\mathbf{x}%
\right\} ,\ldots ,\varepsilon _{n}p_{n}^{\left( f_{n}\right) }\right)  \notag
\\
&=&\int \frac{\mathrm{d}y}{y^{3}}\mathrm{d}^{2}\mathbf{w}\left( \frac{y}{%
y^{2}+\left\vert \mathbf{w}-\mathbf{x}\right\vert ^{2}}\right) ^{1-\mathrm{i}%
\nu }\mathcal{A}_{n}\left( \varepsilon _{1}p_{1}^{\left( f_{1}\right)
},\ldots ,\varepsilon _{i}p_{i}^{\left( \sigma \right) }\left( y,\mathbf{w}%
\right) ,\ldots ,\varepsilon _{n}p_{n}^{\left( f_{n}\right) }\right) .\notag\\
\end{eqnarray}%
The latter representation of $\mathcal{A}_{n}$ simplifies the action of the
operator $D_{i}^{\left( \sigma \right) }$ since%
\begin{eqnarray}
D_{i}^{\left( \sigma \right) }\left( \frac{y_{i}}{y_{i}^{2}+\left\vert 
\mathbf{w}_{i}-\mathbf{x}\right\vert ^{2}}\right) ^{1+\mathrm{i}\nu }
&=&-\left( 1+\mathrm{i}\nu \right) \left( \frac{y_{i}}{y_{i}^{2}+\left\vert 
\mathbf{w}_{i}-\mathbf{x}\right\vert ^{2}}\right) ^{2+\mathrm{i}\nu }\left( 
\frac{y_{i}}{y_{i}^{2}+\left\vert \mathbf{w}_{i}-\mathbf{z}\right\vert ^{2}}%
\right)  \notag \\
&&\times \left[ \left( \frac{y_{i}}{y_{i}^{2}+\left\vert \mathbf{w}_{i}-%
\mathbf{x}\right\vert ^{2}}\right) ^{-1}\left( \frac{y_{i}}{%
y_{i}^{2}+\left\vert \mathbf{w}_{i}-\mathbf{z}\right\vert ^{2}}\right)
^{-1}-2\left\vert \mathbf{z}-\mathbf{x}\right\vert ^{2}\right] ,\notag\\
\end{eqnarray}%
and therefore, the transformation to the celestial sphere needs the
integrals of the type (see Appendix \ref{Appendix_Integrals})%
\begin{eqnarray}
I_{2}^{(i)} &=&\int \frac{\mathrm{d}y_{i}}{y_{i}^{3}}%
\mathrm{d}^{2}\mathbf{w}_{i}\left( \frac{y_{i}}{y_{i}^{2}+\left\vert \mathbf{%
w}_{i}-\mathbf{z}_{i}\right\vert ^{2}}\right) ^{\Delta _{i}}\left( \frac{%
y_{i}}{y_{i}^{2}+\left\vert \mathbf{w}_{i}-\mathbf{x}\right\vert ^{2}}%
\right) ^{1+\mathrm{i}\nu }  \notag \\
&=&\frac{2\pi ^{2}}{\Delta _{i}-1}\delta \left( \Delta _{i}-1-\mathrm{i}\nu
\right) \frac{1}{\left\vert \mathbf{z}_{i}-\mathbf{x}\right\vert ^{2\Delta
_{i}}},
\end{eqnarray}%
and%
\begin{eqnarray}
I_{3}^{(i)}&=&\int \frac{%
\mathrm{d}y_{i}}{y_{i}^{3}}\mathrm{d}^{2}\mathbf{w}_{i}\left( \frac{y_{i}}{%
y_{i}^{2}+\left\vert \mathbf{w}_{i}-\mathbf{z}_{i}\right\vert ^{2}}\right)
^{\Delta _{i}}\left( \frac{y_{i}}{y_{i}^{2}+\left\vert \mathbf{w}_{i}-%
\mathbf{x}\right\vert ^{2}}\right) ^{2+\mathrm{i}\nu }\left( \frac{y_{i}}{%
y_{i}^{2}+\left\vert \mathbf{w}_{i}-\mathbf{z}\right\vert ^{2}}\right) 
\notag \\
&=&\frac{C\left( \Delta _{i},2+\mathrm{i}\nu ,1\right) }{\left\vert \mathbf{x%
}-\mathbf{z}\right\vert ^{3-\Delta _{i}+\mathrm{i}\nu }\left\vert \mathbf{z}%
_{i}-\mathbf{x}\right\vert ^{1+\Delta _{i}+\mathrm{i}\nu }\left\vert \mathbf{%
z}_{i}-\mathbf{z}\right\vert ^{-1+\Delta _{i}-\mathrm{i}\nu }}, 
\end{eqnarray}%
where the coefficient $C(\Delta_1,\Delta_2,\Delta_3)$ is defined by (\ref{I_3}).
The conformal version of the subleading soft dilaton theorem has then the
form%
\begin{eqnarray}
&&\lim_{\Delta \rightarrow 0}\Delta \left\langle O_{\Delta }^{\varepsilon
\left( a \right) }\left( \mathbf{z}\right) O_{\Delta _{1}}^{\varepsilon
_{1}\left( f_{1}\right) }\left( \mathbf{z}_{1}\right) \ldots O_{\Delta
_{n}}^{\varepsilon _{n}\left( f_{n}\right) }\left( \mathbf{z}_{n}\right)
\right\rangle  \notag \\
&=&-\frac{1}{m}\sqrt{\frac{\lambda }{2}}\left[ n-\sum\limits_{i=1}^{n}%
\left( 1-\delta _{f_{i}\sigma }\right) \Delta _{i}\right] \left\langle
O_{\Delta _{1}}^{\varepsilon _{1}\left( f_{1}\right) }\left( \mathbf{z}%
_{1}\right) \ldots O_{\Delta _{n}}^{\varepsilon _{n}\left( f_{n}\right)
}\left( \mathbf{z}_{n}\right) \right\rangle  \notag \\
&&+\frac{2}{m}\sqrt{\frac{\lambda }{2}}\sum\limits_{i=1}^{n}\delta
_{f_{i}\sigma }\int \mathrm{d}\nu \mu \left( \nu \right) \mathrm{d}^{2}%
\mathbf{x}\left( 1+\mathrm{i}\nu \right) \left[ I_{2}^{(i)}-2\left\vert \mathbf{z}-%
\mathbf{x}\right\vert ^{2}I_{3}^{(i)}\right]  \notag \\
&&\times \left\langle O_{\Delta _{1}}^{\varepsilon _{1}\left( f_{1}\right)
}\left( \mathbf{z}_{1}\right) \ldots O_{1-\mathrm{i}\nu }^{\varepsilon
_{i}\left( f_{i}\right) }\left( \mathbf{x}\right) \ldots O_{\Delta
_{n}}^{\varepsilon _{n}\left( f_{n}\right) }\left( \mathbf{z}_{n}\right)
\right\rangle .
\end{eqnarray}%
This can be further simplified, using properties of the shadow transform interpreted in terms of projectors~\cite{Simmons-Duffin:2012juh}
\begin{eqnarray}
&&2\int \mathrm{d}^{2}\mathbf{x}\mathrm{d}\nu \mu \left( \nu \right) I_{2}\left( \left\{ \Delta _{i},%
\mathbf{z}_{i}\right\} ,\left\{ 1+\mathrm{i}\nu ,\mathbf{x}\right\} \right)
O_{1-\mathrm{i}\nu }^{\varepsilon _{i}\left( \sigma \right) }\left( \mathbf{x%
}\right)  \notag \\
&=&\int \mathrm{d}^{2}\mathbf{x}\mathrm{d}\nu \frac{\nu ^{2}}{2\pi ^{3}}%
\frac{2\pi ^{2}}{\Delta _{i}-1}\left( 1+\mathrm{i}\nu \right) \delta \left(
\Delta _{i}-1-\mathrm{i}\nu \right) \frac{1}{\left\vert \mathbf{z}_{i}-%
\mathbf{x}\right\vert ^{2\Delta _{i}}}O_{1-\mathrm{i}\nu }^{\varepsilon
_{i}\left( \sigma \right) }\left( \mathbf{x}\right)  \notag \\
&=&-\int \mathrm{d}^{2}\mathbf{x}\frac{\left( \Delta _{i}-1\right) ^{2}}{\pi 
}\frac{\Delta _{i}}{\Delta _{i}-1}\frac{1}{\left\vert \mathbf{z}_{i}-\mathbf{%
x}\right\vert ^{2\Delta _{i}}}O_{2-\Delta _{i}}^{\varepsilon _{i}\left(
\sigma \right) }\left( \mathbf{x}\right)  \notag \\
&=&-\frac{\left( \Delta _{i}-1\right) \Delta _{i}}{\pi }\int \mathrm{d}^{2}%
\mathbf{x}\frac{1}{\left\vert \mathbf{z}_{i}-\mathbf{x}\right\vert ^{2\Delta
_{i}}}O_{2-\Delta _{i}}^{\varepsilon _{i}\left( \sigma \right) }\left( 
\mathbf{x}\right)  \notag \\
&=&-\frac{\left( \Delta _{i}-1\right) \Delta _{i}}{\pi }\frac{\pi }{1-\Delta
_{i}}O_{\Delta _{i}}^{\varepsilon _{i}\left( \sigma \right) }\left( \mathbf{x%
}\right) =\Delta _{i}O_{\Delta _{i}}^{\varepsilon _{i}\left( \sigma \right)
}\left( \mathbf{x}\right)
\end{eqnarray}
and thus, the subleading conformal soft dilation theorem reads finally
\begin{eqnarray}
&&\lim_{\Delta \rightarrow 0}\Delta \left\langle O_{\Delta }^{\varepsilon
\left( \sigma \right) }\left( \mathbf{z}\right) O_{\Delta _{1}}^{\varepsilon
_{1}\left( f_{1}\right) }\left( \mathbf{z}_{1}\right) \ldots O_{\Delta
_{n}}^{\varepsilon _{n}\left( f_{n}\right) }\left( \mathbf{z}_{n}\right)
\right\rangle  \notag \\
&=&-\frac{1}{m}\sqrt{\frac{\lambda }{2}}\left[ n-\sum\limits_{i=1}^{n}%
\Delta _{i}\right] \left\langle O_{\Delta _{1}}^{\varepsilon _{1}\left(
f_{1}\right) }\left( \mathbf{z}_{1}\right) \ldots O_{\Delta
_{n}}^{\varepsilon _{n}\left( f_{n}\right) }\left( \mathbf{z}_{n}\right)
\right\rangle  \notag \\
&&-\frac{1}{m\pi ^{3}}\sqrt{\frac{\lambda }{2}}\sum\limits_{i=1}^{n}\delta
_{f_{i}\sigma }\int \mathrm{d}\nu \mathrm{d}^{2}\mathbf{x}\frac{\nu
^{2}\left( 1+\mathrm{i}\nu \right) C\left( \Delta _{i},2+\mathrm{i}\nu
,1\right) }{\left\vert \mathbf{x}-\mathbf{z}\right\vert ^{1-\Delta _{i}+%
\mathrm{i}\nu }\left\vert \mathbf{z}_{i}-\mathbf{x}\right\vert ^{1+\Delta
_{i}+\mathrm{i}\nu }\left\vert \mathbf{z}_{i}-\mathbf{z}\right\vert
^{-1+\Delta _{i}-\mathrm{i}\nu }}  \notag \\
&&\times \left\langle O_{\Delta _{1}}^{\varepsilon _{1}\left( f_{1}\right)
}\left( \mathbf{z}_{1}\right) \ldots O_{1-\mathrm{i}\nu }^{\varepsilon
_{i}\left( f_{i}\right) }\left( \mathbf{x}\right) \ldots O_{\Delta
_{n}}^{\varepsilon _{n}\left( f_{n}\right) }\left( \mathbf{z}_{n}\right)
\right\rangle.
\end{eqnarray}
Note that again, the limit ${\bf{z}}\to{\bf{z}}_i$ is regular. 

%=============================
\subsection*{Derivation of the reparameterized subleading dilaton soft theorem \label{Appendix_subleading_soft_dilaton}}
%=============================

In this last section of the appendix, we will derive~\eqref{dil_sub_fin}. Let us parameterize massive
momenta as 
\begin{equation}
p=\frac{m}{y}\left( q\left( \mathbf{w}\right) +y^{2}n\right) ,
\end{equation}%
where 
\begin{eqnarray}
q\left( \mathbf{w}\right)  &=&\frac{1}{2}\left( 1+\left\vert \mathbf{w}%
\right\vert ^{2},2\mathbf{w,}\left\vert \mathbf{w}\right\vert ^{2}-1\right) ,
\notag \\
n &=&\frac{1}{2}\left( 1,0,0,1\right) ,
\end{eqnarray}%
and $\mathbf{w}$ is treated as a two-dimensional real vector with components 
$w^{a}$. It then follows%
\begin{equation}
\frac{\partial }{\partial y}p=-\frac{m}{y^{2}}\left( q\left( \mathbf{w}%
\right) -y^{2}n\right) ,~~~\frac{\partial }{\partial m}p=\frac{1}{y}\left(
q\left( \mathbf{w}\right) +y^{2}n\right) ,~~~\frac{\partial }{\partial w^{a}}%
p=\frac{m}{y}\varepsilon _{a}\left( \mathbf{w}\right) ,
\end{equation}%
where%
\begin{equation}
\varepsilon _{a}\left( \mathbf{w}\right) =\frac{\partial }{\partial w^{a}}%
q\left( \mathbf{w}\right) 
\end{equation}%
is the polarization vector corresponding to the null momentum $q\left( 
\mathbf{w}\right) $. Note that the vectors $q\left( \mathbf{w}\right)$, $\varepsilon _{a}\left( \mathbf{w}\right) $ and $n$ form a basis. For their
scalar products we have%
\begin{equation}
q\left( \mathbf{w}\right) \cdot \varepsilon _{a}\left( \mathbf{w}\right)
=0,~~\varepsilon _{a}\left( \mathbf{w}\right) \cdot \varepsilon _{b}\left( 
\mathbf{w}\right) =-\delta _{ab},~~q\left( \mathbf{w}\right) \cdot n=\frac{1%
}{2},~\ n\cdot \varepsilon _{a}\left( \mathbf{w}\right) ~=0.
\label{scalar_product_of_basis}
\end{equation}%
The on-shell amplitude is then a function of $y_{i}$, $\mathbf{w}_{i}$ and the
mass $m$. The latter dependence is either explicit (from vertices and
propagators) and also implicit (from the dependence of the massive momenta).

In the subleading soft theorem (cf. (\ref{soft_dilaton_theorem})), we need the operators $p\cdot \partial
/\partial p$ and $q\left( \mathbf{z}\right) \cdot \partial /\partial p$,
where $q\left( \mathbf{z}\right) $ refers to the soft dilaton and $p$ is
the massive momentum. Using the above basis, we can write%
\begin{eqnarray}
q\left( \mathbf{z}\right)  &=&2\left( q\left( \mathbf{z}\right) \cdot
n\right) q\left( \mathbf{w}\right) +2\left( q\left( \mathbf{z}\right) \cdot
q\left( \mathbf{w}\right) \right) n-\left( q\left( \mathbf{z}\right) \cdot
\varepsilon _{a}\left( \mathbf{w}\right) \right) \varepsilon _{a}\left( 
\mathbf{w}\right)  \notag\\
&=&q\left( \mathbf{w}\right) +\left\vert \mathbf{z-w}\right\vert
^{2}n-\left( w^{a}-z^{a}\right) \varepsilon _{a}\left( \mathbf{w}\right) \,,
\end{eqnarray}%
where we have used (\ref{scalar_product_of_basis}) and 
\begin{equation}
q\left( \mathbf{z}\right) \cdot q\left( \mathbf{w}\right) =\frac{1}{2}%
\left\vert \mathbf{z-w}\right\vert ^{2}~,~~q\left( \mathbf{z}\right) \cdot
\varepsilon _{a}\left( \mathbf{w}\right) =\frac{\partial }{\partial w^{a}}%
\frac{1}{2}\left\vert \mathbf{z-w}\right\vert ^{2}=w^{a}-z^{a}.
\end{equation}%
We also find%
\begin{equation}
\frac{\partial }{\partial y}=-\frac{m}{y^{2}}\left( q\left( \mathbf{w}%
\right) -y^{2}n\right) \cdot \frac{\partial }{\partial p},~~\frac{\partial 
}{\partial w^{a}}=\frac{m}{y}\varepsilon _{a}\left( \mathbf{w}\right) \cdot 
\frac{\partial }{\partial p},~~\left( \frac{\partial }{\partial m}\right)
_{P}=\frac{1}{y}\left( q\left( \mathbf{w}\right) +y^{2}n\right) \cdot \frac{%
\partial }{\partial p},
\end{equation}%
where we denoted $\left( \partial /\partial m\right) _{P}$ the derivative
acting only on the implicit dependence of the amplitude on $m$ through its
dependence on the massive momentum $p$. Using these formulas, we get%
\begin{eqnarray}
q\left( \mathbf{z}\right) \cdot \frac{\partial }{\partial p} &=&q\left( 
\mathbf{w}\right) \cdot \frac{\partial }{\partial p}+\left\vert \mathbf{z-w}%
\right\vert ^{2}n\cdot \frac{\partial }{\partial p}-\left(
w^{a}-z^{a}\right) \varepsilon _{a}\left( \mathbf{w}\right) \cdot \frac{%
\partial }{\partial p}  \notag \\
&=&\frac{1}{2m}\frac{\left\vert \mathbf{z-w}\right\vert ^{2}-y^{2}}{y}y\frac{%
\partial }{\partial y}+\frac{1}{2}\frac{\left\vert \mathbf{z-w}\right\vert
^{2}+y^{2}}{y}\left( ~\frac{\partial }{\partial m}\right) _{P}\notag\\&&-\frac{y}{m}%
\left( w^{a}-z^{a}\right) ~\frac{\partial }{\partial w^{a}}.
\end{eqnarray}%
Let us remind%
\begin{equation}
\frac{1}{2p\cdot q\left( \mathbf{z}\right) }=\frac{1}{m}\frac{y}{%
y^{2}+\left\vert \mathbf{z-w}\right\vert ^{2}}.
\end{equation}%
Similarly%
\begin{equation}
p\cdot \frac{\partial }{\partial p}=\frac{m}{y}\left( q\left( \mathbf{w}%
\right) +y^{2}n\right) \cdot \frac{\partial }{\partial p}=m\left( \frac{%
\partial }{\partial m}\right) _{P}\,.
\end{equation}%
On the right-hand side of the subleading soft theorem, we have the operator%
\begin{equation}
D_{i}^{\left( \sigma \right) }\equiv p_{i}\cdot \frac{\partial }{\partial
p_{i}}-2\frac{m}{2q\left( \mathbf{z}\right) \cdot p_{i}}q\left( \mathbf{z}%
\right) \cdot \frac{\partial }{\partial p_{i}}.  \label{D^sigma}
\end{equation}%
Using the above formulas, we get%
\begin{equation}
D_{i}^{\left( \sigma \right) }=-y_{i}\frac{\partial }{\partial y_{i}}+\frac{%
2y_{i}^{2}}{y_{i}^{2}+\left\vert \mathbf{w}_{i}-\mathbf{z}\right\vert ^{2}}%
\left( y_{i}\frac{\partial }{\partial y_{i}}+\left(
w_{i}^{a}-z_{i}^{a}\right) ~\frac{\partial }{\partial w_{i}^{a}}\right). 
\end{equation}%
 Similarly, we need the operator%
\begin{equation*}
D_{i}^{\left( a,\pi \right) }=p_{i}\cdot \frac{\partial }{\partial p_{i}}\,,
\end{equation*}%
where now $p_{i}$ is a massless momentum. Using the usual parametrization 
\begin{equation}
p_{i}=\omega _{i}q\left( \mathbf{z}_{i}\right),
\end{equation}%
we easily get%
\begin{equation}
D_{i}^{\left( a,\pi \right) }=\omega _{i}\frac{\partial }{\partial \omega
_{i}}.  \label{D^api}
\end{equation}%
Finally, the subleading soft theorem can be written as\footnote{%
Here we assume the momentum conservation $\delta -$functions to be included
in the amplitudes $\mathcal{A}_{n}$.}%
\begin{eqnarray}
\lim_{\omega \rightarrow 0}\left( 1+\omega \frac{\partial }{\partial \omega }%
\right) \mathcal{A}_{n+1}\left( \omega q\left( \mathbf{z}\right)
,\varepsilon _{i}p_{1}^{\left( f_{1}\right) },\ldots ,\varepsilon
_{n}p_{n}^{\left( f_{n}\right) }\right)\notag\\ =-\frac{1}{m}\sqrt{\frac{\lambda }{2}%
}\sum\limits_{i=1}^{n}\left( D_{i}^{\left( f_{i}\right) }+1\right) \mathcal{%
A}_{n}\left( \varepsilon _{i}p_{1}^{\left( f_{1}\right) },\ldots
,\varepsilon _{n}p_{n}^{\left( f_{n}\right) }\right) \,,
\end{eqnarray}%
where the amplitude is assumed to be a function of $\omega _{i},\mathbf{z}_{i}$
for massless particles and $y_{j},\mathbf{w}_{j}$ for massive ones and the
operators $D^{\left( f_{i}\right) }$ are given by (\ref{D^sigma}) and (\ref%
{D^api}).

%==============================================
\section{Truncated amplitude of $Z$-theory} \label{Z_trunc}
%===============================================

Our master formula~\eqref{residue_formula} for computing the Mellin transform was derived using a Hankel contour (see Fig.~\ref{fig:hank_cont}). A key step in the argument was the ability to drop the arc at infinity. However, this assumption is not valid for this model, as can be seen from the asymptotic behavior of the integrand~\eqref{eq:Z_mellin} depicted in the left panel of Fig.~\ref{fig:wedge}.  Since we cannot apply the master formula to compute the Mellin transform $\mathcal{G}_{4}\left( \Delta ,t_{4}\right) 
$, let us specialize to a truncated amplitude for which the assumptions about its asymptotic behavior will be fulfilled. The truncation is
inspired by the product representation of the beta function%
\begin{equation}
B\left( a,b\right) =\frac{1}{a+b-1}\prod\limits_{k=1}^{\infty }\frac{%
k\left( a+b+k-2\right) }{\left( a+k-1\right) \left( b+k-1\right) }\equiv
\lim_{n\rightarrow \infty }B_{n}\left( a,b\right) ,
\label{Beta_approximation}
\end{equation}%
where we denoted%
\begin{equation}
B_{n}\left( a,b\right) =\frac{n!\left[ a+b\right] ^{n-1}}{\left[ a\right]
^{n}\left[ b\right] ^{n}},
\end{equation}%
and where 
\begin{equation}
\left[ x\right] ^{n}=x\left( x+1\right) \ldots \left( x+n-1\right) 
\end{equation}%
is the Pochhammer symbol. Let us note that unlike $B\left( a,b\right) $
which has an infinite number of poles and an essential singularity at
infinity, $B_{n}\left( a,b\right) $ is meromorphic and regular at infinity
with a finite number of simple poles. It grasps the same poles as $B\left(
a,b\right) $ up to and including $a,b=-n+1$.

For simplicity, we concentrate on the example of Mellin transform of just
one of the terms in (\ref{Z_x}), namely the third one
\begin{equation*}
Z_{4}^{\left( 3\right) }\left( s,t_{4}\right) =-2B\left( -s,-u\right) .
\end{equation*}
Note that this term is relevant also in a different context since it is
proportional to the ``string form factor'' $F_{I}( s,u)$ for the type I open superstring four-gluon MHV
amplitude
\begin{equation}
A_{4g}\left( 1^{-},2^{-},3^{+},4^{+}\right) =\frac{\langle 12\rangle ^{3}}{%
\langle 23\rangle \langle 34\rangle \langle 41\rangle }F_{I}\left(
s,u\right) ,
\end{equation}
where
\begin{equation}
F_{I}\left( s,u\right) =-\frac{su}{t}B\left( -s,-u\right) .
\end{equation}
Therefore, calculating
\begin{equation}
\mathcal{G}_{4}\left( \Delta ,t_{4}\right) =\int_{0}^{\infty }\mathrm{d}%
s~s^{\Delta -1}B\left( -s,s\left( 1-\frac{1}{t_{4}}\right) \right) ,
\end{equation}%
we have at the same time calculated also the relevant Mellin transform for
the amplitude $A_{4g}\left( 1^{-},2^{-},3^{+},4^{+}\right) $,(cf. \cite{Stieberger:2018edy})%
\begin{equation}
\int_{0}^{\infty }\mathrm{d}s~s^{\Delta -1}F_{I}\left( s,-s\left( 1-\frac{1}{%
t_{4}}\right) \right) =-\frac{t_{4}^{2}}{t_{4}-1}\mathcal{G}_{4}\left(
\Delta +1,t_{4}\right) .
\end{equation}%
Let us insert instead of the full beta function its truncated approximation $%
B_{n}$. Physically, we have truncated the original amplitude to include only
finite number of resonance poles and modified the corresponding residues
(cf. (\ref{Z_n_4_residue}) and (\ref{Z_n_4_residue_0}) below). In accord
with (\ref{Beta_approximation}), this
truncation satisfies 
\begin{equation}
Z_{4}^{\left( 3\right) }\left( s,t_{4}\right) =\lim_{n\rightarrow \infty
}Z_{4,n}^{\left( 3\right) }\left( s,t_{4}\right) .
\end{equation}%
Note also that for $s\rightarrow \infty $ we have a weaker asymptotic fall
off then for the original amplitude, namely%
\begin{equation}
Z_{4,n}^{\left( 3\right) }\left( s,t_{4}\right) =O\left( \frac{1}{s^{n+1}}%
\right) ,
\label{Z_n_4_asymptotics}
\end{equation}%
while for $s\rightarrow 0$ the asymptotics is the same\footnote{%
Note that in the full amplitude $Z_{4}$, the pole at $s=0$ is absent due to
the presence of additional two terms, while in the formfactor $F_{I}$ it is
present and corresponds to the one-gluon exchange.}%
\begin{equation}
Z_{4,n}^{\left( 3\right) }\left( s,t_{4}\right) =O\left( \frac{1}{s}\right) .
\label{Z_n_4_asymptotics0}
\end{equation}%
The Mellin transform\footnote{%
Here we again suppose the poles to be bypassed according to the prescription 
$s\rightarrow s+\mathrm{i}0$.} of $Z_{4,n}^{\left( 3\right) }\left(
s,t_{4}\right) $ is therefore holomorphic in the fundamental strip $%
\left\langle 1,n+1\right\rangle $. The truncated amplitude $Z_{4,n}\left(
s,t_{4}\right) $ has simple poles at 
\begin{eqnarray}
s &=&k,\text{~~~}k=0,1,\ldots ,n-1,  \notag \\
s &=&-k\frac{t_{4}}{t_{4}-1},\text{~~~}k=1,2,\ldots ,n-1,
\end{eqnarray}%
Therefore, according to the general formula (\ref{residue_formula}), the Mellin transform of $%
Z_{4,n}^{\left( 3\right) }$ on the fundamental strip reads 
\begin{eqnarray}
\mathcal{G}_{4,n}\left( \Delta ,t_{4}\right)  &=&-\frac{\pi }{\sin \pi
\Delta }\mathrm{e}^{\mathrm{i}\pi \Delta }\sum\limits_{k=1}^{n-1}k^{\Delta
-1}\mathrm{res}\left( Z_{4,n}^{\left( 3\right) },k\right)   \notag \\
&&+\frac{\pi }{\sin \pi \Delta }\sum\limits_{k=1}^{n-1}\left( \frac{kt_{4}}{%
t_{4}-1}\right) ^{\Delta -1}\mathrm{res}\left( Z_{4,n}^{\left( 3\right) },-%
\frac{kt_{4}}{t_{4}-1}\right) ,  \label{G_4_n_at_fund_strip}
\end{eqnarray}%
where the residues are evaluated at $k=1,2,\ldots ,n-1$  
\begin{eqnarray}
\mathrm{res}\left( Z_{4,n},k\right)  &=&-2\left( -1\right) ^{k}n\left( 
\begin{array}{c}
n-1 \\ 
k%
\end{array}%
\right) \frac{\left[ -kt_{4}^{-1}\right] ^{n-1}}{\left[ k\left(
1-t_{4}^{-1}\right) \right] ^{n}}  \notag \\
\mathrm{res}\left( Z_{4,n},-k\frac{t_{4}}{t_{4}-1}\right)  &=&-2\left(
-1\right) ^{k}n\left( 
\begin{array}{c}
n-1 \\ 
k%
\end{array}%
\right) \frac{\left[ \frac{k}{t_{4}-1}\right] ^{n-1}}{\left[ \frac{kt_{4}}{%
t_{4}-1}\right] ^{n}}\frac{t_{4}}{t_{4}-1}  \label{Z_n_4_residue}
\end{eqnarray}%
and%
\begin{equation}
\mathrm{res}\left( Z_{4,n},0\right) =-\frac{2}{t_{4}-1}\frac{n}{n-1}.
\label{Z_n_4_residue_0}
\end{equation}%
Outside the fundamental strip, the Mellin transform is obtained by analytic
continuation of the formula (\ref{G_4_n_at_fund_strip}). Apparently, the
resulting $\mathcal{G}_{4,n}\left( \Delta ,t_{4}\right) $ has simple poles
for $\Delta =m\in \mathbb{Z}$ with residues
\begin{equation}
\mathrm{res}\left( \mathcal{G}_{4,n},m\right)
=-\sum\limits_{k=1}^{n-1}k^{m-1}\mathrm{res}\left( Z_{4,n}^{\left( 3\right)
},k\right) +\left( -1\right) ^{m}\sum\limits_{k=1}^{n-1}\left( \frac{kt_{4}%
}{t_{4}-1}\right) ^{m-1}\mathrm{res}\left( Z_{4,n}^{\left( 3\right) },-\frac{%
kt_{4}}{t_{4}-1}\right) ,  \label{fake residue}
\end{equation}%
however, not all of them are nonzero. This can be seen as follows. Observe that
the right hand side of (\ref{fake residue}) can be rewritten in terms of the
residue of the rational (since $m\in \mathbb{Z}$) function $s^{m-1}Z_{4,n}^{\left( 3\right) }$ as
\begin{equation}
\mathrm{res}\left( \mathcal{G}_{4,n},m\right) =-\sum\limits_{k=1}^{n-1}%
\mathrm{res}\left( s^{m-1}Z_{4,n}^{\left( 3\right) },k\right)
-\sum\limits_{k=1}^{n-1}\mathrm{res}\left( s^{m-1}Z_{4,n}^{\left( 3\right)
},-\frac{kt_{4}}{t_{4}-1}\right) \,.  \label{G_n_4_residue_sum}
\end{equation}
Note that according to (\ref{Z_n_4_asymptotics0}) and (\ref{Z_n_4_asymptotics}), for $m>1$%
, the function $s^{m-1}Z_{4,n}^{\left( 3\right) }$ is regular for $%
s\rightarrow 0$ and for $m<n+1$, this function has vanishing residue at
infinity. Therefore, for $1<m<n+1$, the right hand side of (\ref{G_n_4_residue_sum}%
) is a minus sum of residues of all the poles of the function $s^{m-1}Z_{4,n}
$, which is zero as a consequence of the residue theorem. Thus%
\begin{equation}
\mathrm{res}\left( \mathcal{G}_{4,n},m\right) =0,~~~1<m<n+1,
\end{equation}%
which means that $\mathcal{G}_{4,n}$ is holomorphic in the fundamental strip 
$\left\langle 1,n+1\right\rangle $ as expected. However, for $m\leq 1$,
there is one pole more, namely $s=0$, which is not included to the sum (\ref%
{G_n_4_residue_sum}), therefore, according to the residue theorem, $\mathrm{%
res}\left( \mathcal{G}_{4,n},m\right) \neq 0$ and equals%
\begin{equation}
\mathrm{res}\left( \mathcal{G}_{4,n},m\right) =\mathrm{res}\left(
s^{m-1}Z_{4,n}^{\left( 3\right) },0\right) =\frac{1}{\left\vert m\right\vert
!}\frac{\partial ^{\left\vert m\right\vert }}{\partial s^{\left\vert
m\right\vert }}Z_{4,n}^{\left( 3\right) }|_{s=0},~~~~m\leq 1.
\end{equation}%
Similarly, for $m\geq n+1$, there is a nontrivial residue at infinity, which
is missing in the sum  (\ref{G_n_4_residue_sum}), therefore $\mathrm{res}%
\left( \mathcal{G}_{4,n},m\right) \neq 0$ again and 
\begin{equation}
\mathrm{res}\left( \mathcal{G}_{4,n},m\right) =\mathrm{res}\left(
s^{m-1}Z_{4,n}^{\left( 3\right) },\infty \right) =\frac{1}{2\pi \mathrm{i}}%
\lim_{R\rightarrow \infty }\int_{\left\vert s\right\vert =R}\mathrm{d}%
s~s^{m-1}Z_{4,n}^{\left( 3\right) }\left( s,t_{4}\right) .
\end{equation}%
As a result, the celestial dual of truncated (part of the) amplitude $%
Z_{4,n}^{\left( 3\right) }$ shares the poles for $\mathrm{Re}\Delta \leq 1$
with the celestial dual of the original one, however, the truncation of
higher resonances results in additional poles in the region $\mathrm{Re}\Delta
\geq n+1$.

However, we cannot claim that the true function $\mathcal{G}_{4}$
can be obtained as a limit of the truncation $\mathcal{G}_{4,n}$ for $n\rightarrow \infty$, since the right-hand side of (\ref{G_4_n_at_fund_strip}) does not converge.

%%%%%%%%%%%%%%%%%%%%%%%%%%%%%%%%%%%%%%%%%%%%
\section{Details on the celestial recursion}\label{celest_recurs_detail}
%%%%%%%%%%%%%%%%%%%%%%%%%%%%%%%%%%%%%%%%%%%%%
Here we include more details on the BCFW-like recursion presented in section~\ref{sec:7}.
\subsubsection*{$Q_{\mathcal{F}}$ time-like and $u_{\mathcal{F}}$ positive:}
Provided $Q_{\mathcal{F}}^{2}>0$ corresponding to $u_{\mathcal{F}}>0$ (blue poles in Fig.~\ref{fig:hank_cont}), we can insert to the
right-hand side of (\ref{residue}) unity in the form%
\begin{equation}
1=\int_{0}^{\infty }\mathrm{d}u~\delta \left( u-u_{\mathcal{F}}\right)
=\int_{0}^{\infty }\mathrm{d}u~Q_{\mathcal{F}}^{2}\delta \left( uQ_{\mathcal{%
F}}^{2}-M_{\mathcal{F}}^{2}\right) =\int_{0}^{\infty }\mathrm{d}u~\delta
\left( uQ_{\mathcal{F}}^{2}-M_{\mathcal{F}}^{2}\right) \frac{M_{\mathcal{F}%
}^{2}}{u},
\end{equation}%
and rewrite the momentum conservation $\delta -$function in (\ref%
{mellin_transform_complete}) as 
\begin{equation*}
\delta ^{(4) }\biggl( \sum\limits_{i=1}^{n}\varepsilon _{i}\sigma
_{i}q( z_{i},\overline{z}_{i}) \biggr) =\delta ^{(4)
}( Q_{\mathcal{F}}+Q_{\mathcal{F}^{c}}) =u^{2}\int_{-\infty}^{\infty }%
\mathrm{d}^{4}Q\delta^{(4)} ( \sqrt{u}Q_{\mathcal{F}}-Q) \delta^{(4)} ( 
\sqrt{u}Q_{\mathcal{F}^{c}}+Q) ,
\end{equation*}
where we denoted
$$
Q_{\mathcal{F}^{c}}=\sum\limits_{j\in \mathcal{F}^{c}}\varepsilon
_{j}\sigma _{j}q\left( z_{j},\overline{z}_{j}\right).
$$ 
The above factorization of momentum conservation delta functions allows us to transform the integrand (in $\sigma$-variables)~\eqref{mellin_transform_complete} to a recursive form
\begin{eqnarray}
&&\delta ^{(4) }\biggl( \sum\limits_{i=1}^{n}\varepsilon
_{i}\sigma _{i}q\left( z_{i},\overline{z}_{i}\right) \biggr) \theta \left(
u_{\mathcal{F}}\right) u_{\mathcal{F}}^{\Delta -1}\mathrm{res}\left(
A_{n}\left( u\right) ,u_{\mathcal{F}}\right)  \notag \\
&=&-\int \mathrm{d}^{4}Q\delta \left( Q^{2}-M_{\mathcal{F}}^{2}\right)
\int_{0}^{\infty }\mathrm{d}u\delta \left( \sqrt{u}Q_{\mathcal{F}}-Q\right)
\delta \left( \sqrt{u}Q_{\mathcal{F}^{c}}+Q\right) ~u^{\Delta +1}  \notag \\
&&\times \sum\limits_{I}A_{\mathcal{F}}\left( \left\{ \sqrt{u}\varepsilon
_{i}\sigma _{i}q\left( z_{i},\overline{z}_{i}\right) \right\} _{i\in 
\mathcal{F}},-Q,I\right)   A_{\mathcal{F}^{c}}\left( \left\{ \sqrt{u}\varepsilon _{j}\sigma
_{j}q\left( z_{j},\overline{z}_{j}\right) \right\} _{j\in \mathcal{F}%
^{c}},Q,I\right) .  \label{delta x residue}
\end{eqnarray}%
Inserting this into the right-hand side of (\ref{mellin_transform_complete}%
), substituting back $u=s^{2}$, $\sigma _{i}=\omega _{i}/s$, \ we get for
the contribution of a given factorization channel $\mathcal{F}$%
\begin{eqnarray}
&&\frac{\pi }{\sin \pi \Delta }\mathrm{e}^{\pi \mathrm{i}\Delta }\int \left[ 
\mathrm{d}\sigma ,\Delta \right] \delta ^{(4) }\left(
\sum\limits_{i=1}^{n}\varepsilon _{i}\sigma _{i}q\left( z_{i},\overline{z}%
_{i}\right) \right) \theta \left( u_{\mathcal{F}}\right) u_{\mathcal{F}%
}^{\Delta -1}\mathrm{res}\left( A_{n}\left( u\right) ,u_{\mathcal{F}}\right)
\notag \\
&=&2\frac{\pi }{\sin \pi \Delta }\mathrm{e}^{\pi \mathrm{i}\Delta
}\sum\limits_{I}\int \mathrm{d}^{4}Q\delta \left( Q^{2}-M_{\mathcal{F}%
}^{2}\right) \int \prod\limits_{i=1}^{n}d\omega _{i}\omega _{i}^{\Delta
_{i}-1}  \notag \\
&&\times \mathcal{A}_{\mathcal{F}}\left( \left\{ \varepsilon _{i}\omega
_{i}q\left( z_{i},\overline{z}_{i}\right) \right\} _{i\in \mathcal{F}%
},-Q,I\right) \mathcal{A}_{\mathcal{F}^{c}}\left( \left\{ \varepsilon
_{j}\omega _{j}q\left( z_{j},\overline{z}_{j}\right) \right\} _{j\in 
\mathcal{F}^{c}},Q,I\right) ,  \label{full residue contribution}
\end{eqnarray}%
where, as usual, 
\begin{equation}
\mathcal{A}_{\mathcal{F}}\left( \left\{ \varepsilon _{i}\omega _{i}q\left(
z_{i},\overline{z}_{i}\right) \right\} _{i\in \mathcal{F}},-Q,I\right)
=\delta ^{(4) }\left( \widetilde{Q}_{\mathcal{F}}-Q\right) A_{%
\mathcal{F}}\left( \left\{ \varepsilon _{i}\omega _{i}q\left( z_{i},%
\overline{z}_{i}\right) \right\} _{i\in \mathcal{F}},-Q,I\right) ,
\end{equation}%
with%
\begin{equation}
\widetilde{Q}_{\mathcal{F}}=\sum\limits_{i\in \mathcal{F}}\varepsilon
_{i}\omega _{i}q\left( z_{i},\overline{z}_{i}\right) ,
\end{equation}%
and similarly for $\mathcal{A}_{\mathcal{F}^{c}}$. Note that these two
amplitudes have just one particle massive (the one exchanged in the factorization channel $\mathcal{F}$ with momentum $Q$, see Fig.~\ref{fig:fact_channel}). Our goal is to factorize the $Q$-integration and turn it into a celestial transform of the massive leg for each of the lower point amplitudes (notice that all the remaining massless legs are already Mellin transformed). The logic is very similar to the derivation of the celestial optical theorem~\cite{Lam:2017ofc}. It will be done in two steps: 
\begin{enumerate}[(i)]
\item rewriting the $Q$-integration as an integral over the upper sheet of the unit mass-shell hyperboloid
\item treating the $Q$-momenta of the amplitudes $\mathcal{A}_{\mathcal{F}},\;\mathcal{A}_{\mathcal{F}^c}$ as independent while inserting a delta function (forcing momentum conservation) in a factorized form based on a completeness relation for the bulk-to-boundary propagator
\end{enumerate}
Turning to step (i), we can write for a general
function $F( Q) $%
\begin{equation}
\int \mathrm{d}^{4}Q\,\delta( Q^{2}-M_{\mathcal{F}}^{2}) F(Q) =M_{\mathcal{F}}^{2}\int \mathrm{d}^{4}\widehat{Q}\,\delta(\widehat{Q}^{2}-1) F( M_{\mathcal{F}}\widehat{Q}) =M_{\mathcal{F}}^{2}\int_{H_{+}\cup H_{-}}\widetilde{\mathrm{d}}\widehat{Q}F(M_{\mathcal{F}}\widehat{Q}),
\end{equation}%
where 
\begin{equation}
\widetilde{\mathrm{d}}\widehat{Q}=\frac{\mathrm{d}^{3}\mathbf{Q}}{\sqrt{%
\mathbf{Q}^{2}+1}}
\end{equation}%
is a Lorentz invariant measure on the unit mass hyperboloid $\widehat{Q}%
^{2}=1$. The latter consists of two disconnected sheets $H_{\pm }$
corresponding to $\mathrm{sign}\left( \widehat{Q}^{0}\right) =\pm 1$. Note that for a general function $f\left( \widehat{Q}\right) $ on $H_{+}\cup H_{-}$ 
\begin{eqnarray}
\int_{H_{\pm }}\widetilde{\mathrm{d}}\widehat{Q}~f\left( \widehat{Q}\right)
&=&\int_{H_{\pm }}\widetilde{\mathrm{d}}\widehat{Q}~f\left( \left( \pm \sqrt{%
\mathbf{Q}^{2}+1},\mathbf{Q}\right) \right) =\int_{H_{\pm }}\widetilde{%
\mathrm{d}}\widehat{Q}~f\left( \left( \pm \sqrt{\mathbf{Q}^{2}+1},-\mathbf{Q}%
\right) \right)  \notag \\
&=&\int_{H_{\mp }}\widetilde{\mathrm{d}}\widehat{Q}~f\left( -\widehat{Q}%
\right) ,
\end{eqnarray}%
and thus%
\begin{equation}\label{eq:to_upper}
\int_{H_{+}\cup H_{-}}\widetilde{\mathrm{d}}\widehat{Q}~f\left( \widehat{Q}%
\right) =\int_{H_{+}}\widetilde{\mathrm{d}}\widehat{Q}~\left[ f\left( 
\widehat{Q}\right) +f\left( -\widehat{Q}\right) \right] .
\end{equation}%

\begin{figure}[ht]
\centering
\setlength\tabcolsep{1.3em}
\begin{tabular}{lrl}
\begin{adjustbox}{width=.4\linewidth}
\begin{tikzpicture}[baseline=0,auto,node distance=2cm,thick,main node/.style={circle,fill=blue!20,draw,font=\sffamily\Large\bfseries}, scale=2, pin distance=1.5cm]
	    \tikzset{every pin edge/.append style={black, thick}}
	    \node[main node,pin=90:{$p_4,+$},pin=135:{$p_3,+$},pin=-135:{$p_2,-$},pin=-90:{$p_1,-$}] (F) at (0,0) {$\mathcal{F}_{\phantom{c}}$};
	    \node[main node,pin=90:{$p_4,-$},pin=45:{$p_3,-$},pin=-45:{$p_2,+$},pin=-90:{$p_1,+$}] (Fc) at (2,0) {$\mathcal{F}_c$};
	    \path [-] (F) edge node[below] {\scalebox{0.85}{$\begin{aligned} &\\Q=(-p_1-p_2+p_3+p_4)\end{aligned}$}} (Fc);
\end{tikzpicture}
\end{adjustbox}
&
\begin{adjustbox}{width=.2\linewidth}
\begin{tikzpicture}[baseline=0]
\tikzset{>=latex}	
\draw [help lines,line width=2pt,-] (-3,-3) -- (3,3);
\draw [help lines,line width=2pt,-] (-3,3) -- (3,-3);

	\draw [->,line width=3pt,red] (0,0)--(2,2) node[right,scale=2,red] {$p_4$};
\draw [->,line width=3pt,red] (0,0)--(-2,2) node[left,scale=2,red] {$p_3$};
\draw [->,line width=3pt,blue] (0,0)--(-1,-1) node[left,scale=2,blue] {$-p_1$};
\draw [->,line width=3pt,blue] (0,0)--(1,-1) node[right,scale=2,blue] {$-p_2$};
\draw [->,line width=5pt,black] (0,0)--(0,2) node[yshift=.4cm,scale=2,black] {$Q$};
\end{tikzpicture}
\end{adjustbox}
&
\begin{adjustbox}{width=.2\linewidth}
\begin{tikzpicture}[baseline=0]
\tikzset{>=latex}	
\draw [help lines,line width=2pt,-] (-3,-3) -- (3,3);
\draw [help lines,line width=2pt,-] (-3,3) -- (3,-3);

\draw [->,line width=3pt,red] (0,0)--(1,1) node[right,scale=2,red] {$p_4$};
\draw [->,line width=3pt,red] (0,0)--(-1,1) node[left,scale=2,red] {$p_3$};
\draw [->,line width=3pt,blue] (0,0)--(-2,-2) node[xshift=-0.5cm,yshift=0.6cm,scale=2,blue] {$-p_1$};
\draw [->,line width=3pt,blue] (0,0)--(2,-2) node[xshift=0.4cm,yshift=0.6cm,scale=2,blue] {$-p_2$};
\draw [->,line width=5pt,black] (0,0)--(0,-2) node[yshift=-.4cm,scale=2,black] {$Q$};
\end{tikzpicture}
\end{adjustbox}
\end{tabular}
\caption{\emph{Left}: A particular factorization channel $\mathcal{F}=(-,-,+,+)$. \emph{Right}: Mellin transforming massless states attached to the channel $\mathcal{F}$ means ``summing'' (weighted) amplitudes over null momenta on half light-rays (with fixed directions). Two special (kinematically admissible) configurations belonging to the Mellin integration domain are depicted. For the left one, the exchanged momentum $Q$ is future time-like, while for the right one it is directed to the past.}
\label{fig:recursion_sum}
\end{figure}
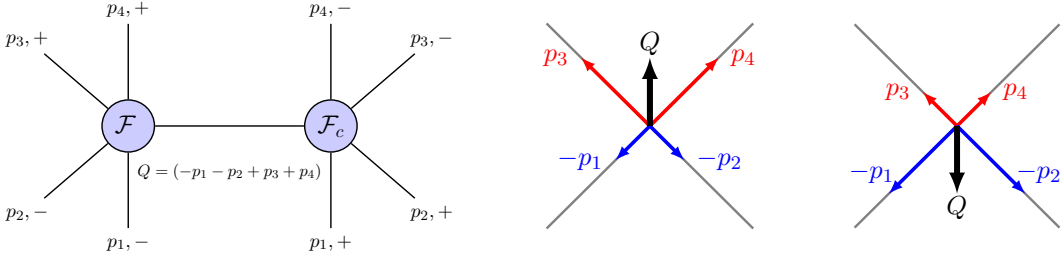

Step (ii) consists of exploiting the completeness relation~\cite{Costa:2014kfa,Pasterski:2017kqt} for the bulk-to-boundary propagator~\eqref{eq:massive_transf},%
\begin{equation}
2\int \mathrm{d}\nu \mu \left( \nu \right) \mathrm{d}^{2}z\left( \frac{1}{2%
\widehat{p}\cdot q\left( z,\overline{z}\right) }\right) ^{1-\mathrm{i}\nu
}\left( \frac{1}{2\widehat{p}^{\prime }\cdot q\left( z,\overline{z}\right) }%
\right) ^{1+\mathrm{i}\nu }=\widehat{p}^{0}\delta ^{\left( 3\right) }\left( 
\widehat{\mathbf{p}}-\widehat{\mathbf{p}}^{\prime }\right) \equiv \widetilde{%
\delta }\left( \widehat{p}-\widehat{p}^{\prime }\right)\,,
\label{completeness}
\end{equation}
where
\begin{equation}
\mu \left( \nu \right) =\frac{\nu ^{2}}{4\pi ^{3}}.
\end{equation}
We now apply steps (i), (ii) explained above to the relevant part of~\eqref{full residue contribution}
\begin{align}
&\int \mathrm{d}^{4}Q\delta ( Q^{2}-M_{\mathcal{F}}^{2}) 
\mathcal{A}_{\mathcal{F}}( \ldots ,-Q,I) \mathcal{A}_{\mathcal{F}%
^{c}}( \ldots ,Q,I)  \notag\\
&\stackrel{\mathclap{\eqref{eq:to_upper}}}{=}\,\,\,M_{\mathcal{F}}^{2}\int_{H_{+}}\widetilde{\mathrm{d}}\widehat{Q}\left[ 
\mathcal{A}_{\mathcal{F}}( \ldots ,-M_{\mathcal{F}}\widehat{Q},I) 
\mathcal{A}_{\mathcal{F}^{c}}( \ldots ,M_{\mathcal{F}}\widehat{Q}%
,I) +( \widehat{Q}\rightarrow -\widehat{Q}) \right]  \notag\\
&=M_{\mathcal{F}}^{2}\int_{H_{+}}\widetilde{\mathrm{d}}\widehat{Q}~%
\widetilde{\mathrm{d}}\widehat{Q^{\prime }}~\widetilde{\delta }( 
\widehat{Q}-\widehat{Q^{\prime }})  \notag\\
&\times \left[ \mathcal{A}_{\mathcal{F}}( \ldots ,-M_{\mathcal{F}}%
\widehat{Q},I) \mathcal{A}_{\mathcal{F}^{c}}( \ldots ,M_{\mathcal{%
F}}\widehat{Q^{\prime }},I) +( \widehat{Q}\rightarrow -\widehat{Q}%
,\widehat{Q^{\prime }}\rightarrow -\widehat{Q^{\prime }}) \right]  \notag\\
&\stackrel{\mathclap{\eqref{completeness}}}{=}\,\,\,2M_{\mathcal{F}}^{2}\int \mathrm{d}\nu \mu ( \nu ) \mathrm{d}^{2}z\Big\{\Big[ \int_{H_{+}}\widetilde{\mathrm{d}}\widehat{Q}~~\Bigl( \frac{1}{2%
\widehat{Q}\cdot q( z,\overline{z}) }\Bigr) ^{1-\mathrm{i}\nu }%
\mathcal{A}_{\mathcal{F}}( \ldots ,-M_{\mathcal{F}}\widehat{Q},I)
\Bigg]  \notag\\
&\times\Big[ \int_{H_{+}}\widetilde{\mathrm{d}}\widehat{Q^{\prime }}%
\Bigl( \frac{1}{2\widehat{Q^{\prime }}\cdot q( z,\overline{z}) }%
\Bigr) ^{1+\mathrm{i}\nu }\mathcal{A}_{\mathcal{F}^{c}}( \ldots ,M_{%
\mathcal{F}}\widehat{Q^{\prime }},I) \Big]  + ( \widehat{Q}\rightarrow -\widehat{Q},\widehat{Q^{\prime }}%
\rightarrow -\widehat{Q^{\prime }}) \Big\}.
\end{align}%
In the final expression, the integrals over $\widehat{Q}$ and $\widehat{Q^{\prime }}$ enclosed within square brackets are the massive celestial transformations~\eqref{eq:massive_transf} of the corresponding on-shell particle to the celestial sphere primary $O_{1-\mathrm{i}\nu }^{\varepsilon ,I}\left( z,%
\overline{z}\right) $ 
\begin{equation}
\int_{H_{+}}\widetilde{\mathrm{d}}\widehat{Q}~~\left( \frac{1}{2\widehat{Q}%
\cdot q\left( z,\overline{z}\right) }\right) ^{1-\mathrm{i}\nu }\mathcal{A}_{%
\mathcal{F}}\left( \ldots ,\varepsilon M_{\mathcal{F}}\widehat{Q},I\right) =%
\widetilde{\mathcal{A}}_{\mathcal{F}}\left( \ldots ,\left\{ \varepsilon
,\left( 1-\mathrm{i}\nu \right) ,z,\overline{z},I\right\} \right) 
\end{equation}%
and similarly for $\mathcal{A}_{\mathcal{F}^{c}}$. Thus we managed to completely transform the lower point amplitudes $\mathcal{A}_{\mathcal{F}}$ and $\mathcal{A}_{\mathcal{F}^c}$ to the celestial sphere. Collecting partial results, the contribution to the full celestial correlator~\eqref{mellin_transform_complete} from a bulk factorization channel $\mathcal{F}$ with internal time-like momentum takes the form
\begin{eqnarray}
&&\frac{\pi }{\sin \pi \Delta }\mathrm{e}^{\pi \mathrm{i}\Delta }\int \left[ 
\mathrm{d}\sigma \right] \delta ^{(4) }\left(
\sum\limits_{i=1}^{n}\varepsilon _{i}\sigma _{i}q\left( z_{i},\overline{z}%
_{i}\right) \right) \theta \left( u_{\mathcal{F}}\right) u_{\mathcal{F}%
}^{\Delta -1}\mathrm{res}\left( A_{n}\left( u\right) ,u_{\mathcal{F}}\right) 
\notag \\
&=&4\frac{\pi }{\sin \pi \Delta }\mathrm{e}^{\pi \mathrm{i}\Delta }M_{%
\mathcal{F}}^{2}\sum\limits_{I}\int \mathrm{d}\nu \mu \left( \nu \right) 
\mathrm{d}^{2}z\left( \widetilde{\mathcal{A}}_{\mathcal{F}}^{-,1-\mathrm{i}%
\nu }\widetilde{\mathcal{A}}_{\mathcal{F}^{c}}^{+,1+\mathrm{i}\nu }+%
\widetilde{\mathcal{A}}_{\mathcal{F}}^{+,1-\mathrm{i}\nu }\widetilde{%
\mathcal{A}}_{\mathcal{F}^{c}}^{-,1+\mathrm{i}\nu }\right) ,
\end{eqnarray}
where we have abbreviated
\begin{eqnarray}
\widetilde{\mathcal{A}}_{\mathcal{F}}^{\varepsilon ,1-\mathrm{i}\nu }
&\equiv &\widetilde{\mathcal{A}}_{\mathcal{F}}\left( \left\{ \varepsilon
_{i},\Delta _{i},z_{i},\overline{z}_{i}\right\} ,\left\{ \varepsilon ,\left(
1-\mathrm{i}\nu \right) ,z,\overline{z},I\right\} \right)   \notag \\
\widetilde{\mathcal{A}}_{\mathcal{F}^{c}}^{\eta ,1+\mathrm{i}\nu } &\equiv &%
\widetilde{\mathcal{A}}_{\mathcal{F}^{c}}\left( \left\{ \varepsilon
_{i},\Delta _{i},z_{i},\overline{z}_{i}\right\} ,\left\{ \eta ,\left( 1+%
\mathrm{i}\nu \right) ,z,\overline{z},I\right\} \right) \,,
\end{eqnarray}%
or in a more compact form%
\begin{eqnarray}
&&\frac{\pi }{\sin \pi \Delta }\mathrm{e}^{\pi \mathrm{i}\Delta }\int \left[ 
\mathrm{d}\sigma ,\Delta \right] \delta ^{(4) }\left(
\sum\limits_{i=1}^{n}\varepsilon _{i}\sigma _{i}q\left( z_{i},\overline{z}%
_{i}\right) \right) \theta \left( u_{\mathcal{F}}\right) u_{\mathcal{F}%
}^{\Delta -1}\mathrm{res}\left( A_{n}\left( u\right) ,u_{\mathcal{F}}\right) 
\notag \\
&=&4\frac{\pi }{\sin \pi \Delta }\mathrm{e}^{\pi \mathrm{i}\Delta }M_{%
\mathcal{F}}^{2}\sum\limits_{I}\int \mathrm{d}\nu \mu \left( \nu \right) 
\mathrm{d}^{2}z\sum\limits_{\varepsilon =\pm }\widetilde{\mathcal{A}}_{%
\mathcal{F}}^{\varepsilon ,1-\mathrm{i}\nu }\widetilde{\mathcal{A}}_{%
\mathcal{F}^{c}}^{-\varepsilon ,1+\mathrm{i}\nu }.
\label{positive_pole_factorization}
\end{eqnarray}%
Let us remark that for a generic factorization channel $\mathcal{F}$ both terms in the last sum contribute, as the $Q^0$ component of the time-like exchanged momentum $Q$ can change a sign when the massless states attached to channel $\mathcal{F}$ are Mellin transformed. This situation is illustrated in Fig.~\ref{fig:recursion_sum}. For special factorization channels, it can however happen that $Q$ is either future or past-oriented (for the whole Mellin integration domain of the massless particles) and thus one of the two terms in the sum vanishes.  

\subsubsection*{$Q_{\mathcal{F}}$ space-like and $u_{\mathcal{F}}$ negative:}
The case $Q_{\mathcal{F}}^2<0$ corresponding to $u_{\mathcal{F}}<0$ (red poles in Fig.~\ref{fig:hank_cont}) can be treated similarly. The residue at $u_{%
\mathcal{F}}$ can be expressed in factorized form as%
\begin{eqnarray}
\mathrm{res}\left( A_{n}\left( u\right) ,u_{\mathcal{F}}\right)  &=&-\frac{1%
}{Q_{\mathcal{F}}^{2}}\sum\limits_{I}A_{\mathcal{F}}\left( \left\{ \mathrm{i}%
\sqrt{\left\vert u_{\mathcal{F}}\right\vert }\varepsilon _{i}\sigma
_{i}q\left( z_{i},\overline{z}_{i}\right) \right\} _{i\in \mathcal{F}},-%
\mathrm{i}\sqrt{\left\vert u_{\mathcal{F}}\right\vert }Q_{\mathcal{F}%
},I\right)   \notag \\
&&\times A_{\mathcal{F}^{c}}\left( \left\{ \mathrm{i}\sqrt{\left\vert u_{%
\mathcal{F}}\right\vert }\varepsilon _{j}\sigma _{j}q\left( z_{j},\overline{z%
}_{j}\right) \right\} _{j\in \mathcal{F}^{c}},\mathrm{i}\sqrt{\left\vert u_{%
\mathcal{F}}\right\vert }Q_{\mathcal{F}},I\right) ,
\end{eqnarray}%
where now $Q_{\mathcal{F}}^{2}<0$. Applying similar manipulations as before,
now with $u_{\mathcal{F}}=-\left\vert u_{\mathcal{F}}\right\vert $, we get
(cf. (\ref{delta x residue}))%
\begin{eqnarray}
&&\delta ^{(4) }\left( \sum\limits_{i=1}^{n}\varepsilon
_{i}\sigma _{i}q\left( z_{i},\overline{z}_{i}\right) \right) \theta \left(
-u_{\mathcal{F}}\right) u_{\mathcal{F}}^{\Delta -1}\mathrm{res}\left(
A_{n}\left( u\right) ,u_{\mathcal{F}}\right)   \notag \\
&=&-e^{\mathrm{i}\pi \Delta }\int \mathrm{d}^{4}Q\delta \left( Q^{2}+M_{%
\mathcal{F}}^{2}\right) \int_{0}^{\infty }\mathrm{d}u\delta \left( \sqrt{u}%
Q_{\mathcal{F}}-Q\right) \delta \left( \sqrt{u}Q_{\mathcal{F}^{c}}+Q\right)
~u^{\Delta +1}  \notag \\
&&\times \sum\limits_{I}A_{\mathcal{F}}\left( \left\{ \mathrm{i}\sqrt{u}%
\varepsilon _{i}\sigma _{i}q\left( z_{i},\overline{z}_{i}\right) \right\}
_{i\in \mathcal{F}},-\mathrm{i}\sqrt{u}Q_{\mathcal{F}},I\right)   \notag \\
&&\times A_{\mathcal{F}^{c}}\left( \left\{ \mathrm{i}\sqrt{u}\varepsilon
_{j}\sigma _{j}q\left( z_{j},\overline{z}_{j}\right) \right\} _{j\in 
\mathcal{F}^{c}},\mathrm{i}\sqrt{u}Q_{\mathcal{F}},I\right) ,
\end{eqnarray}%
and (cf. (\ref{full residue contribution})) 
\begin{eqnarray}
&&-\frac{\pi }{\sin \pi \Delta }\mathrm{e}^{-\pi \mathrm{i}\Delta }\int %
\left[ \mathrm{d}\sigma \right] \delta ^{(4) }\left(
\sum\limits_{i=1}^{n}\varepsilon _{i}\sigma _{i}q\left( z_{i},\overline{z}%
_{i}\right) \right) \theta \left( -u_{\mathcal{F}}\right) u_{\mathcal{F}%
}^{\Delta -1}\mathrm{res}\left( A_{n}\left( u\right) ,u_{\mathcal{F}}\right) 
\notag \\
&=&2\frac{\pi }{\sin \pi \Delta }\sum\limits_{I}\int \mathrm{d}^{4}Q\delta
\left( Q^{2}+M_{\mathcal{F}}^{2}\right) \int \prod\limits_{i=1}^{n}d\omega
_{i}\omega _{i}^{\Delta _{i}-1}  \notag \\
&&\times \mathcal{A}_{\mathcal{F}}\left( \left\{ \mathrm{i}\varepsilon
_{i}\omega _{i}q\left( z_{i},\overline{z}_{i}\right) \right\} _{i\in },-%
\mathrm{i}Q,I\right) \mathcal{A}_{\mathcal{F}^{c}}\left( \left\{ \mathrm{i}%
\varepsilon _{j}\omega _{j}q\left( z_{j},\overline{z}_{j}\right) \right\}
_{j\in \mathcal{F}^{c}},\mathrm{i}Q,I\right) .
\end{eqnarray}%
The latter formula can be understood as an analytic continuation of (\ref%
{full residue contribution}) to purely imaginary momenta.
Note that it can be rewritten in the form, 
\begin{align}
&-\frac{\pi }{\sin \pi \Delta }\mathrm{e}^{-\pi \mathrm{i}\Delta }\int \left[ \mathrm{d}\sigma \right] \delta ^{(4) }\left(
\sum\limits_{i=1}^{n}\varepsilon _{i}\sigma _{i}q\left( z_{i},\overline{z}%
_{i}\right) \right) \theta \left( -u_{\mathcal{F}}\right) u_{\mathcal{F}%
}^{\Delta -1}\mathrm{res}\left( A_{n}\left( u\right) ,u_{\mathcal{F}}\right) \notag\\
&=\frac{\pi }{\sin \pi \Delta }M_{\mathcal{F}}^{2}\sum\limits_{I}%
\int_{H_{dS}}\widetilde{\mathrm{d}}\widehat{Q}\int
\prod\limits_{i=1}^{n}d\omega _{i}\omega _{i}^{\Delta _{i}-1}\delta \left(
Q_{\mathcal{F}}-M_{\mathcal{F}}\widehat{Q}\right) \delta \left( Q_{\mathcal{F%
}^{c}}+M_{\mathcal{F}}\widehat{Q}\right)\notag \\
&\phantom{=}\times A_{\mathcal{F}}\left( \left\{ \mathrm{i}\varepsilon _{i}\omega
_{i}q\left( z_{i},\overline{z}_{i}\right) \right\} _{i\in },-\mathrm{i}M_{%
\mathcal{F}}\widehat{Q},I\right) A_{\mathcal{F}^{c}}\left( \left\{ \mathrm{i}%
\varepsilon _{j}\omega _{j}q\left( z_{j},\overline{z}_{j}\right) \right\}
_{j\in \mathcal{F}^{c}},\mathrm{i}M_{\mathcal{F}}\widehat{Q},I\right),
\label{negative_pole_residue}
\end{align}%
where%
\begin{equation*}
\widetilde{\mathrm{d}}\widehat{Q}~=2\mathrm{d}^{4}\widehat{Q}~\delta \left( 
\widehat{Q}^{2}+1\right) 
\end{equation*}%
is the Lorentz invariant measure concentrated on the $H_{dS}$ hyperboloid
defined by the constraint $\widehat{Q}^{2}=-1$ in momentum space.
Suppose that we can write (as claimed in \cite{Melton:2021kkz}) the following completeness
relation
\begin{equation}
2\int \mathrm{d}\nu \mu \left( \nu \right) \mathrm{d}^{2}z\left( \frac{1}{2%
\widehat{p}\cdot q\left( z,\overline{z}\right) }\right) ^{1-\mathrm{i}\nu
}\left( \frac{1}{2\widehat{p}^{\prime }\cdot q\left( z,\overline{z}\right) }%
\right) ^{1+\mathrm{i}\nu }=-\widetilde{\delta }\left( \widehat{p}-\widehat{p%
}^{\prime }\right) \,,
\end{equation}
for $\widehat{p},\widehat{p}^{\prime }\in H_{dS}$, then we can proceed further. Namely, introducing for the amplitude $A\left( \left\{ \varepsilon _{i}p_{i}\right\}
_{i=1}^{m},\varepsilon Q\right) $ with $n$ massless particles with momenta $p_{i}^{2}=0$
and one massive particle with momentum $Q^{2}=M^{2}$ the following generalized celestial
transform 
\begin{multline}
\widetilde{\mathcal{A}}_{\mathcal{F}}^{C}\left( \left\{ \varepsilon
_{i},\Delta _{i},z_{i},\overline{z}_{i}\right\} _{i=1}^{m},\left\{
\varepsilon ,\Delta ,z,\overline{z}\right\} \right) \equiv\int_{H_{dS}}
\widetilde{\mathrm{d}}\widehat{Q}\int_{0}^{\infty
}\prod\limits_{i=1}^{m}d\omega _{i}\omega _{i}^{\Delta _{i}-1}\delta \biggl(
\sum\limits_{j=1}^{m}\varepsilon _{i}\omega _{i}q\left( z_{i},\overline{z}%
_{i}\right) +\varepsilon M\widehat{Q}\biggr)\\
\times \left( \frac{1}{2\widehat{Q}\cdot q\left( z,\overline{z}\right) }%
\right) ^{\Delta }A\left( \left\{ \mathrm{i}\varepsilon _{i}\omega
_{i}q\left( z_{i},\overline{z}_{i}\right) \right\} _{i=1}^{m},\mathrm{i}%
\varepsilon M\widehat{Q}\right),
\end{multline}%
we can then rewrite (\ref{negative_pole_residue}) formally in the form analogous to (\ref{positive_pole_factorization}), namely%
\begin{eqnarray}
&&-\frac{\pi }{\sin \pi \Delta }\mathrm{e}^{-\pi \mathrm{i}\Delta }\int %
\left[ \mathrm{d}\sigma \right] \delta ^{(4) }\left(
\sum\limits_{i=1}^{n}\varepsilon _{i}\sigma _{i}q\left( z_{i},\overline{z}%
_{i}\right) \right) \theta \left( -u_{\mathcal{F}}\right) u_{\mathcal{F}%
}^{\Delta -1}\mathrm{res}\left( A_{n}\left( u\right) ,u_{\mathcal{F}}\right)
\notag \\
&=&2\frac{\pi }{\sin \pi \Delta }M_{\mathcal{F}}^{2}\sum\limits_{I}\int 
\mathrm{d}\nu \mu \left( \nu \right) \mathrm{d}^{2}z\widetilde{\mathcal{A}}_{%
\mathcal{F}}^{C,-,1-\mathrm{i}\nu }\widetilde{\mathcal{A}}_{\mathcal{F}%
^{c}}^{C,+,1+\mathrm{i}\nu }\,,
\end{eqnarray}%
with%
\begin{eqnarray}
\widetilde{\mathcal{A}}_{\mathcal{F}}^{C,\varepsilon ,\Delta ,I} &\equiv &%
\widetilde{\mathcal{A}}^C_{\mathcal{F}}\left( \left\{ \varepsilon
_{i},\Delta _{i},z_{i},\overline{z}_{i}\right\} _{i\in \mathcal{F}},\left\{
\varepsilon ,\Delta ,z,\overline{z},I\right\} \right)   \notag \\
\widetilde{\mathcal{A}}_{\mathcal{F}^{c}}^{C,\varepsilon ,\Delta ,I} &\equiv
&\widetilde{\mathcal{A}}^C_{\mathcal{F}^{c}}\left( \left\{
\varepsilon _{i},\Delta _{i},z_{i},\overline{z}_{i}\right\} _{j\in \mathcal{F%
}^{c}},\left\{ \varepsilon ,\Delta ,z,\overline{z},I\right\} \right) .
\end{eqnarray}%
To summarize, we get the following formula for the Mellin integral (see last line in~\eqref{mellin_transform_complete}) 
\begin{eqnarray}
&&\int_{0}^{\infty }\mathrm{d}u~u^{\Delta -1}A_{n}\left( \sqrt{u}\varepsilon
_{1}\sigma _{1}q\left( z_{1},\overline{z}_{1}\right) ,\ldots ,\sqrt{u}%
\varepsilon _{n}\sigma _{n}q\left( z_{n},\overline{z}_{n}\right) \right)  
\notag \\
&=&2\frac{\pi }{\sin \pi \Delta }M_{\mathcal{F}}^{2}\int \mathrm{d}\nu \mu
\left( \nu \right) \mathrm{d}^{2}z\sum\limits_{\mathcal{F}}\left[ \theta
\left( Q_{\mathcal{F}}^{2}\right) \sum\limits_{I}\mathrm{e}^{\pi \mathrm{i}%
\Delta }\left( \widetilde{\mathcal{A}}_{\mathcal{F}}^{-,1-\mathrm{i}\nu ,I}%
\widetilde{\mathcal{A}}_{\mathcal{F}^{c}}^{+,1+\mathrm{i}\nu ,I}+\widetilde{%
\mathcal{A}}_{\mathcal{F}}^{+,1-\mathrm{i}\nu ,I}\widetilde{\mathcal{A}}_{%
\mathcal{F}^{c}}^{-,1+\mathrm{i}\nu ,I}\right) \right.   \notag \\
&&\left. +\theta \left( -Q_{\mathcal{F}}^{2}\right) \sum\limits_{I}%
\widetilde{\mathcal{A}}_{\mathcal{F}}^{C,-,1-\mathrm{i}\nu ,I}\widetilde{%
\mathcal{A}}_{\mathcal{F}^{c}}^{C,+,1+\mathrm{i}\nu ,I}\right], 
\end{eqnarray}
which expresses a massless celestial $n$-pt amplitude in terms of lower point celestial amplitudes with one massive external leg corresponding to a massive factorization channel.

%===============================================

\section{Mellin transform of distributions and the complex $\protect\delta -$%
function \label{complex delta}}

Here we give a brief overview concerning the Mellin transform of functions
with empty fundamental strip, which can be understood as a Mellin transform
in the distributional sense. We follow the review articles \cite{Flajolet} and \cite{Bertrand}
where further details can be found.
An alternative discussion of this issue can be found in \cite{Donnay:2020guq}.

Let first remind the inverse formula for the Mellin transform which reads%
\begin{equation}
f\left( t\right) =\frac{1}{2\pi \mathrm{i}}\int_{c-\mathrm{i}\infty }^{c+%
\mathrm{i}\infty }\mathrm{d}s~t^{-s}\mathcal{M}(f,s).  \label{inverse Mellin}
\end{equation}%
Here $c\in 
%TCIMACRO{\U{211d} }%
%BeginExpansion
\mathbb{R}
%EndExpansion
$ satisfies $\alpha <c<\beta $ where $\langle \alpha ,\beta \rangle \equiv
\left\{ s\in 
%TCIMACRO{\U{2102} }%
%BeginExpansion
\mathbb{C}
%EndExpansion
|\alpha <{\rm{Re}}\,s<\beta \right\} $ is the fundamental strip defined by the
asymptotic of the function $f$%
\begin{equation}
f\left( t\right) =\left\{ 
\begin{array}{c}
O\left( t^{-\alpha }\right) ,~t\rightarrow 0 \\ 
O\left( t^{-\beta }\right) ,~t\rightarrow \infty 
\end{array}%
\right. 
\end{equation}%
and we abbreviated $\mathcal{M}(f,s)$ to be the Mellin transform of the
(locally integrable) function~$f$
\begin{equation}
\mathcal{M}(f,s)\equiv \int_{0}^{\infty }\mathrm{d}t~t^{s-1}f\left( t\right)
.  \label{direct Mellin}
\end{equation}%
For $s$ within the fundamental strip, the integral is well-defined and $\mathcal{M}(f,s)$ is holomorphic. Provided for $t\rightarrow 0$ we have an
asymptotic expansion
\begin{equation}
f\left( t\right) =\sum\limits_{-n<-k\leq \alpha }f_{k}^{0}t^{k}+O\left(
t^{n}\right) ,
\end{equation}
then $\mathcal{M}(f,s)$ can be continued outside the fundamental strip to
the strip $\langle -n,\beta \rangle $ to a meromorphic function with simple
poles placed at $s=-k$ with residues%
\begin{equation}
\mathrm{res}\left( \mathcal{M}(f,s),-k\right) =f_{k}^{0}.
\end{equation}%
Similarly, for the asymptotic for $t\rightarrow \infty $%
\begin{equation}
f\left( t\right) =\sum\limits_{\beta \leq -k<-m}f_{k}^{\infty
}t^{k}+O\left( t^{m}\right) 
\end{equation}%
then $\mathcal{M}(f,s)$ can be continued to the strip $\langle \alpha
,-m\rangle $ to a meromorphic function with simple poles at $s=-k$ with
residue%
\begin{equation}
\mathrm{res}\left( \mathcal{M}(f,s),-k\right) =-f_{k}^{\infty }
\end{equation}%
Throughout this paper we always tacitly assumed $\mathcal{M}(f,s)$ to be
identified with such analytic continuation of the integral (\ref{direct
Mellin}) in the maximal possible region $\langle -n,-m\rangle $.

The inverse formula (\ref{inverse Mellin}) allows to prove the Parseval
identity for the Mellin transform in the form
\begin{equation}
\int_{0}^{\infty }\mathrm{d}t~f\left( t\right) g\left( t\right) =\frac{1}{%
2\pi \mathrm{i}}\int_{c-\mathrm{i}\infty }^{c+\mathrm{i}\infty }\mathrm{d}s~%
\mathcal{M}(f,s)\mathcal{M}(g,1-s)  \label{Parseval}
\end{equation}
where real $c$ is supposed to be in the fundamental strip of $f$ while $1-c$
belongs to the fundamental strip of $g$. This formula can be used to define
the Mellin transform also in the cases when $f\left( t\right) $ has an empty
fundamental strip or even when it is a distribution. In this case, we
restrict $g\left( t\right) $ to be an infinitely smooth test function with
compact support in the open interval $\left( 0,\infty \right) $, and
interpret the left-hand side of (\ref{Parseval}) as a pairing of the
distribution $f$ with this test function. Note, that $g\left( t\right) $ has
a Mellin transform with fundamental strip $\langle -\infty ,\infty \rangle $%
, i.e. $\mathcal{M}(g,s)$ is an entire function defined everywhere by the
integral (\ref{direct Mellin}). The right hand side of (\ref{Parseval}) can
be then interpreted as the pairing of the Mellin transform $\mathcal{M}(f,s)$
of the distribution $f$ with a test function $\mathcal{M}(g,1-s)$, which
corresponds to the Mellin transform of the test function $g$.

Let us give some examples relevant to the main text of this article. Let $%
f\left( t\right) =t^{z}$ where $z\in
\mathbb{C}$. 
Clearly, such a function has an empty fundamental strip, i.e. the usual
integral definition of its Mellin transform fails. Nevertheless, the left
hand side of (\ref{Parseval}) is well-defined for the test functions
described above and therefore we can define a pairing of $f$ with the test
function $g$ as%
\begin{equation}
\int_{0}^{\infty }\mathrm{d}t~f\left( t\right) g\left( t\right)
=\int_{0}^{\infty }\mathrm{d}t~t^{z}g\left( t\right) =\mathcal{M}(g,z+1).
\label{delta pairing}
\end{equation}%
On the other hand, this should be the result of the pairing of the
distribution representing the Mellin transform of $f\left( t\right) =t^{z}$
with the test function $\mathcal{G}\left( s\right) \equiv \mathcal{M}(g,1-s)$%
. The result of this pairing is $\mathcal{M}(g,z+1)=\mathcal{G}\left(
-z\right) $. Therefore, the Mellin transform of $f\left( t\right) =t^{z}$
acts effectively as a complex $\delta -$function sitting at $s=-z$. To keep
in touch with the formula (\ref{Parseval}), i.e. writing this statement in
the \textquotedblleft physical\textquotedblright\ notation%
\begin{equation}
\frac{1}{2\pi \mathrm{i}}\int_{c-\mathrm{i}\infty }^{c+\mathrm{i}\infty }%
\mathrm{d}s~\mathcal{M}(t^{z},s)\mathcal{G}\left( s\right) =\mathcal{G}%
\left( -z\right) 
\end{equation}%
it is natural to write down formally%
\begin{equation}
\mathcal{M}(t^{z},s)=\int_{0}^{\infty }\mathrm{d}t~t^{s-1}t^{z}=2\pi \mathrm{%
~}\delta \left( s+z\right) .
\end{equation}%
Namely in this sense, we understand such formulas used in the main text.

Note, that for particular values of $s$ and $z$ this formula makes sense
even in a more traditional way. Namely for $s$ and $z$ pure imaginary, $s=%
\mathrm{i}\sigma $ and $z=\mathrm{i}\zeta $, we get%
\begin{equation}
\int_{0}^{\infty }\mathrm{d}t~t^{\mathrm{i}\sigma -1}t^{\mathrm{i}\zeta
}=\int_{-\infty }^{\infty }\mathrm{d}u~\mathrm{e}^{\mathrm{i}u\left( \sigma
+\zeta \right) }=2\pi \delta \left( \sigma +\zeta \right) ,
\end{equation}%
where we have substituted $t=\mathrm{e}^{u}$.

The complex $\delta -$function in the above sense can be also represented
differently, as a limit of more regular distributions. Namely, we will show,
that in the sense of distributions, formally 
\begin{equation}
2\pi \mathrm{~}\delta \left( s+z\right) =\lim_{\mu \rightarrow \infty
}\left( \frac{\mu ^{2(s+z)}}{s+z}-\frac{\mu ^{-2(s+z)}}{s+z}\right) ,
\label{limit delta}
\end{equation}%
where the action of the distributions $\pm \mu ^{\pm 2(s+z)}/\left(
s+z\right) $ on the test functions $\mathcal{M}(g,1-s)$ is given by the
right hand side of (\ref{Parseval}), i.e. we have to calculate the integrals 
\begin{equation}
I_{\pm }\left( z;c\right) \equiv \pm \lim_{\mu \rightarrow \infty }\frac{1}{%
2\pi \mathrm{i}}\int_{c-\mathrm{i}\infty }^{c+\mathrm{i}\infty }\mathrm{d}s~%
\frac{\mu ^{\pm 2(s+z)}}{s+z}\mathcal{M}(g,1-s)
\end{equation}%
for arbitrary $c\in 
%TCIMACRO{\U{211d} }%
%BeginExpansion
\mathbb{R}
%EndExpansion
$. This can be done as follows. Note, that we can interpret $\pm \mu ^{\pm
2(s+z)}/\left( s+z\right) $ as a result of the Mellin transforms 
\begin{eqnarray}
\frac{\mu ^{2(s+z)}}{s+z} &=&\int_{0}^{\infty }\mathrm{d}t~t^{s-1}\chi
_{\langle 0,\mu ^{2}\rangle }\left( t\right) t^{z} \\
-\frac{\mu ^{-2(s+z)}}{s+z} &=&\int_{0}^{\infty }\mathrm{d}t~t^{s-1}\chi
_{\langle 0,\mu ^{2}\rangle }\left( \frac{1}{t}\right)
t^{z}=\int_{0}^{\infty }\mathrm{d}t~t^{s-1}\chi _{\langle 1/\mu ^{2},\infty
\rangle }\left( t\right) t^{z}
\end{eqnarray}%
where $\chi _{\langle a,b\rangle }\left( t\right) $ is the characteristic
function of the interval $\langle a,b\rangle $. The fundamental strips for
the above Mellin transforms are $\langle -{\rm{Re}}\,z,\infty \rangle $ and $%
\langle -\infty ,-{\rm{Re}}\,z\rangle $ respectively. Therefore, using the
Parseval identity with $c>-{\rm{Re}}\,z$ we get%
\begin{eqnarray}
I_{+}\left( z;c\right)  &=&\lim_{\mu \rightarrow \infty }\frac{1}{2\pi 
\mathrm{i}}\int_{c-\mathrm{i}\infty }^{c+\mathrm{i}\infty }\mathrm{d}s~%
\mathcal{M}\left( \chi _{\langle 0,\mu ^{2}\rangle }\left( t\right)
t^{z},s\right) \mathcal{M}(g,1-s)  \nonumber \\
&=&\lim_{\mu \rightarrow \infty }\int_{0}^{\infty }\mathrm{d}t\chi _{\langle
0,\mu ^{2}\rangle }\left( t\right) t^{z}g\left( t\right) =\lim_{\mu
\rightarrow \infty }\int_{0}^{\mu ^{2}}\mathrm{d}t~t^{z}g\left( t\right) =%
\mathcal{M}(g,z+1).  \label{c>-Rez}
\end{eqnarray}
On the other hand, for $c^{\prime }<-{\rm{Re}}\,z$, we can use the residue
theorem in the form%
\begin{eqnarray}
\frac{1}{2\pi \mathrm{i}}\int_{c^{\prime }-\mathrm{i}\infty }^{c^{\prime }+%
\mathrm{i}\infty }\mathrm{d}s~\frac{\mu ^{2(s+z)}}{s+z}\mathcal{M}(g,1-s) &=&%
\frac{1}{2\pi \mathrm{i}}\int_{c-\mathrm{i}\infty }^{c+\mathrm{i}\infty }%
\mathrm{d}s~\frac{\mu ^{2(s+z)}}{s+z}\mathcal{M}(g,1-s)  \nonumber \\
&&-\mathrm{res}\left( \frac{\mu ^{2(s+z)}}{s+z}\mathcal{M}(g,1-s),-z\right) 
\end{eqnarray}%
where $c>-{\rm{Re}}\,z$. Since $\mathcal{M}(g,1-s)$ is entire function, we get%
\begin{equation}
\mathrm{res}\left( \frac{\mu ^{2(s+z)}}{s+z}\mathcal{M}(g,1-s),-z\right) =%
\mathcal{M}(g,z+1)
\end{equation}%
and using the previous result (\ref{c>-Rez}), we get finally for $c^{\prime
}<-{\rm{Re}}\,z$%
\begin{equation}
I_{+}\left( z;c^{\prime }\right) =0.
\end{equation}%
Finally\footnote{Here $\theta$ is the Heaviside step function.}%
\begin{equation}
I_{+}\left( z;c\right) =\theta \left( c+{\rm{Re}}\,z\right) \mathcal{M}(g,z+1)
\label{I+}
\end{equation}%
Similarly, using Parseval identity for $c<-{\rm{Re}}\,z$ we get%
\begin{eqnarray}
I_{-}\left( z;c\right)  &=&\lim_{\mu \rightarrow \infty }\frac{1}{2\pi 
\mathrm{i}}\int_{c-\mathrm{i}\infty }^{c+\mathrm{i}\infty }\mathrm{d}s~%
\mathcal{M}\left( \chi _{\langle 1/\mu ^{2},\infty \rangle }\left( t\right)
t^{z},s\right) \mathcal{M}(g,1-s)  \nonumber \\
&=&\lim_{\mu \rightarrow \infty }\int_{0}^{\infty }\mathrm{d}t\chi _{\langle
1/\mu ^{2},\infty \rangle }\left( t\right) t^{z}g\left( t\right) =\lim_{\mu
\rightarrow \infty }\int_{1/\mu ^{2}}^{\infty }\mathrm{d}t~t^{z}g\left(
t\right) =\mathcal{M}(g,z+1)\nonumber\\
\end{eqnarray}%
while using the same arguments as before, for $c^{\prime}>-{\rm{Re}}\,z$, we
arrive at%
\begin{equation}
I_{-}\left( z;c^{\prime }\right) =0
\end{equation}%
and therefore%
\begin{equation}
I_{-}\left( z;c\right) =\theta \left( -c-{\rm{Re}}\,z\right) \mathcal{M}%
(g,z+1)
\label{I-}
\end{equation}%
Therefore, independently on the choice of $c\in 
%TCIMACRO{\U{211d} }%
%BeginExpansion
\mathbb{R}
%EndExpansion
$%
\begin{equation}
\lim_{\mu \rightarrow \infty }\frac{1}{2\pi \mathrm{i}}\int_{c-\mathrm{i}%
\infty }^{c+\mathrm{i}\infty }\mathrm{d}s\left( \frac{\mu ^{2(s+z)}}{s+z}-%
\frac{\mu ^{-2(s+z)}}{s+z}\right) \mathcal{M}(g,1-s)=\mathcal{M}(g,z+1),
\label{eq:limit_delta}
\end{equation}%
which is exactly the result of the pairing of the complex delta function
with $\mathcal{M}(g,1-s)$, cf. (\ref{delta pairing}).

The results (\ref{I+}) and (\ref{I-}) can be also formally interpreted as follows (cf. also \cite{Donnay:2020guq}). Let us
define the distributions $\delta _{\mu ,>}(s+z)$ and $\delta _{\mu ,<}(s+z)$
the pairing of which with the entire test functions $\mathcal{M}(g,z+1)=%
\mathcal{G}\left( -z\right) $ is in the physical notation%
\begin{equation}
\frac{1}{\mathrm{i}}\int_{c-\mathrm{i}\infty }^{c+\mathrm{i}\infty }\mathrm{d%
}s~\delta _{\mu ,\gtrless }(s+z)\mathcal{G}\left( s\right) \equiv \pm \frac{1%
}{2\pi \mathrm{i}}\int_{c-\mathrm{i}\infty }^{c+\mathrm{i}\infty }\mathrm{d}%
s~\frac{\mu ^{\pm 2(s+z)}}{s+z}\mathcal{M}(g,1-s)
\label{delta><}
\end{equation}%
where now the integration contour is fixed as $c\gtrless -{\rm{Re}}z$. Then,
using the results discussed above, we can write the following limit in the
sense of distributions 
\begin{equation}
\lim_{\mu \rightarrow \infty }\delta _{\mu ,\gtrless }(s+z)=\delta \left(
s+z\right) .
\end{equation}
Note however, that the fixing of the contour on the right-hand side of (\ref{delta><}) is an integral part of the definition of the distributions $\delta _{\mu ,\gtrless }(s+z)$. Without such a prescription,  the formal limit \begin{equation}
 \pm\lim_{\mu \rightarrow \infty } \frac{\mu ^{\pm 2(s+z)}}{s+z}=2\pi \delta \left(
s+z\right) 
\label{formal delta limit}
\end{equation} 
does not make rigorous sense. 
Whenever we use this formal limit for general complex $z$ in the main text, we  tacitly assume the appropriate choice of the integration contour.

Again, for purely imaginary $s$ and $z$, the formal formula (\ref%
{limit delta}) can be interpreted more traditionally (cf. eq. (3.7) in \cite{Atanasov:2021cje}), e.g. %
\begin{eqnarray}
\lim_{\mu \rightarrow \infty }\frac{\mu ^{2\left( s+z\right) }}{s+z} &=&%
\frac{1}{\Gamma \left( s+z+1\right) }\lim_{\mu \rightarrow \infty}\lim_{\varepsilon\rightarrow 0}
\int_{0}^{\infty }\mathrm{d}t~t^{s+z-1+\varepsilon}\mathrm{e}^{-t/\mu ^{2}}  \nonumber
\\
&=&\frac{1}{\Gamma \left( \mathrm{i}(\sigma +\zeta )+1\right) }%
\int_{0}^{\infty }\mathrm{d}t~t^{\mathrm{i}(\sigma +\zeta )-1}=2\pi \mathrm{~%
}\delta \left( \sigma +\zeta \right) .
\end{eqnarray}
This formally corresponds to taking $c$ infinitesimally above $-{\rm{Re}}\,z$ in the formula (\ref{eq:limit_delta}).

%==================================================================

\bibliography{ref}
\bibliographystyle{JHEP}
\end{document}